\documentclass[fleqn,usenatbib]{config/mnras}
\usepackage{bm}
\usepackage{fix-cm}
\usepackage{natbib}
\usepackage[utf8]{inputenc}
\usepackage{hyperref}
\usepackage{array} 
\usepackage{arydshln}
\usepackage{booktabs}
\usepackage{subcaption}
\defcitealias{pozzetti2010}{P10}
\usepackage{newtxtext,newtxmath}
\usepackage{placeins}
\usepackage[T1]{fontenc}
\usepackage{supertabular,booktabs}
\usepackage{caption} 
\DeclareRobustCommand{\VAN}[3]{#2}
\let\VANthebibliography\thebibliography
\def\thebibliography{\DeclareRobustCommand{\VAN}[3]{##3}\VANthebibliography}

\usepackage{graphicx}	
\usepackage{amsmath}	
\usepackage{tcolorbox}

\newcommand{\kevin}{\citet{Hainline2026}}
\newcommand{\prosp}{{\texttt{Prospector}} }

\usepackage{adjustbox}
\usepackage{textgreek}
\usepackage{xargs}
\usepackage{xspace}

\newcommand{\orcidsymb}[2]{\href{http://orcid.org/#2}{#1\adjustbox{trim={-.15\width} {0\height} {-.15\width} {0\height},clip}{\includegraphics[height=10pt]{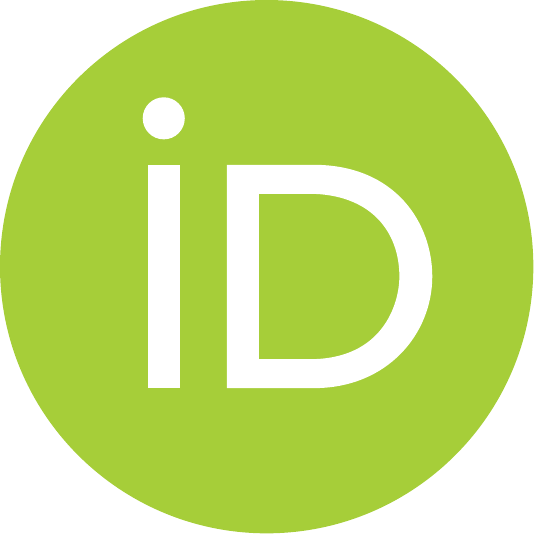}}}}
\usepackage{etoolbox}
\makeatletter
\newcommand\sendemail[4]{
\edef\@tempa{mailto:#1?subject=#2&body=#3 }%
\edef\@tempb{\expandafter\html@spaces\@tempa\@empty}%
\href{\@tempb}{#4}}

\catcode\%=11
\def\html@spaces#1 #2{#1
\catcode\%=14
\makeatother

\hypersetup{colorlinks = true,linkcolor  = blue}
\renewcommand{\sectionautorefname}{Section}
\renewcommand{\subsectionautorefname}{Section}
\renewcommand{\subsubsectionautorefname}{Section}

\title[JADES Stellar Population]{JWST Advanced Deep Extragalactic Survey (JADES) Data Release 5: stellar population catalogue for galaxies in GOODS-N and GOODS-S}

\author[Q. Duan et al.]{
\parbox{\textwidth}{
\orcidsymb{Qiao Duan}{0009-0009-8105-4564}$^{\hyperlink{aff1}{1},\hyperlink{aff2}{2}}$\thanks{E-mail: qd231@cam.ac.uk},
\orcidsymb{Sandro Tacchella}{0000-0002-8224-4505}$^{\hyperlink{aff1}{1},\hyperlink{aff2}{2}}$,
\orcidsymb{Benjamin D.~Johnson}{0000-0002-9280-7594}$^{\hyperlink{aff3}{3}}$,
\orcidsymb{Brant Robertson}{0000-0002-4271-0364}$^{\hyperlink{aff4}{4}}$, \newline
\orcidsymb{Charlotte Simmonds}{0000-0003-4770-7516}$^{\hyperlink{aff1}{1},\hyperlink{aff5}{5}}$,
\orcidsymb{William M. Baker}{0000-0003-0215-1104}$^{\hyperlink{aff6}{6}}$,
\orcidsymb{Andrew J.\ Bunker}{0000-0002-8651-9879}$^{\hyperlink{aff7}{7}}$,
\orcidsymb{Stefano Carniani}{0000-0002-6719-380X}$^{\hyperlink{aff8}{8}}$, \newline
\orcidsymb{Courtney Carreira}{0000-0001-6301-3667}$^{\hyperlink{aff4}{4}}$,
\orcidsymb{Stéphane Charlot}{0000-0003-3458-2275}$^{\hyperlink{aff9}{9}}$,
\orcidsymb{Jacopo Chevallard}{0000-0002-8651-9879}$^{\hyperlink{aff7}{7}}$,
\orcidsymb{Emma Curtis-Lake}{0000-0002-9551-0534}$^{\hyperlink{aff10}{10}}$, \newline
\orcidsymb{A. Lola Danhaive}{0000-0002-9708-9958}$^{\hyperlink{aff1}{1},\hyperlink{aff2}{2}}$, 
\orcidsymb{Francesco D'Eugenio}{0000-0003-2388-8172}$^{\hyperlink{aff1}{1},\hyperlink{aff2}{2}}$,
\orcidsymb{Daniel J.~Eisenstein}{0000-0002-2929-3121}$^{\hyperlink{aff4}{4}}$,
\orcidsymb{Sophia Geris}{0000-0003-2709-2913}$^{\hyperlink{aff1}{1},\hyperlink{aff2}{2}}$, \newline
\orcidsymb{Kevin N. Hainline}{0000-0003-4565-8239}$^{\hyperlink{aff11}{11}}$,
\orcidsymb{Ryan Hausen}{0000-0002-8543-761X}$^{\hyperlink{aff12}{12}}$,
\orcidsymb{Jakob M. Helton}{0000-0003-4337-6211}$^{\hyperlink{aff13}{13}}$,
\orcidsymb{Patricia Iglesias-Navarro}{0009-0009-8959-2404}$^{\hyperlink{aff14}{14},\hyperlink{aff15}{15}}$, \newline
\orcidsymb{Yuki Isobe}{0000-0001-7730-8634}$^{\hyperlink{aff1}{1},\hyperlink{aff2}{2},\hyperlink{aff16}{16}}$, 
\orcidsymb{Zhiyuan Ji}{0000-0001-7673-2257}$^{\hyperlink{aff11}{11}}$,
\orcidsymb{Maria Koller}{0009-0000-1950-9112}$^{\hyperlink{aff1}{1},\hyperlink{aff2}{2}}$,
\orcidsymb{Tobias J. Looser}{0000-0002-3642-2446}$^{\hyperlink{aff3}{3}}$,
\orcidsymb{Roberto Maiolino}{0000-0002-4985-3819}$^{\hyperlink{aff1}{1},\hyperlink{aff2}{2},\hyperlink{aff17}{17}}$, \newline
\orcidsymb{Robert G. Pascalau}{0000-0001-9820-5773}$^{\hyperlink{aff1}{1},\hyperlink{aff2}{2}}$,
\orcidsymb{Pablo G. P\'erez-Gonz\'alez}{0000-0003-4528-5639}$^{\hyperlink{aff18}{18}}$,
\orcidsymb{D\'avid Pusk\'as}{0000-0001-8630-2031}$^{\hyperlink{aff1}{1},\hyperlink{aff2}{2}}$, 
\orcidsymb{Marcia Rieke}{0000-0002-7893-6170}$^{\hyperlink{aff11}{11}}$, \newline
\orcidsymb{Bruno Rodríguez Del Pino}{0000-0001-5171-3930}$^{\hyperlink{aff18}{18}}$,
\orcidsymb{Pierluigi Rinaldi}{0000-0002-5104-8245}$^{\hyperlink{aff19}{19}}$,
\orcidsymb{Jan Scholtz}{0000-0001-6010-6809}$^{\hyperlink{aff1}{1},\hyperlink{aff2}{2}}$,
\orcidsymb{Amanda Stoffers}{0009-0009-5464-3558}$^{\hyperlink{aff1}{1},\hyperlink{aff2}{2}}$, \newline
\orcidsymb{Yang Sun}{0000-0001-6561-9443}$^{\hyperlink{aff11}{11}}$, 
\orcidsymb{James A. A. Trussler}{0000-0002-9081-2111}$^{\hyperlink{aff3}{3}}$,
\orcidsymb{Hannah \"Ubler}{0000-0003-4891-0794}$^{\hyperlink{aff20}{20}}$,
\orcidsymb{Christina C. Williams}{0000-0003-2919-7495}$^{\hyperlink{aff21}{21}}$,
\orcidsymb{Lily Whitler}{0000-0003-1432-7744}$^{\hyperlink{aff1}{1},\hyperlink{aff2}{2}}$,
\orcidsymb{Zihao Wu}{0000-0002-8876-5248}$^{\hyperlink{aff3}{3}}$, 
\orcidsymb{Yongda Zhu}{0000-0003-3307-7525}$^{\hyperlink{aff11}{11}}$
}
\vspace{0.4cm}
\\
\parbox{\textwidth}{
\hypertarget{aff1}{$^{1}$}Kavli Institute for Cosmology, University of Cambridge, Madingley Road, Cambridge, CB3 0HA, United Kingdom\\
\hypertarget{aff2}{$^{2}$}Cavendish Laboratory - Astrophysics Group, University of Cambridge, 19 JJ Thomson Avenue, Cambridge, CB3 0HE, United Kingdom\\
\hypertarget{aff3}{$^{3}$}Center for Astrophysics $|$ Harvard \& Smithsonian, 60 Garden St., Cambridge, MA 02138, USA\\
\hypertarget{aff4}{$^{4}$}Department of Astronomy and Astrophysics, University of California, Santa Cruz, 1156 High Street, Santa Cruz, CA 95064, USA \\
\hypertarget{aff5}{$^{5}$} Departamento de Astronomía, Universidad de Chile, Camino El Observatorio 1515, Las Condes, Santiago \\
\hypertarget{aff6}{$^{6}$} DARK, Niels Bohr Institute, University of Copenhagen, Jagtvej 155A, DK-2200 Copenhagen, Denmark \\
\hypertarget{aff7}{$^{7}$} Department of Physics, University of Oxford, Denys Wilkinson Building, Keble Road, Oxford OX1 3RH, UK \\
\hypertarget{aff8}{$^{8}$} Scuola Normale Superiore, Piazza dei Cavalieri 7, I-56126 Pisa, Italy \\
\hypertarget{aff9}{$^{9}$} Sorbonne Universit\'e, CNRS, UMR 7095, Institut d'Astrophysique de Paris, 98 bis bd Arago, 75014 Paris, France \\
\hypertarget{aff10}{$^{10}$} Centre for Astrophysics Research, Department of Physics, Astronomy and Mathematics, University of Hertfordshire, Hatfield AL10 9AB, UK \\
\hypertarget{aff11}{$^{11}$} Steward Observatory, University of Arizona, 933 N. Cherry Ave., Tucson, AZ, 85721, USA\\
\hypertarget{aff12}{$^{12}$} Department of Physics and Astronomy, The Johns Hopkins University, 3400 N. Charles St., Baltimore, MD 21218, USA \\
\hypertarget{aff13}{$^{13}$} Department of Astronomy \& Astrophysics, The Pennsylvania State University, University Park, PA 16802, USA\\
\hypertarget{aff14}{$^{14}$} Instituto de Astrofísica de Canarias, C/ Vía Láctea s/n, 38205 La Laguna, Tenerife, Spain \\
\hypertarget{aff15}{$^{15}$} Departamento de Astrofísica, Universidad de La Laguna, 38200 La Laguna, Tenerife, Spain \\
\hypertarget{aff16}{$^{16}$} Waseda Research Institute for Science and Engineering, Faculty of Science and Engineering, Waseda University, 3-4-1, Shinjuku, Tokyo 169-8555, Japan \\
\hypertarget{aff17}{$^{17}$} Department of Physics and Astronomy, University College London, Gower Street, London WC1E 6BT, UK \\
\hypertarget{aff18}{$^{18}$} Centro de Astrobiolog\'ia (CAB), CSIC–INTA, Cra. de Ajalvir Km.~4, 28850- Torrej\'on de Ardoz, Madrid, Spain \\
\hypertarget{aff19}{$^{19}$} Space Telescope Science Institute, 3700 San Martin Drive, Baltimore, Maryland 21218, USA \\
\hypertarget{aff20}{$^{20}$} Max-Planck-Institut f\"ur extraterrestrische Physik (MPE), Gie{\ss}enbachstra{\ss}e 1, 85748 Garching, Germany \\
\hypertarget{aff21}{$^{21}$} NSF National Optical-Infrared Astronomy Research Laboratory, 950 North Cherry Avenue, Tucson, AZ 85719, USA \\
}}

\date{Accepted XXX. Received YYY; in original form ZZZ}

\pubyear{\the\year{}}

\begin{document}
\label{firstpage}
\pagerange{\pageref{firstpage}--\pageref{lastpage}}
\maketitle
\begin{abstract}
We present the galaxy stellar population catalogue from the JWST Advanced Deep Extragalactic Survey (JADES) Data Release 5 (DR5), providing homogeneous Bayesian inference of physical galaxy properties in GOODS-N and GOODS-S. Using deep JWST/NIRCam and MIRI imaging combined with ancillary multi-wavelength data, we model the spectral energy distributions of $\sim500{,}000$ sources with the \texttt{Prospector} framework. Our modelling incorporates flexible non-parametric star-formation histories (SFHs), nebular emission, dust attenuation, metallicities, and mid-infrared AGN and dust emission. We adopt an evolving star-forming main sequence (SFMS) prior for modelling the SFHs, which provides a physically-motivated long-term shape of SFHs while retaining non-parametric flexibility. The prior links stellar mass growth and star-formation rate (SFR) through the observed redshift-dependent SFMS, shaping the global behaviour of the inferred SFHs but allowing substantial deviations and scatters wherever supported by the data. We derive posterior distributions for stellar masses, SFRs, SFHs, dust attenuation, metallicities, and AGN contributions. The depth and wavelength coverage of JADES enable robust stellar mass measurements down to low-mass limits, as well as improved constraints on recent star-formation activity for $\sim 350{,}000$ galaxies at $z = 1 - 9$. The adoption of a physically motivated prior mitigates unphysical solutions and reduces degeneracies between redshift, age, dust, and metallicity, particularly for faint sources. We validate the catalogue through internal consistency checks and comparison to spectroscopic redshifts where available. The resulting value-added catalogue provides a uniform set of stellar population parameters suitable for statistical studies of galaxy growth, quenching, and the build-up of stellar mass across cosmic time. The full catalogue and posterior summaries are publicly released as part of JADES DR5.
\end{abstract}



 \clearpage
\section{Introduction}
Understanding how galaxies assemble their stellar mass across cosmic time remains a central goal of extragalactic astronomy \citep[e.g.,][]{conselice2014,somerville_2015, madau_dickinson_2014, naab2017}. Key physical quantities such as stellar mass, star-formation rate (SFR), star-formation history (SFH), metallicity, morphology, and active galactic nucleus (AGN) activity provide essential insight into the processes governing galaxy growth, quenching, and feedback \citep[e.g.,][]{noeske2007,peng2010, fabian2012, hopkins2014,speagle2014}. Robustly constraining these properties across large galaxy samples is therefore critical for connecting observations to theoretical models of galaxy formation \citep[e.g.,][]{Vogelsberger2014,Schaye2015,sijacki2015_illustris,tacchella2016,Pillepich2018,mcclymont2025, McClymont2026}.

Spectral energy distribution (SED) modelling has become the primary tool for inferring galaxy physical properties from multi-wavelength photometry. Early implementations relied on pre-computed grids of stellar population templates and $\chi^2$ minimization techniques, as implemented in codes such as \texttt{HyperZ} \citep{bolzonella2000_hyperz}, \texttt{Le Phare} \citep{arnouts_2002_lephare, ilbert_2006_lephare, arnouts2011}, \texttt{pPXF} \citep{cappellari2012, cappellari_2023}, and \texttt{EAZY} \citep{brammer2008eazy}. These approaches typically constrained a limited set of parameters within discretely sampled template libraries. While computationally efficient, coarse sampling and restricted model flexibility limited their ability to explore complex degeneracies or incorporate additional physical processes.

Subsequent generations of SED modelling frameworks expanded both the physical realism and dimensionality of the models. Energy-balance codes such as \texttt{MAGPHYS} \citep{decunha2008_magphys, decunha_2011_magphys} and later versions of \texttt{CIGALE} \citep{burgarella2005_cigal, noll2009_cigal, boquien2020} self-consistently linked ultraviolet–optical attenuation to infrared dust emission, introducing additional parameters describing dust heating and re-radiation. Fully Bayesian forward-modelling codes, including \texttt{BEAGLE} \citep{chevallard2016_beagle}, \texttt{Bagpipes} \citep{carnall2018, carnall2019_bagpipes}, \texttt{Synference} \citep{harvey2026_synference}, and \texttt{Prospector} \citep{johnson2021}, further generalized the inference framework by treating galaxy properties as continuous variables and generating model photometry on-the-fly using stellar population synthesis such as \texttt{galaxev} \citet[][hereafter BC03]{bruzual2003_BC03} and \textsc{FSPS} \citep{conroy2009, conroy2010}. These modern frameworks incorporate increasingly sophisticated physical ingredients, including parametric and non-parametric SFHs and population models \citep{leja2019sfh, carnall2019, alsing2024_popcosmos,2024OJAp....7E..54H, 2025arXiv250615811R, 2025OJAp....8E.152L, thorp2025_popcosmos}, independent stellar and gas-phase metallicities \citep{leja2017sfh}, flexible multi-component dust attenuation laws \citep{calzetti1994,calzetti2000,charlot2000, noll2009_cigal, keiek2013}, nebular emission modelling \citep{ferland1998_cloudy,conroy2010, ferland2013_cloudy, byler2017, ferland2017_cloudy,li_cue}, and AGN contributions based on clumpy torus models \citep{nenkova2008a, nenkova2008b, leja2018sfh}.

Modern SED modelling frameworks now routinely explore parameter spaces with more than a dozen free parameters. While this flexibility enables more physically realistic reconstructions of galaxy assembly histories, it also increases computational cost and amplifies sensitivity to prior assumptions, particularly when adopting non-parametric SFHs or hierarchical constraints \citep{leja2017sfh,leja2019sfh,tacchella2022,wang2023,wang2025outshining,jenny_2024_stochastic, turner_2025}. 

The launch of the \textit{James Webb Space Telescope} (\textit{JWST}) in 2021 \citep{rigby_jwst} marked a milestone for extragalactic astronomy. Operating across the near- and mid-infrared wavelength range ($0.6$–$28.8\,\mu$m), \textit{JWST} carries four state-of-the-art scientific instruments: NIRCam \citep[Near-Infrared Camera]{rieke2005_nircam}, NIRSpec \citep[Near-Infrared Spectrograph]{jakobsen2022_nirspec, ferruit2022_nirspec}, MIRI \citep[Mid-Infrared Instrument]{rieke2015_miri, wright2023_miri}, and NIRISS \citep[Near-Infrared Imager and Slitless Spectrograph]{willott2022_niriss_2,albert_2023_niriss_3, doyon2023_niriss_1, sivaramakrishnan2023_niris_4}. Together, these instruments provide unprecedented imaging resolution (down to $0.031''$/pixel) and spectroscopic capabilities with resolving powers up to $R \sim 2{,}700$.  In particular, the exquisite near-infrared sensitivity of \textit{JWST} enables robust constraints on galaxy stellar masses by directly probing rest-frame optical emission out to $z > 10$, overcoming the limitations of earlier surveys that relied primarily on rest-frame ultraviolet observations. \textit{JWST} observations have revealed a larger population of galaxies at high redshifts than previously expected, together with a more abundant population of massive and dusty galaxies compared to earlier optical and Lyman-break selected samples.\citep[e.g.,][]{weibel2024, barrufet2023, gottumukkala2024, labbe2023, xiao2024nature, casey2024, harikane2022, atek2022, castellano2022, naidu2022, adams2024_uv_luminosity, carniani2024spectroscopic, conselice2025, carniani2025,finkelstein2023ceers,Robertson2024,naidu2025,  napolitano2025_highz_agn, whitler2025_uv,constellano2025,  Hainline2026,pascalau26b}.

Combining deep \textit{JWST} near- and mid-infrared imaging with extensive \textit{HST} \citep[Hubble Space Telescope]{Illingworth_2016, Whitaker_2019} data, several major \textit{JWST} surveys have now released public catalogues of inferred galaxy physical properties across multiple extragalactic fields. For example, the Cosmic Evolution Early Release Science Survey \citep[CEERS; PI: S.~Finkelstein; PID: 1345;][]{finkelstein2023ceers} has produced comprehensive galaxy property catalogue \citep{cox_ceers_catalog_2025}, derived using the \texttt{LePhare}  template SED modelling code. Similarly, COSMOS-Web \citep[COSMOS-Webb; PI: Jeyhan Kartaltepe; PID: 1727;][]{cosmos_webb_2023} has provided large-area catalogue with derived stellar population properties \citep{shuntov_cosmosweb_stellar_pop}, based on \texttt{LePhare} code. The UNCOVER program \citep[UNCOVER; PI: Ivo Labb\'e, Rachel Bezanson; PID: 2561;][]{bezanson2024_uncover} has released stellar population measurements based on deep cluster-field observations \citep{wang2023, wang_2024_uncover_stellar_pop}, inferred with \texttt{Prospector}.  In addition, the EPOCHS team has delivered uniformly reduced catalogue across multiple \textit{JWST} fields, enabling consistent comparisons of galaxy properties over diverse environments and survey depths \citep{adams2024_uv_luminosity,austin2024, Harvey2025, conselice2025}, with physical parameters inferred using \texttt{Bagpipes}.

Among the various major \textit{JWST} surveys, the JWST Advanced Deep Extragalactic Survey \citep[JADES; PIs: Marcia Rieke, Kate Isaak;][]{rieke2020_iau,bunker_2020_iau,Rieke2023_jades, eisenstein2023_jades} stands out as one of the deepest and most comprehensive extragalactic programs. JADES targets the legacy Cosmic Assembly Near-infrared Deep Extragalactic Legacy Survey (CANDELS) Great Observatories Origins Deep Survey-South (GOODS-S) and Great Observatories Origins Deep Survey-North (GOODS-N) \citep{Giavalisco2004_candels, candels_gs_2013, candels_gn_2013}, leveraging their unparalleled multi-wavelength ancillary data and extensive spectroscopic coverage. JADES has recently released its latest data release (Data Release 5; \citealt{alberts2026,Hainline2026,johnson2026,robertson2026,carreira2026}), and this stellar population study is part of the DR5 data products. Over the past four data releases, JADES has provided initial NIRCam imaging mosaics and photometric catalogue in GOODS-N and GOODS-S \citep{Rieke2023_jades,Eisenstein2023_jof,deugenio_2025_dr3}, as well as NIRSpec prism and medium-resolution grating spectroscopy for 5,190 objects \citep{bunker2024_jades,deugenio_2025_dr3,emma2025_dr4,scholtz2025_dr4}. Using data from these earlier releases, the JADES mission has already enabled a wide range of groundbreaking scientific results, providing new insights into high-redshift galaxy formation and evolution \citep[e.g.,][]{tacchella2023_gnz11,bunker2023,curtis_lake2023_metalpoor,carniani2024spectroscopic,2025_witstok,lyu2024_miri_agn,2024_helton,2025helton_nature,2025wu_z14,maiolino_2024,Robertson2024,whitler2025_uv,simmonds_2024_lowmass_bursty,simmonds_2024_ionising, baker_2025_nature,baker_2025_AA,baker_2025_quiescent,danhaive2025,danhaive2026,puskas2025,puskas2025_sf_agn,geris2026,ignas2026_type1}.

JADES Data Release 5 combines NIRCam imaging observations from all JADES programs together with other \textit{JWST} programs within the JADES fields. With a total of 1,250 hours of NIRCam exposure time across GOODS-S and GOODS-N, covering approximately 500 arcmin$^2$, JADES DR5 provides imaging in up to 18 JWST/NIRCam wide and medium filters, 8 JWST/MIRI filters, and 9 \textit{HST} ACS/WFC3 filters \citep{robertson2026,johnson2026}. These features make JADES one of the most compelling datasets for studying galaxy formation and evolution, providing contiguous imaging coverage from the UV to the mid-infrared (MIR), together with an extensive set of \textit{JWST}/NIRSpec spectroscopic observations.

The unprecedented depth and extensive multi-filter coverage from HST through NIRCam to MIRI, and large spectroscopic observations together make JADES uniquely positioned for precise stellar population inference. The combination of broad wavelength coverage and robust spectroscopic calibration enables the simultaneous and self-consistent constraint of redshift, stellar mass, SFH, dust attenuation, dust emission, and AGN contributions across a vast dynamic range in redshift and stellar mass. 

In this work, we build a stellar population catalogue for all $\sim 500{,}000$ galaxies in GOODS-S and GOODS-N. We employ the \prosp SED modelling framework to infer galaxy physical properties, adopting the \texttt{Parrot} emulator to accelerate \texttt{FSPS} \citep{mathews2023}. In addition, we introduce a physically motivated prior on the non-parametric SFH, referred to as the star-forming main sequence prior (SFMS prior), designed to balance physical realism with flexibility in high-dimensional Bayesian inference.

The structure of this paper is as follows. In \autoref{sec: data sampes}, we describe the datasets used in this work. In \autoref{sec:inferring_galaxy_properties}, we introduce our \prosp SED modelling framework and the SFMS prior. In \autoref{sec: photometric sed quality}, we assess the quality of the SED fits and present the stellar population analysis. In \autoref{sec: compare with spec}, we compare several photometrically derived quantities with those obtained from spectroscopic measurements. In \autoref{sec: systematic effects: wavelength coverage and sfh priors}, we  investigate systematic effects on stellar mass inference, including the impact of NIRCam medium bands, MIRI photometry, and the choice of SFH prior. Finally, we present and describe the stellar population catalogue in \autoref{sec: prospector catalog}, and summarise our conclusions in \autoref{sec: conclusions}. Throughout this work, we adopt the Planck 2018 cosmology \citep{planck18_cosmology} with $H_0 = 67.4 \pm 0.5\,\mathrm{km}\,\mathrm{s}^{-1}\,\mathrm{Mpc}^{-1}$, $\Omega_{\rm M} = 0.315 \pm 0.007$, and $\Omega_{\Lambda} = 0.685 \pm 0.007$. All magnitudes are reported in the AB magnitude system \citep{Oke1974,Oke1983}.

\section{Data Samples}
\label{sec: data sampes}
\subsection{JWST/NIRCam and MIRI Imaging Data}
Our analysis is based on the JWST Advanced Deep Extragalactic Survey (JADES) Data Release~5 (DR5) imaging and photometric catalogue in the GOODS-South and GOODS-North fields. The DR5 imaging products provide deep, uniformly processed mosaics spanning optical-to-mid-infrared wavelengths. The dataset combines HST imaging with the full available JWST/NIRCam and JWST/MIRI observations from all Cycle~1--4 programs, totalling 1,250 hours of exposure time and covering 500~arcmin$^2$ over GOODS-S and GOODS-N \citep{alberts2026,Hainline2026,johnson2026,robertson2026,carreira2026}.

The reduction, calibration, astrometric alignment, and mosaicking of the DR5 NIRCam imaging are described by \citet{johnson2026}. The DR5 NIRCam dataset is based on the primary JADES programs (PIDs 1180, 1181, 1210, 1286, 1287, and 4540) and is complemented by additional public NIRCam imaging in the GOODS fields from a range of community surveys and programs, including the JADES Origins Field (PID~3215), PEARLS (PID~1176), NGDEEP (PID~2079), MIDIS (PID~1283), PANORAMIC (PID~2514), CONGRESS (PID~3577), BEACONS (PID~3990), POPPIES (PID~5398), SAPPHIRES (PID~6434), FRESCO (PID~1895), OASIS (PID~5997), and the Director's Discretionary transient follow-up program (PID~6541). The DR5 catalogue further incorporates HST imaging from the Hubble Legacy Fields mosaics, including ACS/F435W, F606W, F775W, F814W, F850LP, and WFC3/F105W, F125W, F140W, and F160W.

The construction of the DR5 source catalogue, including source detection, deblending, matched multi-band photometry, and photometric redshifts, is presented by \citet{robertson2026}. The catalogue provides multi-band photometry across the full DR5 imaging set, alongside common-PSF products and calibrated uncertainty estimates.

Mid-infrared coverage is provided by a combination of JADES MIRI parallel imaging and the public SMILES mosaics. The reduction and mosaicing of the JADES MIRI parallel component are described by \citet{alberts2026}, while the DR5 catalogue combine these data with SMILES MIRI imaging products following the DR5 mosaicking strategy. In this work, we adopt the DR5 photometric catalogue directly as the observational basis for our stellar population inference.

Galaxy morphologies are analysed and presented in \citet{carreira2026}, while \citet{Hainline2026} present a robustly selected sample of the highest-redshift sources ($z \geq 8$).

\subsection{JWST/NIRCam and MIRI Photometry}
\label{sec: imaging data}
The JADES DR5 photometry catalogue provides extensive multi-band photometric coverage over a broad wavelength range. In the GOODS-S field, the catalogue includes a total of 35 filters, comprising 9 HST/ACS and WFC3 bands, 18 JWST/NIRCam bands, and 8 JWST/MIRI bands. In GOODS-N, the catalogue contains 26 filters, including 9 HST bands, 15 NIRCam bands, and 2 MIRI bands. For each galaxy, we adopt the photometric measurements from the \texttt{KRON\_CONV} header, using \texttt{\{BAND\}\_KRON} as the flux measurement and the corresponding \texttt{\{BAND\}\_KRON\_e} values as the associated uncertainties \citep{robertson2026}.

These fluxes are measured on common-PSF mosaics, in which all imaging data are convolved to match the PSF of the JWST/NIRCam F444W band. This ensures consistent spatial resolution across all filters and minimizes color biases arising from wavelength-dependent PSF variations (see section~6 of \citealt{robertson2026} for a detailed description of the construction of the common-PSF mosaics). Source photometry is performed using forced ellipsoidal Kron apertures centered on each galaxy’s detection centroid (see section~9.2 of \citealt{robertson2026}). The Kron apertures are defined based on Gaussian profile regression fits to the multi-band detection image, with a Kron parameter of $k = 2.5$, and are applied uniformly across all PSF-matched images. Flux uncertainties are interpolated from the DR5 uncertainty-model mosaics (see section~8.1 of \citealt{robertson2026}) and are provided in the \texttt{\{BAND\}\_KRON\_e} columns.

We adopt the \texttt{KRON\_CONV} photometry for our analysis because these apertures are designed to capture galaxy’s total light while remaining robust to variations in galaxy morphology and surface brightness profiles. Compared to fixed circular apertures, Kron apertures provide a more reliable estimate of total flux for extended and irregular sources, particularly at high redshift where galaxies often exhibit complex morphologies \citep[e.g.,][]{Kartaltepe2015, bail2024, ormerod2024,carreira2026}. This is especially important for stellar population modelling, as incomplete flux measurements can bias inferred stellar masses and SFRs.

\begin{figure*}
    \centering
    \includegraphics[width=\linewidth]{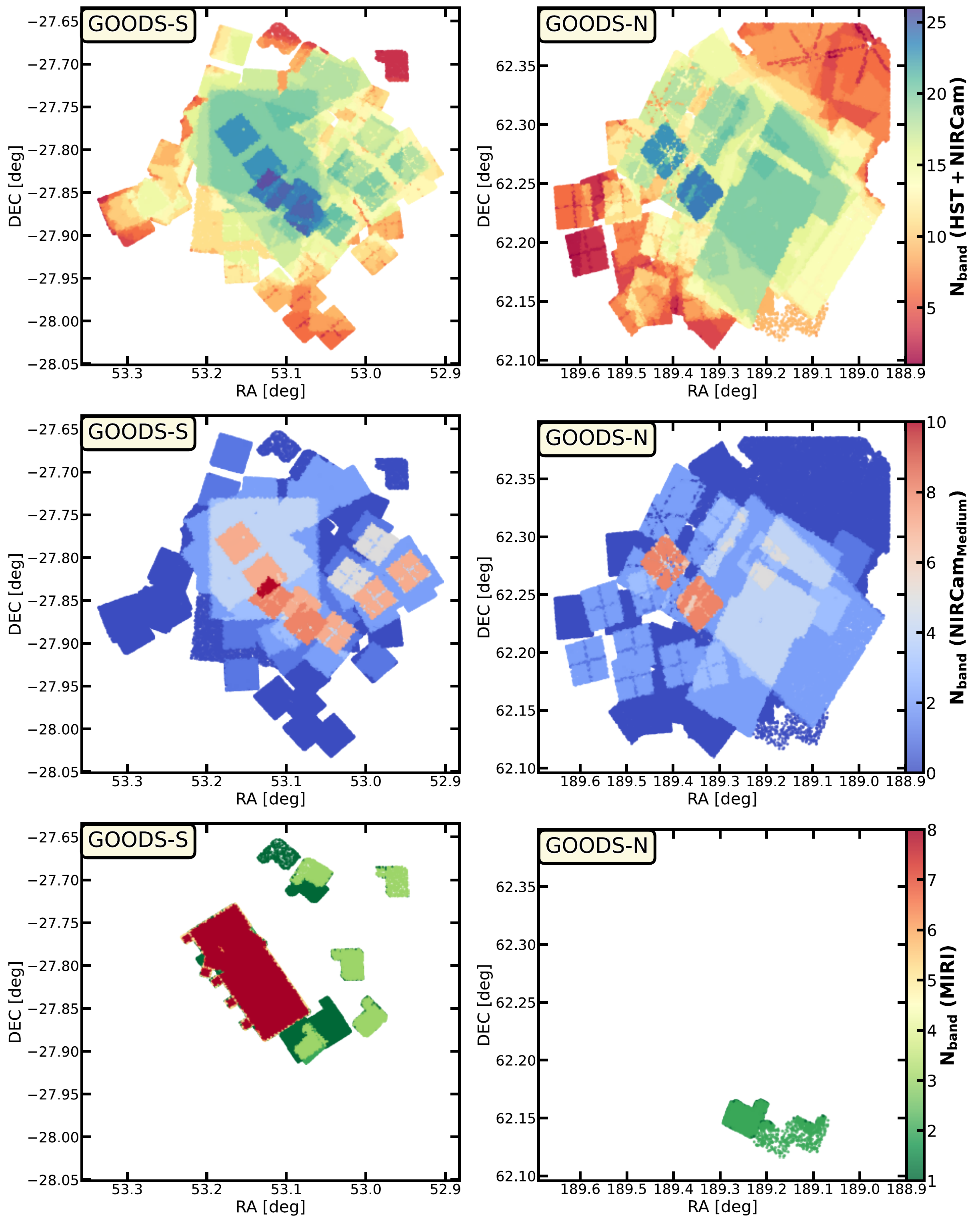}
    \caption{Number of available filter bands in the JADES DR5 release across the GOODS-S and GOODS-N fields. A total of $469$ arcmin$^2$ is covered, with $245$ arcmin$^2$ in GOODS-S and $224$ arcmin$^2$ in GOODS-N. We plot each galaxy as a point, with the color indicating the number of available photometric bands used for the SED fitting. \textit{\textbf{Top panels:}} Combined HST and JWST/NIRCam coverage, with GOODS-S observed in 27 bands (9 HST and 18 NIRCam) and GOODS-N in 24 bands (9 HST and 15 NIRCam). \textit{\textbf{Middle panels:}} JWST/NIRCam medium-band coverage, with a maximum of 10 and 8 medium-band filters available in GOODS-S and GOODS-N, respectively. \textit{\textbf{Bottom panels:}} JWST/MIRI coverage, up to 8 bands in GOODS-S and 2 bands in GOODS-N.}
    \label{fig: nbands all}
\end{figure*}

In \autoref{fig: nbands all}, we present the number of available HST, JWST/NIRCam, and MIRI filter bands across GOODS-S and GOODS-N. The top panel shows the total number of HST + NIRCam bands, the middle panel presents the total number of NIRCam medium bands, and the bottom panel displays the number of MIRI bands. The DR5 imaging covers a total area of $469$~arcmin$^2$, including $245$~arcmin$^2$ in GOODS-S and $224$~arcmin$^2$ in GOODS-N. Of this area, $323$~arcmin$^2$ ($209$~arcmin$^2$ in GOODS-S and $114$~arcmin$^2$ in GOODS-N) is covered by the six wide-band NIRCam filters F115W, F150W, F200W, F277W, F356W, and F444W. Furthermore, $240$~arcmin$^2$ ($151$~arcmin$^2$ in GOODS-S and $89$~arcmin$^2$ in GOODS-N) is covered by the eight core JADES NIRCam filters: F090W, F115W, F150W, F200W, F277W, F356W, F410M, and F444W.

There are also many medium-band observations from various primary and parallel programs, bringing the total number of NIRCam medium-band filters available in the JADES fields to as many as ten. For further details, including the relationship between the number of filters and the survey area, see section 2.5 and figure 3 of \citet{johnson2026}.

\subsection{Spectroscopic Redshifts}
\label{sec: spec-z}
We compile all available spectroscopic redshift catalogues in the JADES fields \citep[e.g.,][]{deugenio_2025_dr3, scholtz2025_dr4, emma2025_dr4}, drawing from 14 spectroscopic \textit{JWST} and non-\textit{JWST} surveys in GOODS-S and 8 in GOODS-N. A complete list of the spectroscopic catalogues used, along with a detailed description of the construction and homogenisation of these samples, can be found in \citet{puskas2025} and \citet{puskas2025_sf_agn}. For DR5, following the same methodology, we extend this compilation by adding spectroscopic redshifts for objects in newly covered areas. In brief, we cross-match the spectroscopic catalogues with our JADES DR5 catalogue using a sky-position tolerance of $<0.25$~arcsec (assigned a \texttt{sky-flag} of 1) and $0.25$--$0.50$~arcsec (\texttt{sky-flag} of 2). In addition, each spectroscopic redshift is assigned a reliability \texttt{spec-flag}, with values of 1 indicating highly secure redshifts, 2 indicating less certain measurements, and 3 denoting unreliable estimates. 

In total, $12{,}128$ galaxies in GOODS-S and $6{,}972$ galaxies in GOODS-N have robust spectroscopic redshifts with \texttt{sky-flag} and \texttt{spec-flag} $\leq 2$, which we used in our analysis.

\section{Inferring Galaxy Physical Properties}
\label{sec:inferring_galaxy_properties}

Inferring galaxy properties through state-of-the-art spectral energy distribution (SED) modelling plays a critical role in connecting new observations to theories of galaxy formation and evolution. In this work, we employ \texttt{Prospector} \citep{johnson2021}, a fully bayesian forward-modelling SED modelling framework that combines on-the-fly model generation with efficient nested sampling. This approach enables robust constraints on stellar populations, dust attenuation/emission, and potential AGN contributions to the mid-infrared, while providing reliable bayesian posterior distributions that accurately quantify parameter uncertainties and allow assessment of degeneracies. We present an overview of the full set of runs in \autoref{sec: Input Photometry and SED-Modelling Setup}, and detail the \prosp modelling methodology in the remaining subsections.

\subsection{Input Photometry and SED-Modelling Setup}
\label{sec: Input Photometry and SED-Modelling Setup}
For each galaxy, we include all available photometric bands (\autoref{sec: imaging data}), excluding only those bands for which the flux is measured but the associated uncertainty cannot be reliably estimated \citep{robertson2026}. After this band selection, 459 galaxies in GOODS-S and 83 galaxies in GOODS-N are left without any usable photometric bands. These objects are primarily located near the edges of the mosaics, where they suffer from limited filter coverage and insufficient exposure depth. A further 108 galaxies remain with only a single usable photometric band, for which we nevertheless conduct fits and flag accordingly in the catalog. Furthermore, for each band flux ($f_\nu$) and uncertainty, we apply a $5\%$ error floor; that is, if the uncertainty is smaller than $5\%$ of the flux, it is increased to $5\%$ of the flux.

We perform stellar population inference using four complementary filterbands combinations, summarised below.  
Importantly, these configurations are \emph{not mutually exclusive}: a given galaxy may be analysed under multiple configurations depending on the availability of additional observational constraints (e.g., MIRI coverage or spectroscopic redshifts). The number of galaxies in each configuration, as well as the sample sizes after applying different $\mathrm{SNR}_{\mathrm{F444W}}$ quality cuts, are summarised in \autoref{tab: number of galaxies snr}.

\begin{enumerate}
    \item \textbf{HST + NIRCam}: This configuration is applied to \textit{all} galaxies in the JADES DR5 catalogue, providing a uniform set of stellar population constraints for 303{,}907 galaxies in GOODS-S and 181{,}061 galaxies in GOODS-N.
    
    \item \textbf{HST + NIRCam + MIRI}: For galaxies with at least one available MIRI band, we perform an additional SED modelling run that includes the MIRI photometry. This complementary configuration applies to 96{,}730 galaxies in GOODS-S and 4{,}861 galaxies in GOODS-N.
    
    \item \textbf{Spec-$z$ + HST + NIRCam}: For galaxies with secure spectroscopic redshifts, we carry out an additional run in which the redshift is fixed to the spectroscopic value. This configuration includes 12{,}128 galaxies in GOODS-S and 6{,}972 galaxies in GOODS-N.
    
    \item \textbf{Spec-$z$ + HST + NIRCam + MIRI}: For galaxies that have both reliable spectroscopic redshifts and MIRI coverage, we perform a further complementary run that fixes the redshift and includes the MIRI bands. This configuration applies to 6{,}508 galaxies in GOODS-S and 301 galaxies in GOODS-N.
\end{enumerate}
Throughout this paper, the \textbf{HST + NIRCam} configuration is adopted as the fiducial run, as it provides homogeneous stellar population constraints across the full galaxy sample. The additional configurations described above are used as complementary analyses, enabling us to incorporate richer observational information where available and to assess its impact on the inferred galaxy properties.

\begin{table*}
\centering
\caption{Number of galaxies, and the number above different $\mathrm{SNR_{F444W}}$ thresholds, for each of the four \texttt{Prospector} SED modelling configurations in GOODS-S and GOODS-N. The HST+NIRCam configuration is adopted as the fiducial run and provides stellar population constraints for the full galaxy sample. For subsamples with additional MIRI imaging or secure spectroscopic redshifts, supplementary fits are performed by including the MIRI band or by fixing the redshift to the spectroscopic value, respectively.}
\label{tab: number of galaxies snr}
\begin{tabular}{l c r r r r r}
\toprule \toprule
\textbf{Configuration} & \textbf{Field} 
& $\boldsymbol{N_\mathrm{all}}$ 
& $\boldsymbol{N(\mathrm{SNR_{F444W}}>3)}$ 
& $\boldsymbol{N(\mathrm{SNR_{F444W}}>5)}$ 
& $\boldsymbol{N(\mathrm{SNR_{F444W}}>10)}$ \\
\midrule
HST + NIRCam & GOODS-S & $303{,}907$ & $157{,}467$ & $117{,}184$ & $72{,}507$ \\[2.0pt]
HST + NIRCam & GOODS-N & $181{,}061$ & $90{,}588$  & $67{,}229$  & $42{,}012$  \\[2.0pt]
\midrule
HST + NIRCam + MIRI & GOODS-S & $96{,}730$ & $48{,}385$ & $35{,}905$ & $21{,}995$  \\[2.0pt]
HST + NIRCam + MIRI & GOODS-N & $4{,}861$  & $2{,}168$  & $1{,}599$  & $997$  \\[2.0pt]
\midrule
Spec-$z$ + HST + NIRCam & GOODS-S & $12{,}128$ & $11{,}857$ & $11{,}200$ & $10{,}746$   \\[2.0pt]
Spec-$z$ + HST + NIRCam & GOODS-N & $6{,}972$ & $6{,}423$ & $6{,}195$ & $5{,}597$ \\[2.0pt]
\midrule
Spec-$z$ + HST + NIRCam + MIRI & GOODS-S & $6{,}508$ & $5{,}836$ & $5{,}495$ & $4{,}916$ \\[2.0pt]
Spec-$z$ + HST + NIRCam + MIRI & GOODS-N & $301$ & $159$ & $154$ & $113$  \\[2.0pt]

\bottomrule \bottomrule
\end{tabular}
\end{table*}

\begin{table*}
    \centering
    \caption{Free parameters in our \texttt{Prospector} model, grouped into four components: redshift and stellar populations, dust emission, dust attenuation, and AGN emission. The model contains a total of 18 free parameters, including 6 parameters describing the SFH and 12 additional parameters for the remaining components. We adopt a combination of uniform, Gaussian ($\mathcal{N}(\mu, \sigma)$), and non-uniform priors for these parameters.}
    \label{tab:prospector_priors}

    \footnotesize 
    \begin{tabular}{>{\raggedright\arraybackslash}p{2.0cm} >{\raggedright\arraybackslash}p{7.0cm} >{\raggedright\arraybackslash}p{5.8cm}}
        \toprule \toprule
        \textbf{Parameter} & \textbf{Description} & \textbf{Prior} \\ 
        \midrule
         $z$ & Redshift & Uniform in $[10^{-2},\, 20]$ 
         \\[4.5pt]
        $\log_{10}(M_*/\mathrm{M_\odot})$ & Total stellar mass formed & Mass function prior from \citet{wang2023}, truncated to $[6.25,\, 13]$ \\[4.5pt]
        $\log_{10}(Z_*/\mathrm{Z_\odot})$ & Galaxy stellar metallicity & $\mathcal{N}(Z(M_*),\, \sigma(M_*))$ from the mass--metallicity relation in \citet{gallazzi2005}, truncated to $[-1.98,\, 0.19]$ \\[4.5pt]
        SFH & Adjacent SFR ratios in $\log_{10}\left(\frac{\mathrm{SFR}_i}{\mathrm{SFR}_{i+1}}\right)$, 7 age bins & Star-forming main sequence prior (this work, see \autoref{sec: rising sfh prior}) \\ [4.5pt] 
        $\log_{10}(Z_\mathrm{gas} / Z_\odot)$ & Gas-phase metallicity & Uniform in $[-2.0, 0.5]$ \\ [4.5pt]\\ \hdashline \\
         $Q_{\mathrm{PAH}}$ & PAH mass fraction \citep{draine2007}& $\mathcal{N}(2,\, 2)$, truncated to $[0,\, 7]$ \\
        $U_{\mathrm{min}}$ & Minimum radiation field for dust emission & $\mathcal{N}(1,\, 10)$, truncated to $[0.1,\, 25]$ \\[4.5pt]
        $\log_{10}(\gamma_e)$ & Fraction of dust mass exposed to $U_{\mathrm{min}}$ & $\mathcal{N}(-2,\, 1)$, truncated to $[-4,\, 0]$ \\ [4.5pt]
         \\ \hdashline \\
         $\tau_{\mathrm{dust,2}}$ & Diffuse dust optical depth \citep{charlot2000} & $\mathcal{N}(0.3,\, 1)$, truncated to $[0,\, 4]$ \\[4.5pt]
        $n$ & Power-law modifier to shape of the \citet{calzetti2000} attenuation curve of the diffuse dust & Uniform in $[-1.2,\, 0.4]$ \\[4.5pt]
        $\tau_{\mathrm{dust,1}} / \tau_{\mathrm{dust,2}}$ & Birth cloud to diffuse dust optical depth ratio \citep{charlot2000} & $\mathcal{N}(1,\, 0.3)$, truncated to $[0,\, 2]$ \\[4.5pt]
        \\ \hdashline \\
         $\log_{10}(f_{\mathrm{AGN}})$ & AGN luminosity to bolometric luminosity ratio & Uniform in $[-5,\, \log_{10}(3)]$ \\[4.5pt]
        $\log_{10}(\tau_{\mathrm{AGN}})$ & AGN torus dust optical depth & Uniform in $[\log_{10}(5),\, \log_{10}(150)]$ \\
        \bottomrule \bottomrule
    \end{tabular}
\end{table*}

\subsection{Constraining Galaxy Properties with \texttt{Prospector}}
\label{sec: inferring galaxy properties}

In this subsection, we describe our \texttt{Prospector} model setup, including the model parameters and their associated priors. \texttt{Prospector} is built on the Flexible Stellar Population Synthesis (\texttt{FSPS}) framework \citep{conroy2009, conroy2010}, accessed via the \texttt{python-fsps} \citep{johnson2021}. In this work, within \texttt{FSPS}, we adopt the MESA Isochrones and Stellar Tracks (\texttt{MIST}) \citep{choi2016, dotter2016} and the \texttt{MILES} stellar spectral library \citep{sanchez2006, miles_2011}. The \citet{chabrier2003imf} initial mass function is used throughout this work.

Our \texttt{Prospector} model comprises 18 free parameters describing galaxy properties including redshift, stellar mass, stellar and gas-phase metallicities, SFH, dust attenuation, dust emission, and AGN contributions. The model configuration and prior choices are motivated by a broad range of previous studies \citep[e.g.,][]{leja2017sfh, leja2018sfh, leja2019sfh, leja2020sfh, tacchella2022,tacchella2022_quench, Tacchella2023,mathews2023, wang2023, wang_2024_uncover_stellar_pop, duan2024_addingvalue, jenny_2024_stochastic, turner_2025, Harvey2025}. All parameters and their associated priors are summarised in \autoref{tab:prospector_priors}.

We use a uniform redshift prior over the interval $\left[0.01,20\right]$. For galaxy stellar mass, we implement the stellar mass function prior from \citet{wang2023}. This mass function prior is the normalized stellar mass function at each redshift, such that lower-mass galaxies have higher probability density in the prior, which helps avoid spurious high-mass solutions. Stellar metallicity is linked to stellar mass using the empirical mass--metallicity relation of \citet{gallazzi2005}, adopting a Gaussian prior with a mass-dependent mean and scatter that increases toward lower stellar masses. For the SFH, we adopt the non-parametric, continuity SFH model \citep{leja2019sfh}. We use 7 age bins to describe the SFH, and fit the logarithmic ratios of the SFRs between adjacent bins. In this work, a novel star-forming main sequence (SFMS) prior is introduced for the base of the continuity SFH, around which SFH varies. We describe this in detail in the next subsection (\autoref{sec: rising sfh prior}). 

Furthermore, the gas-phase metallicity is treated independently from the stellar metallicity, allowing it to vary freely over the range $-2 < \log(Z_{\rm gas}/Z_\odot) < 0.5$. The ionisation parameter for the nebular emission model is fixed to $\log(U) = -1$. Our dust modelling includes both absorption and emission. Dust emission is modelled using the framework from \citet{draine2007}, essential for accurate star formation rate measurements, particularly in galaxies with lower specific SFRs where older stellar populations significantly heat dust \citep{utomo2014}. The \citet{draine2007} dust model relies on the silicate-graphite-PAH model \citep[see also][]{draine1984, mathis1977} and features three free parameters: $Q_{\mathrm{PAH}}$, $U_{\mathrm{min}}$, and $\log_{10}(\gamma_e)$. Here, $Q_{\mathrm{PAH}}$ specifies the fraction of dust mass in polycyclic aromatic hydrocarbons (PAHs), crucial for mid-infrared emission features.
$U_{\mathrm{min}}$ sets the minimum interstellar radiation field intensity relative to the local Milky Way radiation field, and $\log_{10}(\gamma_e)$ control  the fraction of dust mass exposed to this minimum field. These two parameters control the shape and location of the  thermal dust emission bump in the IR SED.

We adopt a two-component dust attenuation model \citep{charlot2000, noll2009_cigal}, consisting of a birth-cloud component ($\tau_{\mathrm{dust,1}}$) and a diffuse interstellar medium component ($\tau_{\mathrm{dust,2}}$). The birth-cloud component accounts for the additional attenuation experienced by stars younger than 10 Myr that remain embedded in molecular clouds and H~\textsc{ii} regions, while the diffuse ISM component describes attenuation affecting the entire stellar population. The wavelength dependence of the diffuse attenuation is parameterized by $n$, which is a power-law modifier to shape of the \citet{calzetti2000} attenuation curve. We tie the strength of the UV dust absorption bump to the best-fit diffuse dust attenuation index, following the results of ~\citet{keiek2013}. We treat $\tau_{\mathrm{dust,2}}$, $n$, and the ratio $\tau_{\mathrm{dust,1}}/\tau_{\mathrm{dust,2}}$ as free parameters. For a detailed discussion on dust attenuation impacts on the galaxy SED, see \citet[Section 3.1.3]{leja2017sfh}.

Finally, we incorporate AGN emission \citep{leja2018sfh} using templates from the \texttt{CLUMPY} models of \citet{nenkova2008a, nenkova2008b}. These models simulate AGN emission by radiative transfer through a clumpy dust torus illuminated by an AGN power-law spectrum. The AGN component introduces two additional free parameters: $\log_{10}(f_{\mathrm{AGN}})$, the fraction of the total $4-20 \,\mu$m luminosity contributed by the AGN, and $\log_{10}(\tau_{\mathrm{AGN}})$, the optical depth of individual dust clumps at 5500 \AA.

\subsection{The Star-forming Main Sequence Prior}
\label{sec: rising sfh prior}

\subsubsection{Non-parametric SFHs and the Continuity Prior}

Observational constraints on galaxy SFHs provide a crucial window into galaxy growth and evolution, offering insight into key physical processes such as the duration of star formation episodes and the mechanisms driving quenching. Accurate reconstruction of SFHs is also essential for obtaining reliable estimates of stellar masses and SFRs.

To infer SFHs from SED modelling, two broad classes of approaches are commonly used: parametric and non-parametric \citep[e.g.,][]{carnall2019, leja2019sfh}. Parametric approaches assume analytic forms for the SFR as a function of time, $\mathrm{SFR}(t, \theta)$, such as exponential, log-normal \citep{gladders2013, abramson2015, diemer2017} or double power-law functions \citep{carnall2018}. These models are computationally efficient and conceptually straightforward, but they often fail to capture the diversity and complexity of SFHs seen in both observations and simulations. Many galaxies exhibit features such as bursty star formation or rapid quenching, which are poorly represented by smooth, fixed-form SFHs \citep{pacifici_2013}. Relying on overly rigid assumptions can therefore introduce systematic biases in the inferred stellar population properties \citep[e.g.,][]{dominguez2015, iyer2017}.

To address this limitation, non-parametric SFH models have been developed to provide greater flexibility \citep{leja2019sfh, iyer2019_non_parametric}. Within the \texttt{Prospector} SED modelling framework, we discretize the SFH into a series of age bins, assuming a constant SFR within each bin while allowing it to vary between bins. In this work, we adopt 7 age bins, which are defined as follows. For galaxies at a redshift $z_{\mathrm{obs}} < 3$, the first bin spans 0--30 Myr (lookback time), the second spans 30--100 Myr, the subsequent bins are logarithmically spaced in time between 100 Myr and $0.90 \times t_{\mathrm{univ}}(z_{\mathrm{obs}})$, and the final bin extends from $0.90 \times t_{\mathrm{univ}}(z_{\mathrm{obs}})$ to $t_{\mathrm{univ}}(z_{\mathrm{obs}})$. For galaxies at $z_{\mathrm{obs}} \geq 3$, the bin edges are logarithmically spaced between $\log_{10}(t/\mathrm{yr}) = 0$ and $0.90 \times t_{\mathrm{univ}}(z_{\mathrm{obs}})$, with the final bin again spanning $0.90 \times t_{\mathrm{univ}}(z_{\mathrm{obs}})$ to $t_{\mathrm{univ}}(z_{\mathrm{obs}})$.

In \texttt{Prospector}, different priors for the SFRs in each bin can be chosen \citep{johnson2021}. We adopt the \textit{Continuity SFH} model \citep{leja2017sfh, leja2019sfh, johnson2021, tacchella2022, Tacchella2023,wang2023, simmonds_2024_lowmass_bursty, simmonds_2024_ionising}, in which the logarithmic SFR ratios between adjacent bins are modelled. For $N$ age bins, this results in $N-1$ free parameters of $\log_{10}\mathrm{SFR}_{i}/{\rm SFR}_{i+1}$. Each $\log_{10}\mathrm{SFR}_{i}/{\rm SFR}_{i+1}$ ratio follows a Student's $t$ distribution, written as
\begin{equation}
\label{eq: students't}
\mathrm{PDF}(x\mid \nu,\mu,\sigma)=
\frac{\Gamma\!\left(\frac{\nu+1}{2}\right)}
{\sqrt{\nu\pi}\,\sigma\,\Gamma\!\left(\frac{\nu}{2}\right)}
\left[
1+\frac{1}{\nu}\left(\frac{x-\mu}{\sigma}\right)^2
\right]^{-\frac{\nu+1}{2}},
\end{equation}
where $x$ is the SFR ratio in two adjacent bins. The parameter $\nu$ controls the heaviness of the tails and therefore the allowed frequency of large variations in SFR between neighbouring bins (with smaller $\nu$ permitting more extreme, bursty changes), $\sigma$ sets the typical amplitude of these variations, determining how strongly the SFR can fluctuate from one bin to the next, and $\mu$ defines the expected value of the SFR ratio, thereby encoding the baseline trend of the SFH, usually assumed to be constant. 

For the standard \textit{Continuity SFH} model, as used in the \texttt{Prospector}$-\alpha$ framework \citep{leja2017sfh}, $\log_{10}\mathrm{(SFR_\mathrm{ratio})}$ is assigned a Student's $t$ distribution with $\nu = 2$, $\sigma = 0.3$, and $\mu = 0$, corresponding to a heavy-tailed yet relatively concentrated distribution. The choice of $\mu = 0$ implies an expectation of equal SFRs in adjacent time bins, i.e., a constant SFH in the absence of strong data constraints. The parameters $\nu$ and $\sigma$ are motivated by the variability of SFHs seen in hydrodynamical simulations such as Illustris \citep{Vogelsberger2014, genel_2014_illustris, nelson2015_illustris, sijacki2015_illustris}, allowing for moderate stochasticity while suppressing extreme, unphysical fluctuations. Later, \citet{tacchella2022} introduced the \textit{Bursty Continuity} model, in which the scale parameter is increased to $\sigma = 1.0$, thereby allowing substantially larger variations in SFR between adjacent time bins. This modification is motivated by the expectation that high-redshift ($z \gtrsim 8$) galaxies may exhibit highly bursty star formation, driven by rapid gas accretion, mergers, and strong feedback processes, and thus require greater flexibility in their SFHs \citep[e.g.,][]{brooks2009, dekel2006, dekel2009, duan2025, duan_2026_merger_2, duncan2019, kerevs2005, puskas2025, puskas2025_sf_agn,putman2017, waterval2025}. In practice, adopting a broader prior typically leads to inferred SFHs with enhanced recent star formation, resulting in younger stellar ages and correspondingly lower stellar masses.

Furthermore, \citet{jenny_2024_stochastic} proposed a \textit{stochastic prior}, in which the $\log_{10}\mathrm{(SFR_\mathrm{ratio})}$ values across all age bins are linked and quantified via the power spectral density (PSD) \citep{caplar2019, tacchella2020_stochastic, iyer2024_stochastic}. In contrast to continuity-based models, where the prior is defined locally between adjacent time bins, the stochastic prior characterises the SFH as a correlated process across all timescales. This distinction is important because the interpretation of $\log_{10}(\mathrm{SFR}i/\mathrm{SFR}{i+1})$ depends on the chosen binning scheme, and therefore does not directly map onto physical variability timescales. By instead specifying the SFH in terms of its PSD, the stochastic prior encodes how variability is distributed as a function of timescale, providing a more physically motivated description of star formation driven by processes such as gas accretion, feedback, and mergers. This approach has been shown to more accurately recover recent star formation activity, while enabling future extensions to hierarchical modelling of galaxy populations.

The flat SFH prior implemented in \texttt{Prospector}-$\alpha$ (\citealt{leja2017sfh}, where $\mu = 0$), while not very physically motivated and subject to outshining issues, remains readily interpretable and straightforward to understand.

\subsubsection{Global Constraints on the SFH}

Variations in the scale parameter $\sigma$ of the Student’s $t$ distribution on $\log_{10}\mathrm{(SFR_\mathrm{ratio})}$ primarily control the level of stochasticity between adjacent time bins and have been explored extensively in the aforementioned works. However, the overall shape of the SFH is also critically important. In practice, the SFH at large lookback times is often weakly constrained by observational data due to outshining, whereby young stellar populations dominate the observed light and obscure older stellar components. As a result, the inferred SFH in these regimes can be strongly prior-dominated \citep{tacchella2022, whitler2023, Tacchella2023, wang2025outshining, harvey2025_outshining}.

This issue has been highlighted in recent work by \citet{turner_2025}, who demonstrated that different prior assumptions on the global SFH shape can lead to significant shifts in inferred stellar ages, particularly for quiescent galaxies. In their \textit{rising prior} model, the expected SFR in each age bin is tied to the average halo mass accretion rate as a function of redshift \citep{dekel2013, tacchella2018}, favouring SFHs that rise toward later times. This modification can substantially alter the inferred formation histories and ages of galaxies compared to standard continuity models. Similarly, \citet{2025wu_z14} adopts a \textit{rising prior} in which the SFR scales as $(1+z)^{-4.5}$, motivated by halo mass abundance matching based on the Abacus N-body simulations \citep{maksimova2021abacussummit,carniani2024spectroscopic}.

Furthermore, \citet{wang2023} introduced the \texttt{Prospector}$-\beta$ model, which adopts non-uniform priors on redshift, stellar mass, metallicity, and SFH. For the SFH, while retaining $\sigma = 0.3$ as in the $\alpha$ model, the expected values $\mu$ are set to follow the evolution of the cosmic star formation rate density (SFRD) from \citet{Behroozi2019}, with an additional dependence on stellar mass. While this approach provides a reasonable global trend for galaxy growth, it implicitly assumes that individual galaxies evolve in a manner broadly consistent with the cosmic average, which may instead be the result of an ensemble of galaxies with more complicated SFHs.

\begin{figure}
    \centering
    \includegraphics[width=\linewidth]{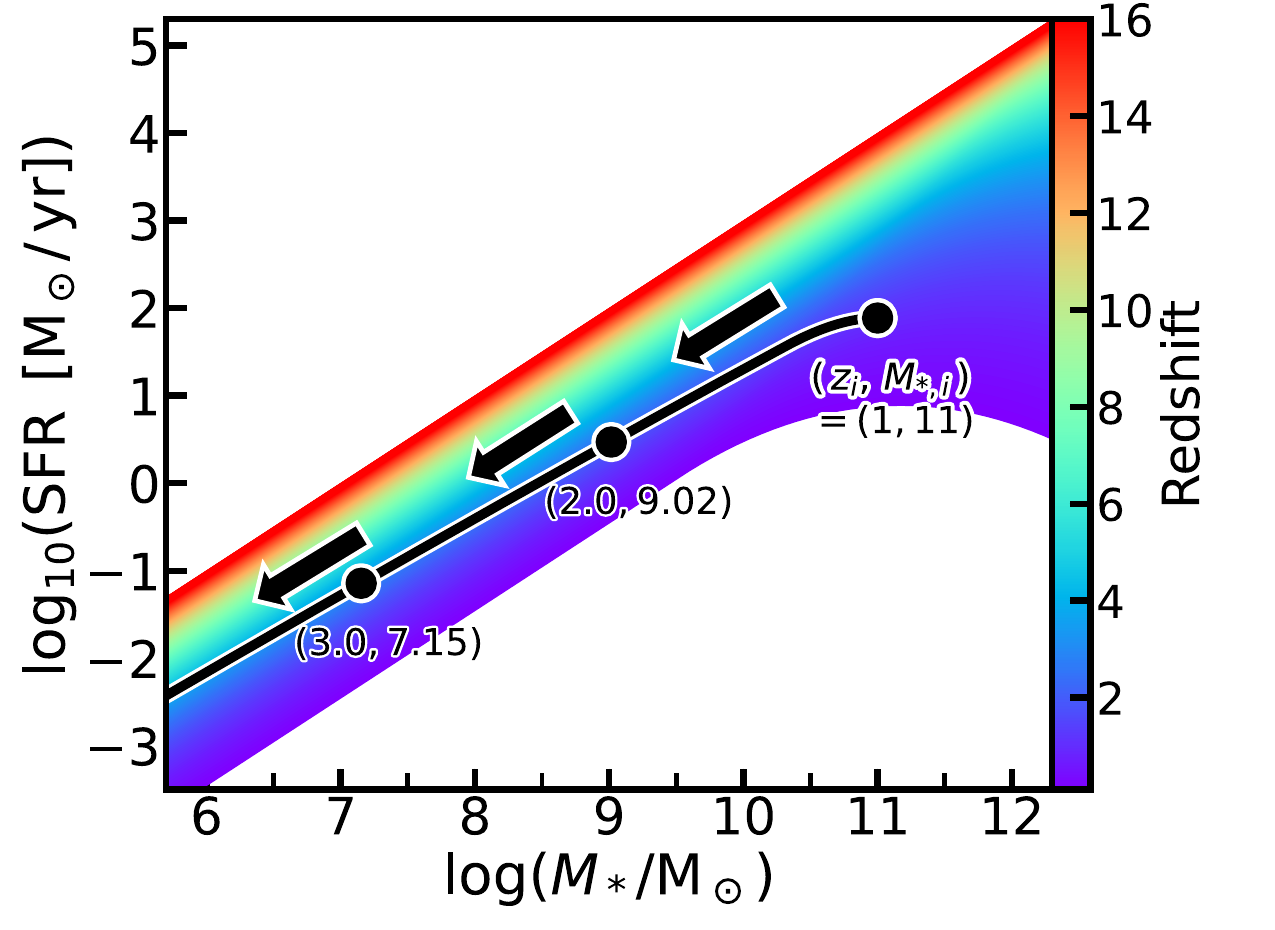}
    \caption{SFH trajectories derived from the star-forming main sequence (SFMS). The background rainbow curves represent the SFMS fitted to the data from \citet{simmonds_2025}, using the parameterisation described in \autoref{eq: schreiber}, with redshift indicated by the colour bar. While the observational constraints from \citet{simmonds_2025} span $3 \leq z \leq 9$, the SFMS is extrapolated beyond this range following the adopted parameterisation. The black points and curves illustrate an SFMS-based track, constructed by integrating the SFMS backward in time from a galaxy observed at $(z_i, M_{*,i})$. As an illustrative example, for a galaxy observed at $(z, \log (M_*/\mathrm{M_\odot})) = (1, 11)$, the stellar mass at $z = 2.0$ is $\log (M_*/\mathrm{M_\odot}) = 9.02$, decreasing further toward higher redshift. This SFMS-based trajectory serves as the baseline for the non-parametric SFH prior, with the inferred SFHs allowed to deviate from it within the continuity framework as supported by the data.}
    \label{fig: sfms}
\end{figure}

\begin{figure*}
    \centering
    \includegraphics[width=\linewidth]{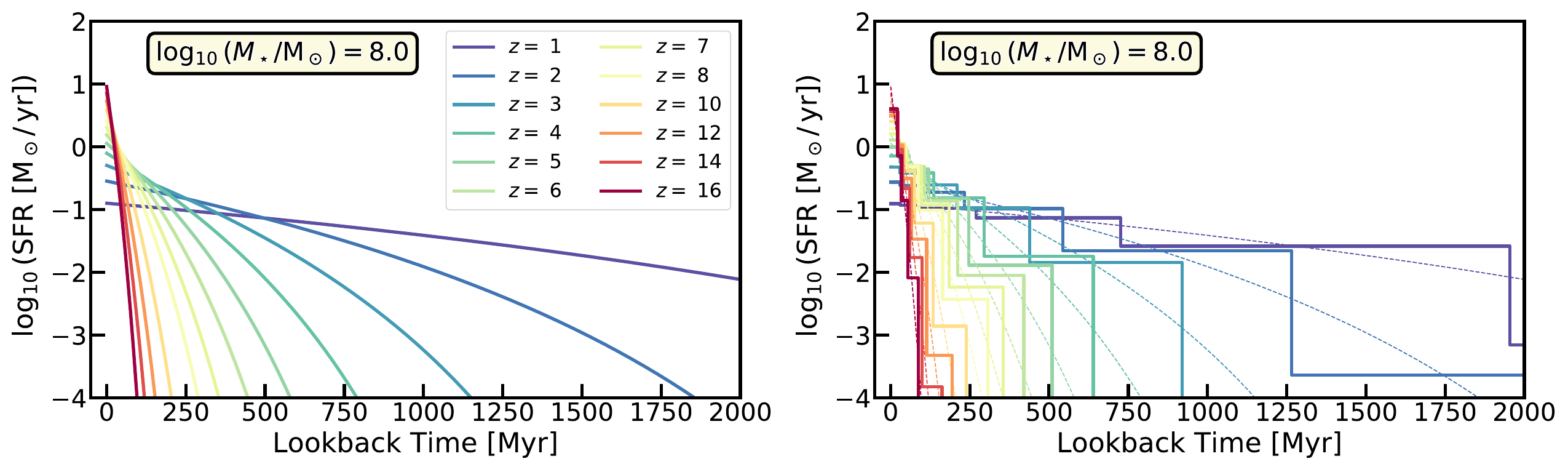}
    \caption{Base of the star-forming main sequence (SFMS) prior introduced in this work, around which the SFHs are allowed to vary. \textbf{\textit{Left Panel:}} Example base SFHs derived from the SFMS for galaxies at stellar mass $\log_{10}(M_*/\mathrm{M}_\odot)=8.0$ at 12 discrete redshifts spanning $1 \le z \le 16$, with curves colour-coded by redshift. These SFHs represent the baseline trajectories implied by the SFMS prior, rather than the final inferred histories. \textbf{\textit{Right Panel:}} Translation of the smooth SFMS trajectories (dashed lines) into binned SFHs (solid lines), following the age bin definitions described in \autoref{sec: rising sfh prior}. The binned SFHs define the expected values of the non-parametric prior, around which the inferred SFHs are allowed to vary within the continuity framework as constrained by the data.}
    \label{fig:sfh prior combine}
\end{figure*}

\subsubsection{The Star-Forming Main Sequence Prior}

Building on the \textit{rising prior}, we introduce a new prior for the continuity SFH, referred to as the \textbf{Star-Forming Main Sequence prior (SFMS prior)}. While the rising prior links SFHs to the average dark matter halo mass accretion rate, the SFMS prior instead anchors galaxy growth to the empirically observed SFMS. These two perspectives are closely connected, as the evolution of the SFMS is thought to reflect, to first order, the underlying baryon accretion driven by halo growth \citep{lilly2013,dekel2013,tacchella2016_gas_depletion}. The SFMS prior is therefore physically motivated by the assumption that galaxies predominantly grow along the SFMS, while exhibiting scatter around it. In practice, this prior modifies both the scale parameter $\sigma$ and the expected value $\mu$ of the Student’s $t$ distribution for $\log_{10}\mathrm{(SFR_\mathrm{ratio})}$ (\autoref{eq: students't}), thereby encoding both the typical growth trajectory and the allowed variability around it. We describe the construction and implementation of this prior in detail in the following paragraphs.

To construct this prior, we first establish an empirical parameterization of the SFMS. Numerous studies have characterized SFMS evolution across cosmic time using a range of SED modelling methodologies \citep[e.g.,][]{speagle2014, rinaldi2022_sfms,popesso2023sfms, clarke2024sfms, cole2025sfms, rinaldi2025_sfms,simmonds_2025}. We adopt the dataset and SFMS measurements presented by \citet{simmonds_2025}, who employs the \texttt{Prospector} modelling framework on galaxies from JADES GOODS-S and GOODS-N fields. The extensive medium-band coverage in these fields provides strong constraints on emission-line strengths, enabling robust and well-constrained SFR measurements. Using this dataset, we describe the SFMS with the following prescription:
\begin{equation}
\log_{10}(\mathrm{SFR}~[\mathrm{M_\odot/yr}]) = m - m_0 + a_0 r - a_1 \left[\max(0, m - m_1 - a_2 r)\right]^2,
\label{eq: schreiber}
\end{equation}
where $r = \log_{10}(1 + z)$ and $m = \log_{10}(M_* / 10^9 \mathrm{M_\odot})$ (parametric form from equation 9 of \citealt{Schreiber2015}). We obtain best-fit parameters of $m_0 = 0.5$, $m_1 = 0.4$, $a_0 = 2.0$, $a_1 = 0.3$, and $a_2 = 2.5$. The resulting fitted SFMS is shown in \autoref{fig: sfms}, color-coded by redshift.

Having constructed a parametric form of the SFMS across redshift, we next derive the corresponding SFH prior. We assume that galaxies grow along the SFMS and construct SFMS-based SFHs by integrating the relation backward in time. For a galaxy observed at $(z_i, M_{*,i})$, we evaluate its SFR on the SFMS and iteratively infer its past stellar mass and SFR at earlier times, thereby tracing its growth history toward higher redshift. This procedure yields a smooth SFMS-based trajectory that defines the expected evolution of the SFH. We illustrate this procedure in \autoref{fig: sfms}. For an example galaxy with $(z_i, \log_{10}(M_{*} / \mathrm{M_\odot})) = (1, 11)$, we iteratively compute the stellar mass formed over successive earlier time intervals, thereby tracing its evolution toward progressively lower stellar masses and SFRs at higher redshift. Importantly, we do not fix the SFH to this trajectory; instead, it provides only the baseline expectation, with the inferred SFHs allowed to deviate from it with significant scatter ($\sigma = 0.5$ in Student's distributions) as constrained by the data.

\subsubsection{Computational Implementation}

In the left panel of \autoref{fig:sfh prior combine}, we present example SFHs derived by backward integration of the SFMS for galaxies with $\log_{10}(M_*/\mathrm{M_\odot}) = 8.0$, shown at different observed redshifts ranging from $z = 1$ to $z = 16$. The resulting SFHs exhibit a characteristic rising behavior, with the SFR increasing from near zero at early cosmic times toward the observed redshift. However, this rising behavior represents the expected prior trajectory rather than necessarily the inferred SFH; inferred SFHs can span a wide range of forms, including starburst episodes or rapid quenching events.

Since the continuity SFH discretizes the SFH into age bins, the smooth SFH obtained from the backward integration must be translated into a binned representation. This procedure is illustrated in the right panel of \autoref{fig:sfh prior combine}. The SFR assigned to each age bin is computed as the mean SFR within that age bin interval, evaluated from the smooth SFH. In the figure, smooth SFHs are shown as dashed lines, while the corresponding binned SFHs are shown as solid lines. From these binned values, $\log_{10}\mathrm{(SFR_\mathrm{ratio})}$ between adjacent age bins are then calculated.  These $\log_{10}\mathrm{(SFR_\mathrm{ratio})}$ define the expected values, $\mu$, of the Student’s $t$ distribution. We adopt $\nu = 2$ and $\sigma = 0.5$ to allow substantial scatter around the SFMS predicted evolution, thereby accommodating bursty star formation or rapid quenching. 

Finally, evaluating the SFMS prior requires recursively integrating a galaxy's SFR and stellar mass evolution backward along the SFMS. A single evaluation has a median runtime of $0.0408^{+0.0024}_{-0.0041}$ seconds, which is computationally expensive within \texttt{Prospector} sampling. To mitigate this, we precompute the prior on a grid spanning redshifts $0.01 < z < 20$ and stellar masses $4 < \log_{10}(M_*/\mathrm{M}_\odot) < 15$, with a resolution of 0.01 in both dimensions. For all $2.2 \times 10^6$ grid points, we evaluate the full set of $\log_{10}(\mathrm{SFR}_{\mathrm{ratio}})$ values (corresponding to $N_{\mathrm{bins}} - 1 = 6$ ratios for $N_{\mathrm{bins}} = 7$ age bins), and construct a set of \texttt{scipy.interpolate.RegularGridInterpolator} objects, one for each ratio, defined over the $(z, \log_{10}(M_*/\mathrm{M}_\odot))$ grid. This enables near-instantaneous retrieval of all $\log_{10}(\mathrm{SFR}_{\mathrm{ratio}})$ values for arbitrary $(z, \log_{10}(M_*/\mathrm{M}_\odot))$, with a median runtime of $0.000125^{+0.000024}_{-0.000008}$ seconds, corresponding to a $\sim 330\times$ speed-up. The interpolated values remain highly accurate, with deviations relative to the direct backward-integration results of only $0.000000^{+1.33\times10^{-6}}_{-4.44\times10^{-16}}$ seconds.

\begin{figure*}
    \centering
    \includegraphics[width=\linewidth]{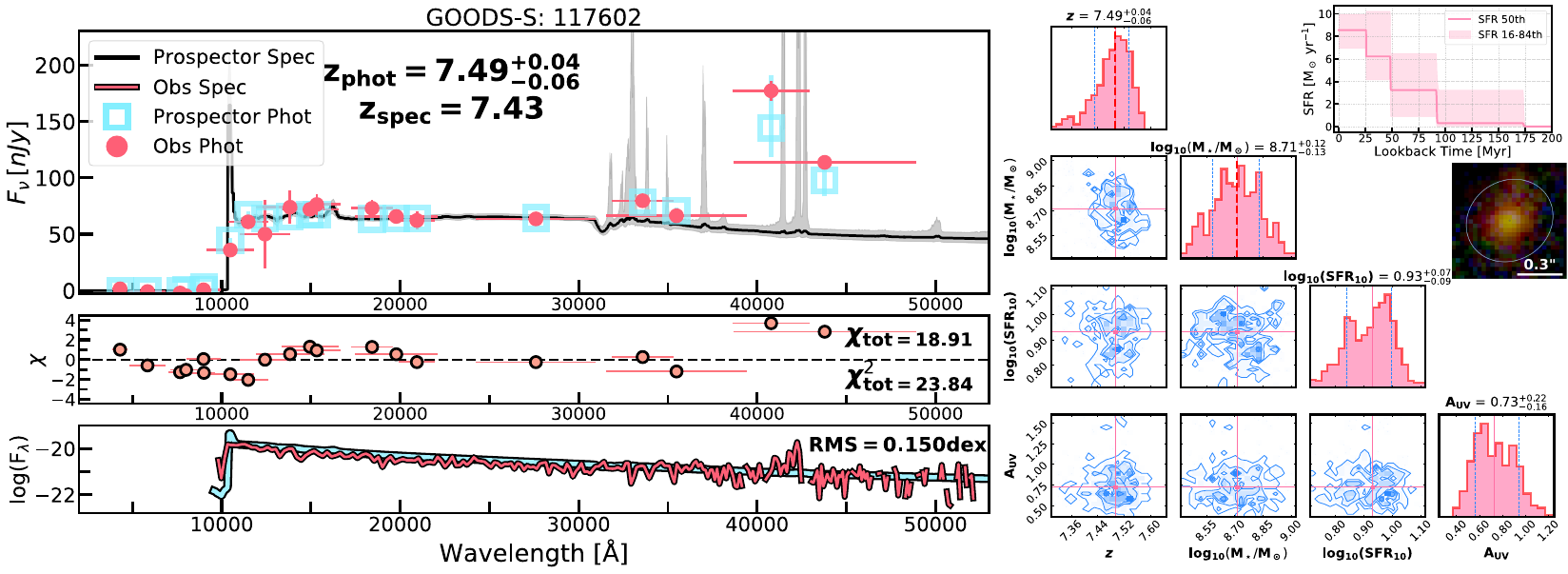}
    \caption{\texttt{Prospector} SED fit for a galaxy at $z_\mathrm{spec} = 7.43$, using 28-band photometry from HST/ACS+WFC3 (9 bands), JWST/NIRCam (11 bands), and MIRI (8 bands). The photometric coverage spans key spectral features, including the Lyman break, optical emission-lines, dust continuum, and potential AGN contributions. \textbf{\textit{Left panels:}} The modelled photometry and spectrum are shown as blue square and a black curve, respectively, while the observed photometry and NIRSpec/PRISM spectrum from JADES DR4 are shown in red. The best-fit photometric redshift is $z_\mathrm{phot} = 7.49^{+0.04}_{-0.06}$, in excellent agreement with the spectroscopic value. The middle panel of the left part displays the $\chi$ values for each photometric band. Total $\chi$ and $\chi^2$ values, computed using only HST and NIRCam bands, are 18.91 and 23.84, respectively. The bottom-left panel compares the modelled (blue) and observed NIRSpec/PRISM spectra (red), yielding a root mean square (RMS) residual of 0.150 dex (masking strong emission-lines) and 0.178 dex (full spectrum), illustrating the tight constraints provided by the deep, multi-band photometry (especially the medium-band filters) in JADES DR5. \textbf{\textit{Right panels:}} Corner plot, SFH, and imaging cutout for the same galaxy. The corner plot only displays the inferred redshift, stellar mass, SFR averaged over the past 10 Myr, and $A_\mathrm{UV}$. All parameters are well constrained.}
    \label{fig: sed z7 galaxy}
\end{figure*}

\subsection{Posterior Inference and Uncertainty Quantification}
\label{sec: posterior inference and uncertainty quantification}
As discussed above, our \prosp model consists of 18 free parameters that constrain the galaxy redshift, stellar mass, SFH, dust attenuation and emission, and the AGN contribution. The high dimensionality of the parameter space, together with the non-linear nature of the model, requires a statistically robust method to efficiently explore the posterior distributions. To infer these parameters, we use the nested sampling algorithm implemented in \texttt{dynesty} \citep{speagle2020}. \texttt{dynesty} is a public, open-source Python package that estimates Bayesian posterior distributions and evidences using dynamic nested sampling. By adaptively allocating samples based on the structure of the posterior distribution, dynamic nested sampling combines the advantages of Markov Chain Monte Carlo methods, which efficiently explore the posterior, with the ability to estimate evidences and to sample complex, multimodal distributions robustly.

For each free parameter, we summarise the posterior constraints using percentile-based statistics. Specifically, we draw 500 samples from the full posterior distribution according to their posterior weights. These samples are then treated as equally weighted to compute the 16th, 50th, and 84th percentiles of the resulting distribution. The 50th percentile is adopted as the \textit{best-fit value}, while the 16th and 84th percentiles define the lower and upper uncertainties, respectively. The upper uncertainty is computed as $(p_{84} - p_{50})$, and the lower uncertainty as $(p_{50} - p_{16})$.

To further accelerate the inference process, we employ the \texttt{Parrot} neural network emulator \citep{mathews2023}, which replaces computationally expensive stellar population synthesis (SPS) calculations with an artificial neural network (ANN). The \texttt{Parrot} emulator accurately reproduces SPS predictions while achieving execution speeds approximately $10^{3}$--$10^{4}$ times faster than standard SPS routines, with typical model evaluation times of $\sim 1$--$100\,\mu\mathrm{s}$ compared to $\sim 10$--$100\,\mathrm{ms}$ for conventional SPS calls. This substantial performance gain reduces the typical sampling runtime per galaxy to $\sim 1300\,\mathrm{s}$ when executed on a single core of an Intel Xeon Platinum~8380 CPU, enabling efficient and scalable analysis of large photometric datasets. In total, the sampling and post-processing analyses across all runs in the GOODS-S and GOODS-N fields require approximately $1.5$ million CPU hours, performed on the EPSRC Tier-2 National HPC Service, Cambridge Service for Data Driven Discovery (CSD3), operated by the University of Cambridge, as well as the DiRAC High Performance Computing facility.

\begin{figure*}
    \centering
    \includegraphics[width=\linewidth]{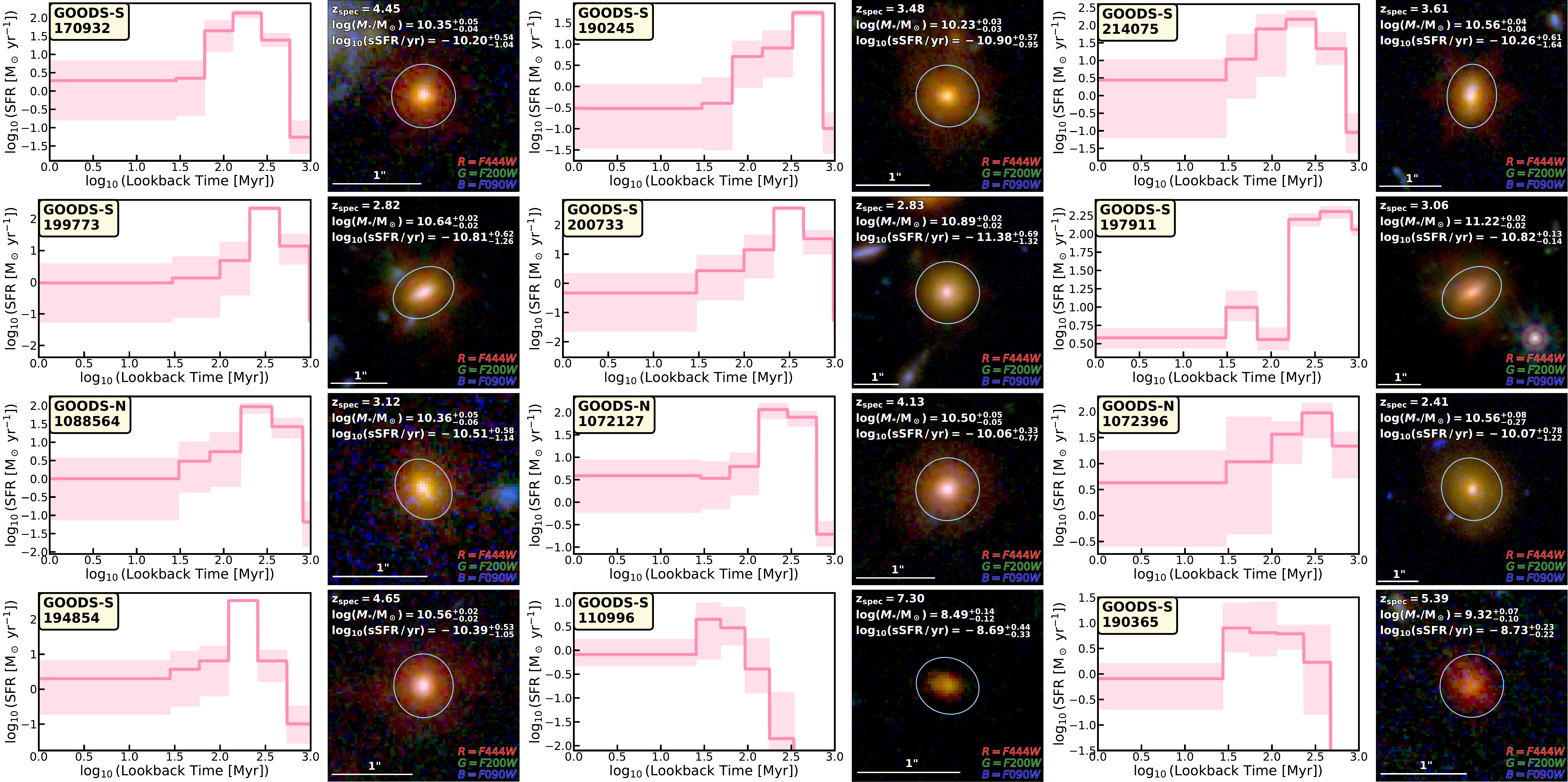}
    \caption{SFHs inferred using the non-parametric SFMS prior based on photometric data alone for 16 spectroscopically confirmed quiescent galaxies from the literature \citep{2023_carnall,deugenio_2024_quiescent,looser2024,baker_2025_quiescent,pascalau2026,baker2026}. For each galaxy, we show the modelled SFH on the left, and the corresponding RGB cutout, together with the stellar mass and sSFR, on the right. It is clear that our SFMS prior is able to accurately and robustly capture quiescent behaviour when the relevant features are present in the data.}
    \label{fig: combined quiescent sfh}
\end{figure*}

\section{Validation of the SED Fitting Framework} 
\label{sec: photometric sed quality}

In this section, we evaluate the robustness and reliability of the galaxy properties inferred from our photometric SED modelling framework. We begin by presenting a representative SED modelling example to illustrate the overall quality of the fits (\autoref{sec:sed fitting example}). We then demonstrate the powerful ability of the SFMS SFH prior to model quiescent galaxies SFH (\autoref{sec: quiescent galaxies sfh}). We in addition assess the agreement between observed and model-predicted photometry (\autoref{sec: Observed and Modelled Photometry}). Next, we quantify how strongly the inferred parameters are constrained by the data relative to the priors using the KL divergence metric (\autoref{sec: Kl divergence}). We further investigate the robustness of stellar mass measurements and their associated uncertainties, exploring their dependence on photometric flux and signal-to-noise ratio (\autoref{sec: mass f444w}, \autoref{sec: stellar mass uncertainty f444}). Unless otherwise stated, all demographic analyses use \texttt{Prospector} modelling results from all galaxies in the HST+NIRCam run, which correspond to all detected sources in the DR5 photometric catalogue.

\subsection{Illustrative Fits}
\label{sec:sed fitting example}
In \autoref{fig: sed z7 galaxy}, we present a randomly selected example of an SED fit for a galaxy that has NIRSpec/Prism spectra but is fitted using photometry only. We find a photometric redshift of $7.49^{+0.04}_{-0.06}$, closely matching its spectroscopic redshift of $7.43$ from NIRSpec PRISM data. The analysis uses 9 HST bands, 11 \textit{JWST} bands, and 8 MIRI bands. The left panel displays the observed photometry along with the predicted photometry and model spectra, while the middle panel shows the photometric residuals, quantified by the $\chi$ values, for HST and NIRCam bands only. As the degrees of freedom (DOF) are not clearly defined in nonlinear models, we refrain from calculating a reduced $\chi^2$ value \citep{andrae2010}. In the bottom panel, we evaluate the root mean square (RMS) residual between the modeled SED and the observed NIRSpec/PRISM spectrum. When masking regions with strong emission-lines, we find an RMS value of 0.150 dex, compared to 0.178 dex for the entire unmasked spectrum. The most significant residuals occur near strong emission-lines, particularly $\left[\rm O \, \textsc{ii}\right]\lambda\lambda3726,3729$, $\left[\rm O \, \textsc{iii}\right]\lambda5007$, and $\mathrm{H}\beta$, due to their strong sensitivity to variations in the ionization parameter \citep{kewley2013}. However, this typically has a negligible impact on the fit quality for most galaxies, as forbidden emission-lines generally contribute only a modest fraction of the flux in broad photometric bands. In some cases, however, particularly in galaxies with very strong nebular emission or extremely high equivalent widths, emission-lines can significantly affect, or even dominate, the broadband photometry \citep[e.g.,][]{shaerer2009,schaerer2010,pacifici2015, endsley2023, boyett2024, davis2024}. In addition, the presence of AGN can further complicate the interpretation, as AGN driven emission can enhance line fluxes and contribute non-stellar continuum emission, thereby contaminating broadband measurements.

On the right side of the plot, we show the corner plot, the SFH, and an image cutout displaying the \texttt{KRON} aperture around the galaxy. We present corner-plot distributions only for the redshift, stellar mass, SFR averaged over 10 Myr, and $A_\mathrm{UV}$. All parameters are well constrained. We do not show $\mathrm{SFR_{\rm 100 Myr}}$ because, by construction of the SFH, most of the stellar mass is formed within the past 100 Myr, leading to a strong correlation between $\mathrm{SFR_{\rm 100 Myr}}$ and $\log_{10}(M_*/\mathrm{M_\odot})$. This interpretation is further supported by the spectrum, which exhibits strong UV emission and a weak Balmer break, indicating that the galaxy is dominated by short-lived O- and B-type stars, with A-type stars contributing only a minor fraction of the total light.

\subsection{SFHs of Quiescent Galaxies}
\label{sec: quiescent galaxies sfh}
One potential concern of the SFMS prior (\autoref{sec: rising sfh prior}) is its ability to recover quiescent galaxies, as its expected behaviour assumes that galaxies lie on the SFMS. In principle, this introduces a penalty term in the likelihood during sampling when the inferred SFH deviates from the main sequence. However, we emphasise that, for galaxies that are genuinely quiescent with data comparable to JADES (e.g., what we consider here), this penalty does not impact the overall inference. In such cases, the likelihood is predominantly driven by the data rather than the prior, and the identification of quiescent galaxies under the SFMS prior is therefore particularly informative.

By adopting the SFMS prior, we eliminate solutions in which intrinsically star-forming galaxies are erroneously inferred to have quiescent SFH due to stochasticity in the sampling process, as can occur with uninformative priors. This leads to a more physically consistent recovery of stellar populations while still allowing truly quiescent systems to be identified when supported by the data.

In \autoref{fig: combined quiescent sfh}, we demonstrate the ability of our SFMS prior to recover quiescent galaxy SFHs. We present the SFHs for 16 example galaxies identified as quiescent based on spectroscopic analyses from \citet{2023_carnall,deugenio_2024_quiescent,looser2024,baker_2025_quiescent,pascalau2026,baker2026}. Our SFHs are inferred purely from photometry, with redshifts fixed to their spectroscopic values. We also show RGB cutouts (F090W, F200W, F444W), with the stellar mass and specific SFR overlaid. It is clear that all of these galaxies exhibit quiescent behaviour in the SFHs recovered using the SFMS prior. This provides strong evidence for the ability of the SFMS prior to model and recover genuine quiescent behaviour, which may not be reliably captured when using uninformative priors.

\begin{figure*}
    \centering
    \includegraphics[width=\linewidth]{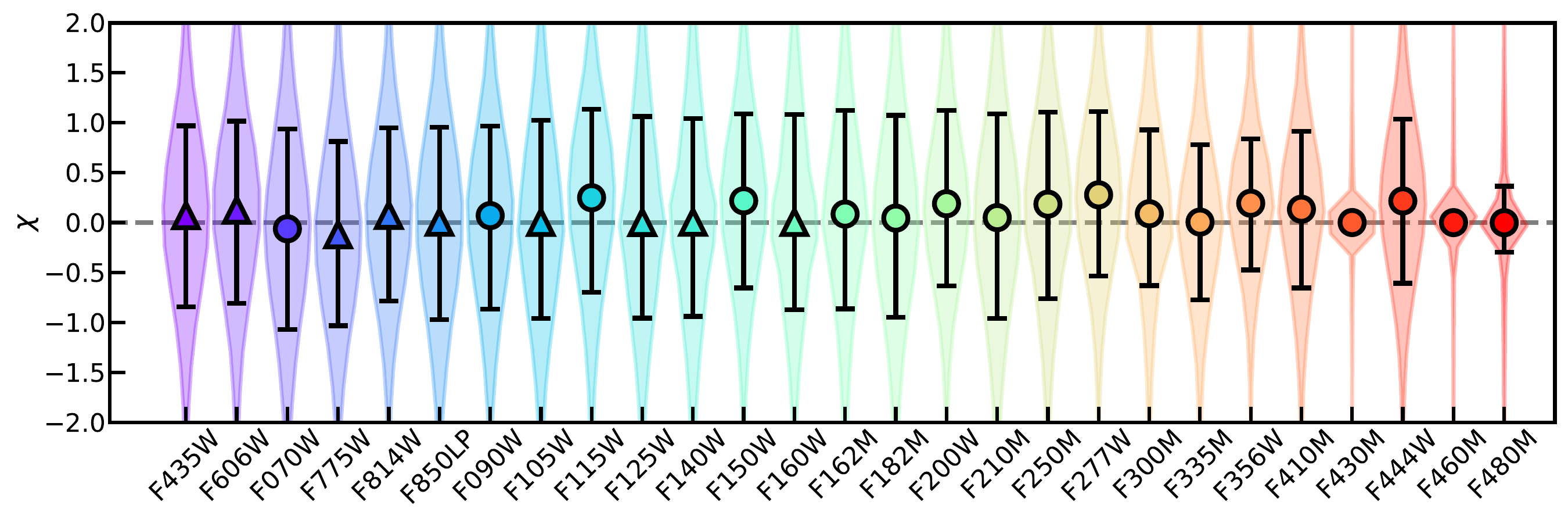}

    \vspace{1.0em}
 
    \includegraphics[width=\linewidth]{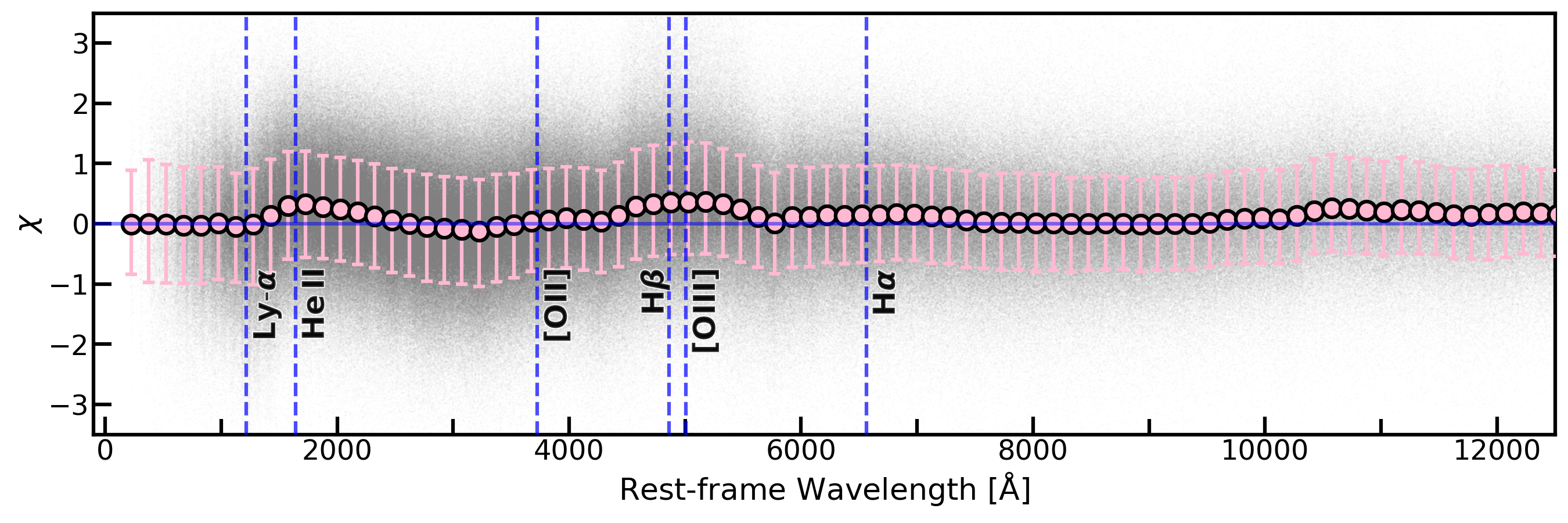}

    \caption{Distribution of $\chi$ values across all HST and NIRCam filter bands from all galaxies, computed as $\chi = (f_{\mathrm{obs}} - f_{\mathrm{Prospector}}) / \sigma_{\mathrm{obs}}$. \textbf{\textit{Top panel:}} $\chi$ distributions for individual bands, shown as violin plots. Circles indicate NIRCam bands and triangles indicate HST bands; vertical error bars denote the $1\sigma$ spread. The low $\chi$ values across all filters reflect the high fidelity of the SED fits. \textbf{\textit{Bottom panel:}} $\chi$ as a function of rest-frame wavelength, computed by dividing each filter’s pivot wavelength by the galaxy’s best-fit redshift. Median values are shown in bins of $150$\,\AA, with errorbars representing 16th and 84th percentiles. We find no significant residual trends across the spectrum. A minor excess is seen near nebular emission-line wavelengths, though this has negligible impact on the inferred stellar population and SFR parameters, as discussed in \autoref{sec: Observed and Modelled Photometry} and \autoref{sec: sfr and prosp comparison}.}
    \label{fig: chi all}
\end{figure*}

\subsection{Goodness-of-Fit Statistics}
\label{sec: Observed and Modelled Photometry}
Since \texttt{Prospector} is a forward-modelling Bayesian SED modelling framework, it infers galaxy physical properties and subsequently predicts the corresponding model photometry using either \texttt{FSPS} or the \texttt{Parrot} emulator adopted in this work. In this section, we assess the quality of our SED fits by comparing the observed fluxes with the \texttt{Prospector}-predicted photometry.

Specifically, we compute the $\chi$ value for each photometric band, defined as
\begin{equation}
    \label{eq: chi formulae}
    \chi = \frac{f_{\mathrm{obs}} - f_{\mathrm{Prospector}}}{\sigma_{\mathrm{obs}}},
\end{equation}
where $f_{\mathrm{obs}}$ and $f_{\mathrm{Prospector}}$ are the observed and modelled fluxes, respectively, and $\sigma_{\mathrm{obs}}$ is the uncertainty on the observed photometry. We assess the quality of the \texttt{Prospector}-modelled photometry in \autoref{fig: chi all}.

In the top panel of \autoref{fig: chi all}, we present the $\chi$ distributions for each photometric band from all galaxies. Triangles denote the median $\chi$ values for HST ACS/WFC3 and WFC3/IR bands, while circles indicate the median $\chi$ values for \textit{JWST} bands. The error bars represent the 16th--84th percentile range (1$\sigma$), and the full $\chi$ distributions for each band are visualised using violin plots. Overall, we find that our \texttt{Prospector} fits perform well across all photometric bands, with median $\chi$ values tightly clustered around zero, indicating excellent agreement between the observed and modelled photometry.

In the bottom panel of \autoref{fig: chi all}, we present the $\chi$ values as a function of the rest-frame wavelength of each filter band from all galaxies. The rest-frame wavelength is computed by dividing the pivot wavelength of each filter by $(1 + z_{\mathrm{Prospector}})$. Individual measurements are shown as small grey points, and we further bin the $\chi$ values in intervals of 150~\AA. The binned medians are indicated by pink circles, with error bars representing the 16th and 84th percentiles ($1\sigma$).

In this representation, we find $\chi$ to also be very small overall. We find only small residuals near the Balmer emission lines. This not only shows good SED modelling in our framework, but also indirectly demonstrates the suitability of the SFMS prior. If the SFMS prior were very strong, then galaxy SFHs would be forced toward the SFMS prior, thereby biasing the predicted photometry, which is not what we observe. Of course, dust and metallicity may also contribute to this.

For the minor positive offset in $\chi$, these offset happened mostly at rest-frame wavelengths corresponding to prominent nebular emission-lines. In particular, excesses are evident near the Balmer recombination line H$\beta$ ($\lambda4861$) and the strong oxygen doublet [O\,\textsc{iii}]$\lambda\lambda4959,5007$, but not near H$\alpha$ ($\lambda6563$). In these regions, the binned median $\chi$ values reach $\sim 0.3$ above zero. According to \autoref{eq: chi formulae}, a positive $\chi$ offset implies that the \texttt{Prospector} models systematically \emph{underestimate} the observed fluxes in these bands.

These systematic offsets indicate that our current SED modelling approach, which combines \texttt{Prospector} with the \texttt{Parrot} neural-network emulator, slightly underpredicts nebular emission-line fluxes. There are several reasons for this. First, we use seven age bins for the SFH, and the first age bin spans 0–30 Myr. Very recent starbursts (less than 5–10 Myr) are therefore smoothed out by this broader bin, leading to an underprediction of the nebular emission-line strength from \ion{H}{ii} regions associated with recent star formation. Second, in the \texttt{Parrot}-based inference adopted in this work, nebular emission is effectively regulated only through the gas-phase metallicity parameter, while all other nebular parameters, most notably the ionization parameter ($\log U$), gas density, and the detailed shape of the ionizing radiation field are held fixed at their default values in the underlying \texttt{FSPS} implementation. This restricted parameterisation limits the model’s ability to independently adjust nebular line strengths in response to varying physical conditions, particularly in galaxies with high specific SFRs or extreme emission-line equivalent widths. Thirdly, while our model includes an AGN component, it primarily contributes in the mid-infrared component through the parameters $f_{\mathrm{AGN}}$ and $\tau_{\mathrm{AGN}}$, and does not account for AGN-driven ionising radiation that could enhance nebular emission lines.

Because \texttt{Parrot} is trained to emulate the baseline \texttt{FSPS} models used within \texttt{Prospector}, it necessarily inherits the limitations of the adopted nebular emission prescriptions. If the default \texttt{FSPS} nebular models (e.g.\ \texttt{CloudyFSPS}) underpredict the strengths of prominent emission-lines under certain physical regimes, such as elevated ionization parameters, harder ionizing spectra, or non-solar abundance patterns, then \texttt{Parrot} will reproduce these same underestimates by construction. This provides a natural explanation for the systematic positive residuals observed around H$\beta$, [O\,\textsc{iii}], and H$\alpha$ in the broadband photometry.

A promising avenue for mitigating this limitation is the incorporation of a more flexible nebular emission model, such as the \texttt{Cue} photoionization emulator \citep{li_cue}. \texttt{Cue} is a neural-network emulator trained to reproduce the output of the photoionization code \texttt{Cloudy} \citep{ferland1998_cloudy, ferland2013_cloudy, ferland2017_cloudy}, enabling fast and accurate predictions of nebular line and continuum emission across a broad range of ionizing spectra and gas conditions. By emulating \texttt{Cloudy}, \texttt{Cue} achieves orders-of-magnitude improvements in computational efficiency while retaining high physical fidelity. Crucially, \texttt{Cue} allows key nebular parameters, including the ionizing spectral shape, gas-phase metallicity, ionization parameter, and elemental abundance ratios, to vary independently. As a result, it can reproduce strong nebular emission lines with substantially improved flexibility and accuracy compared to standard fixed-template nebular emission prescriptions.

However, \texttt{Cue} is specifically designed to model nebular emission and does not independently infer stellar population properties such as stellar masses, SFHs, dust attenuation, or AGN components. To fully exploit \texttt{Cue} within a Bayesian SED modelling framework, it must therefore be coupled to a stellar population model that provides the ionizing radiation field. In practice, this would require explicitly generating the stellar continuum using \texttt{FSPS}, followed by the application of \texttt{Cue} to compute nebular line and continuum emission based on the resulting ionizing spectrum.

While this approach provides substantially greater physical flexibility in modelling nebular emission than the standard \texttt{FSPS} prescriptions, it is inherently more computationally expensive than the \texttt{Parrot} emulator. Unlike \texttt{Parrot}, which directly maps galaxy physical parameters to photometry, a combined \texttt{FSPS}+\texttt{Cue} forward model requires explicit stellar spectral synthesis followed by photoionization emulation at each likelihood evaluation. 

Although \texttt{Cue} is more efficient than direct photoionization calculations, the overall runtime of an \texttt{FSPS}+\texttt{Cue} implementation remains orders of magnitudes higher than that of \texttt{Parrot}. Such a framework is therefore better suited to detailed analyses of individual systems or modest-sized samples. Given the large size of our galaxy sample, we adopt \texttt{Parrot} for its computational efficiency, which enables scalable Bayesian SED inference across $\sim 500{,}000$ galaxies in JADES DR5 release. Consequently, we employ the standard \texttt{FSPS}-based nebular emission model within \texttt{Prospector}, and interpret the small residual excesses ($\chi \sim 0.3$) near strong emission lines as well-understood features that are not significant enough to affect the SED modelling.

\begin{figure*}
    \centering
    \includegraphics[width=0.99\linewidth]{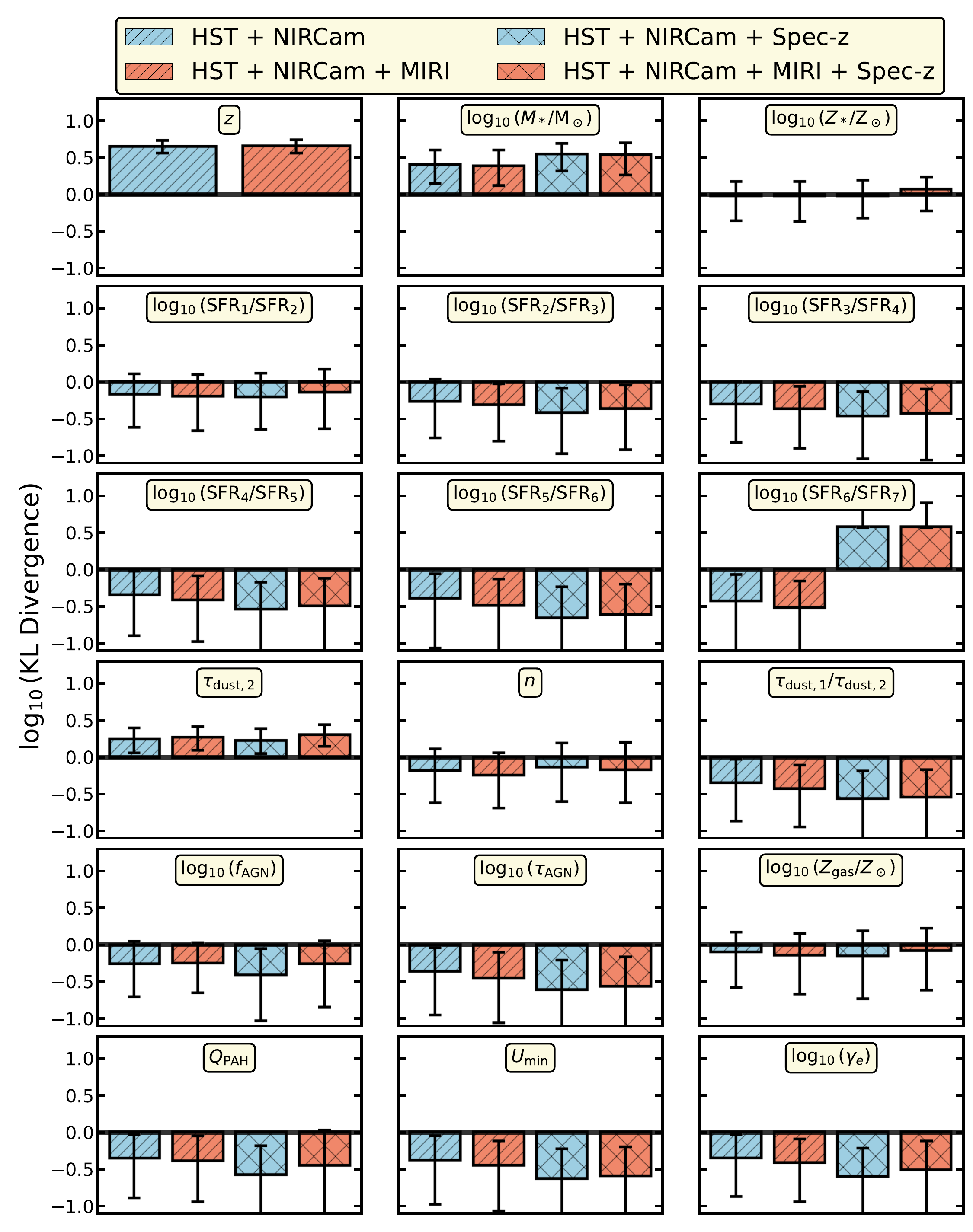}
    \caption{Kullback–Leibler (KL) divergence for the 18 free parameters in our \prosp SED model. For each parameter, we compute the KL divergence across the full galaxy sample, with the bar height indicating the median value and the error bars showing the 16th and 84th percentiles. A strongly data-driven constraint corresponds to $\mathrm{KL} > 1.0$, i.e., $\log_{10}(\mathrm{ KL}) > 0.0$, as indicated by the horizontal line. Each parameter is shown for four different \prosp configurations. Blue bars correspond to runs using HST + NIRCam photometry, while red bars indicate runs that additionally include MIRI data. Configurations with spectroscopic redshifts fixed are marked with cross hatching, whereas runs without spectroscopic redshift constraints are shown with diagonal hatching.}
    \label{fig: kl divergence}
\end{figure*}

\subsection{Information Gain and Posterior Constraining Power (KL divergence)}
\label{sec: Kl divergence}
The inference of our 18 free parameters in the \texttt{Prospector} SED model (\autoref{sec: inferring galaxy properties}) involves both uniform and non-uniform priors, which can significantly influence the inferred posterior distributions through the combination with the likelihood function. It is therefore important to quantify the extent to which each parameter is constrained by the data, as opposed to being dominated by the prior.

To assess this, we employ the Kullback–Leibler (KL) divergence \citep{kl_divergence}, which quantifies the information gain when updating from a prior distribution $Q(x)$ to a posterior distribution $P(x)$. In our context, $P(x)$ denotes the posterior distribution of a given parameter, while $Q(x)$ denotes its prior distribution. For continuous distributions, the KL divergence is defined as
\begin{equation}
    \label{eq:kl_divergence_continuous}
    D_{\mathrm{KL}}(P \| Q)
    =
    \int P(x)\,\ln\!\left(\frac{P(x)}{Q(x)}\right)\,dx.
\end{equation}

\noindent This expression can be interpreted as the expectation value under the posterior distribution:
\begin{equation}
    D_{\mathrm{KL}}(P \| Q)
    =
    \mathbb{E}_{x \sim P}
    \left[
        \ln\!\left(\frac{P(x)}{Q(x)}\right)
    \right].
\end{equation}

\noindent In practice, the posterior distribution is represented by a set of Monte Carlo samples $\{x_i\}_{i=1}^N$ drawn from $P(x)$. By the law of large numbers, the expectation value can therefore be approximated as
\begin{equation}
    \label{eq: kl final eq}
    D_{\mathrm{KL}}(P \| Q)
    \approx
    \frac{1}{N}
    \sum_{i=1}^{N}
    \ln\!\left(\frac{P(x_i)}{Q(x_i)}\right),
    \quad
    x_i \sim P(x).
\end{equation}
Thus, the KL divergence measures the average logarithmic difference between posterior and prior, weighted by the posterior probability density. $D_{\mathrm{KL}}(P \| Q)$ always $\geq 0$. A larger KL divergence values indicating greater information gain from the data. A value close to zero implies that the posterior remains similar to the prior, indicating that the parameter is weakly constrained by the observations. In general, according to \autoref{eq: kl final eq}, a $D_{\mathrm{KL}}$ of 1 indicates a very strongly data-constrained inference, as this means that posterior constraints from observed data are better by a factor of $e$ than those of the prior.

In \autoref{fig: kl divergence}, we present the KL divergence for all 18 free parameters in the four different \prosp configurations, combining results from the GOODS-S and GOODS-N fields (see \autoref{tab: number of galaxies snr}). For each parameter, the height of the bar indicates the median (50th percentile) KL value across galaxies, while the error bars denote the 16th and 84th percentiles of the distribution. Runs including additional MIRI photometry are shown in red (otherwise blue), while runs in which the redshift is fixed to spectroscopic values are indicated with cross-hatching (otherwise diagonal hatching). We show the $\log_{10}$ of the KL divergence metric, and a horizontal line at a KL divergence of $1.0$ is drawn for each free parameter.

We begin with redshift, which exhibits the largest KL divergence. This is expected, as the redshift prior is uniform over a wide range ($0.01 < z < 20$), whereas the posterior is tightly constrained by the photometric and/or spectroscopic data. Consequently, the inferred redshift is strongly data-dominated rather than prior-driven. We only show redshifts KL Divergence for runs without fixing redshifts to spectroscopic values.

In addition, a similarly large KL divergence is observed for stellar mass, indicating that the stellar mass is also strongly constrained by the data. Metallicity is moderately constrained by the data, since we do not have spectra and it is inferred from photometry alone. A higher KL divergence is found for galaxies with more than five NIRCam medium bands, with which we can more effectively capture emission-line strengths.

The $\log_{10}(\mathrm{SFR})$ ratios show moderate KL values. Despite applying the physically motivated SFMS prior introduced in this work to the SFH (\autoref{sec: rising sfh prior}), the data still provide substantial additional constraints, leading to clear information gain relative to the prior. Interestingly, the sixth $\log_{10}(\mathrm{SFR})$ ratio exhibits enhanced KL values when spectroscopic redshifts are fixed, suggesting that breaking the redshift–SFH degeneracy significantly improves constraints on the oldest SFH bin (which is also the longest and therefore contains a large fraction of the older stellar population mass).

Dust attenuation parameters are also generally well constrained by the data. Among them, $\tau_{\mathrm{dust},2}$ (the diffuse dust optical depth) exhibits the highest KL divergence compared to $n$ and $\tau_{\mathrm{dust},1}/\tau_{\mathrm{dust},2}$. This behaviour is expected, as $\tau_{\mathrm{dust},2}$ directly controls the overall normalization of the attenuation and therefore has a strong and immediate impact on the observed SED. In contrast, $n$ (the attenuation curve slope) and the ratio $\tau_{\mathrm{dust},1}/\tau_{\mathrm{dust},2}$ primarily modulate the shape and relative scaling of the attenuation within already well-understood and physically motivated prior bounds, leading to comparatively smaller KL values.

Furthermore, we find that the inclusion of MIRI photometry increases the KL divergence for $\log f_{\mathrm{AGN}}$, $\log \tau_{\mathrm{AGN}}$, and also dust emission ($Q_\mathrm{PAH}$, $U_\mathrm{min}$, $\log_{10}(\gamma_e)$). This behaviour is expected, as the longer-wavelength MIRI data directly probe hot dust emission and therefore provide stronger constraints on AGN-related parameters, reducing prior dominance.

Overall, we find that, provided the relevant filter bands are available to probe each parameter, all free parameters are predominantly constrained by the observational data rather than by the prior, indicating that the fits are robust and of high quality.

\begin{figure*}
    \centering
    \includegraphics[width=\linewidth]{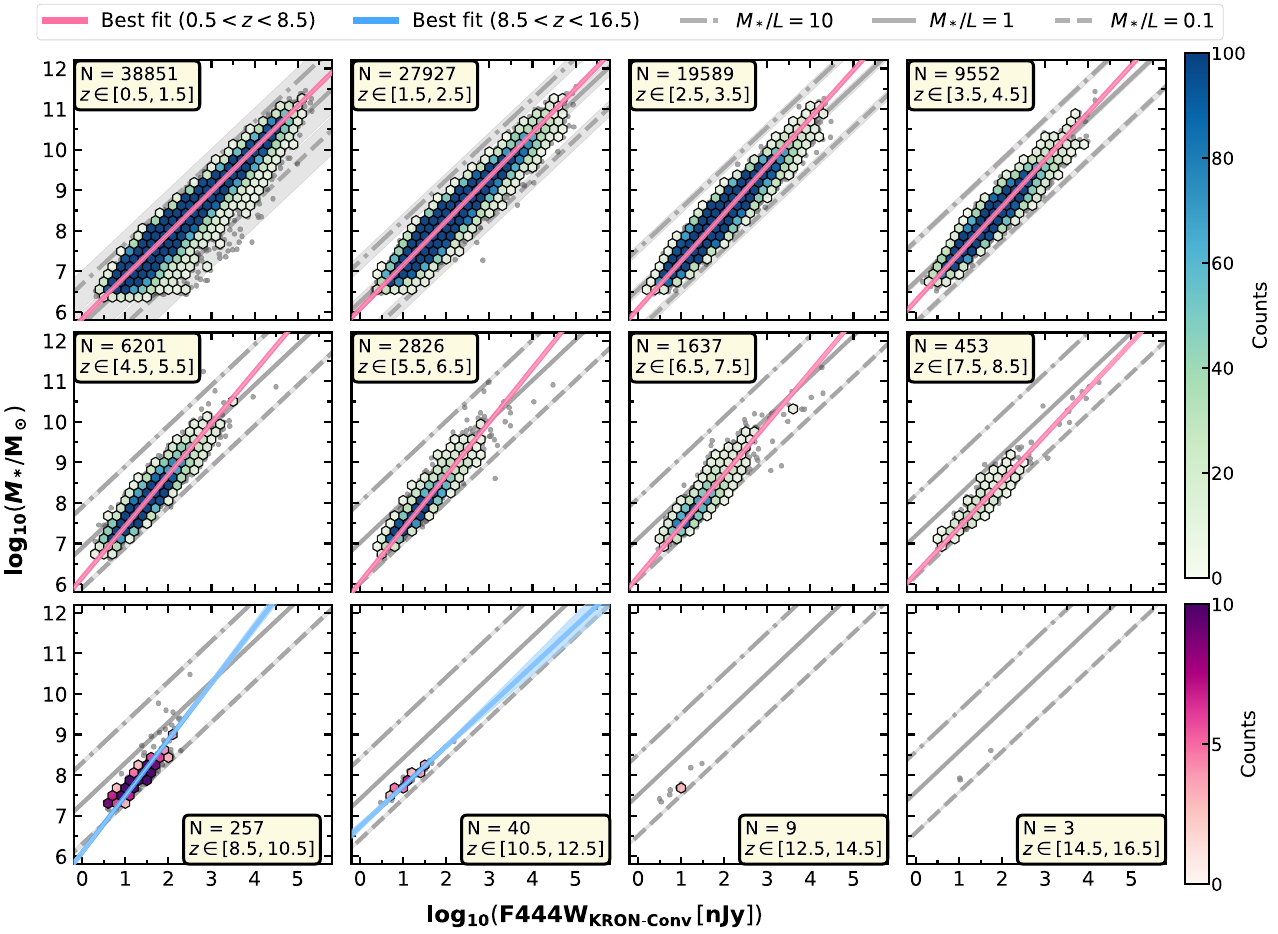}
    \caption{Hexbin plot showing $\log_{10}(M_*/\mathrm{M}_\odot)$ as a function of $\log_{10}(\mathrm{F444W}_{\mathrm{KRON\text{-}Conv}}\,\mathrm{[nJy]})$, across 12 redshift bins. Hexbins are shown wherever at least three data point is present. Galaxies with bin centers in the range $0.5 \leq z \leq 8.5$ are shown using a green–blue colormap, while those in the range $8.5 \leq z \leq 16.5$ are shown with a purple–pink colormap. A linear model with intrinsic scatter (see \autoref{sec: linear with scatter fit}) is fit to each redshift bin, with the best-fit relations overplotted as blue and pink lines. Constant mass-to-light ratios of 10, 1, and 0.1 are overplotted, shown as dash-dotted, solid, and dashed lines. The increasingly steeper best-fit slope arises because the filter bands sample progressively bluer rest-frame wavelengths, which provide less constraints on the underlying older stellar populations. Analyses are performed using galaxies that satisfy the following criteria: $\mathrm{SNR}_{\mathrm{F444W}} \geq 3$, $\mathrm{N}_{\mathrm{bands,\,NIRCam}} \geq 6$, and redshift uncertainties such that $(z_{84} - z_{16}) < 0.5 \times z_{50}$.}
    \label{fig: mass f444w}
\end{figure*}

\subsection{Mass-to-Light Ratios and Flux Scaling}
\label{sec: mass f444w}
Galaxy stellar mass is dominated by long-lived, low-mass stars that accumulate over a galaxy’s lifetime \citep[e.g.,][]{bell_2001, bell2003, conroy2009}. Although these stars contribute only modestly to the instantaneous luminosity at short wavelengths, they account for the majority of the total stellar mass and dominate the emission at rest-frame near-infrared wavelengths. In contrast, rest-frame ultraviolet and blue optical emission is dominated by short-lived massive stars (O- and B-type stars), making it highly sensitive to recent star-formation activity and dust attenuation. 

\textit{JWST}’s F444W filter provides a particularly effective probe of stellar mass across a wide redshift range. At low and intermediate redshifts, F444W samples rest-frame near-infrared and optical wavelengths that are dominated by evolved stars and characterised by relatively stable mass-to-light ratios \citep[e.g.,][]{bell_2001, zibetti2009, meidt2014, mcgaugh_2014}. Toward higher redshifts, where the band progressively shifts into the rest-frame optical, F444W remains the reddest available NIRCam filter and therefore continues to trace the longest-wavelength stellar emission accessible. This makes F444W especially valuable for anchoring stellar mass estimates in high-redshift galaxies, where shorter-wavelength bands are increasingly dominated by young stars and emission line, as has been demonstrated by various studies \citep[e.g.,][]{simmonds_2024_lowmass_bursty, simmonds_2024_ionising, Harvey2025}.

An additional advantage of F444W is its reduced sensitivity to dust attenuation compared to ultraviolet and optical bands. At wavelengths around 4.4~$\mu$m, extinction is substantially lower, allowing F444W to probe the underlying stellar population even in dusty star-forming systems \citep[e.g.,][]{calzetti2000, meidt2014, whitaker2017_dust}. For these reasons, F444W serves as a physically motivated and empirically validated proxy for constraining stellar mass. Motivated by these considerations, we examine the empirical relationship between stellar masses inferred with \texttt{Prospector} and observed F444W photometry over a wide redshift range, and investigate how this relation evolves with cosmic time.

In \autoref{fig: mass f444w}, we present the relationship between stellar mass and observed $\log_{10}(\mathrm{F444W}_{\mathrm{KRON\text{-}Conv}})$ flux for our samples. These comparisons are performed using galaxies that satisfy the following criteria: $\mathrm{SNR}_{\mathrm{F444W}} \geq 3$, $\mathrm{N}_{\mathrm{bands,\,NIRCam}} \geq 6$, and redshift uncertainties such that $(z_{84} - z_{16}) \leq 0.5 \times z_{50}$. At ultra-high redshift, $z > 8$, we present our galaxy sample that overlaps with the $z > 8$ high-redshift galaxies in \citet{Hainline2026}, selecting those with redshift differences of $\Delta z < 1.0$. The reason for making this choice is that, at such high redshifts, galaxies are very faint. We are using the \texttt{KRON\_CONV} aperture to capture galaxies' light, which biases our results by including background noise that would not affect photometric measurements at lower redshifts, but does affect our case because the sources are faint. \citet{Hainline2026} uses a smaller circular aperture, effectively capturing only the light from the galaxies themselves and excluding background noise, leading to better photometry.

We divide these samples into 12 redshift bins spanning $z=1$–16. Galaxies in the range $0.5<z<8.5$ are shown using a blue–green colourmap, while those at $8.5<z<16.5$, where the sample size decreases rapidly, are displayed using a purple–red colourmap. Within each redshift bin, we fit a linear relation between stellar mass and F444W flux, explicitly accounting for an intrinsic scatter term that captures the dispersion about the main trend. The resulting best-fit relations are overplotted in \autoref{fig: mass f444w} as straight lines (pink for $0.5<z<8.5$ and blue for $8.5<z<16.5$), with shaded regions indicating the 16th–84th percentile uncertainty. Full details of the modelling methodology are provided in Appendix \ref{sec: linear with scatter fit}.

Furthermore, we additionally overplot lines of constant stellar mass-to-light ratio ($M_*/L$) in each redshift bin to better interpret the mass–flux relations. Specifically, the three gray lines correspond to $M_*/L = 10$, $1$, and $0.1$ in solar units, shown as dash–dotted, solid, and dashed lines, respectively. For each panel, the gray line is computed at the central redshift of the bin, while the gray shaded region reflects the variation expected across the bin width, arising from the redshift dependence of the luminosity distance and the $(1+z)$ cosmological correction.

The constant $M_*/L$ lines are constructed by converting the observed F444W flux density into a rest-frame monochromatic luminosity and adopting a fixed dimensionless mass-to-light ratio in solar units. We start from
\begin{equation}
    M_*=C_\nu\,L_\nu,
\end{equation}
where $C_\nu$ is the stellar mass-to-light ratio at the relevant rest-frame frequency. The rest-frame monochromatic luminosity is related to the observed flux density $f_\nu$ by
\begin{equation}
    L_\nu=\frac{4\pi D_L^2(z)}{1+z}\,f_\nu,
\end{equation}
with $D_L(z)$ is the luminosity distance.

\noindent We define the constant-$M_*/L$ lines by expressing $C_\nu$ in solar units,
\begin{equation}
    C_\nu \equiv \left(\frac{M_*}{L_\nu}\right)= C\,\left(\frac{M_\odot}{L_{\nu,\odot}}\right),
\end{equation}
where $C$ is a dimensionless constant (e.g. $C=0.1,\,1,\,10$) and $L_{\nu,\odot}$ is the solar monochromatic luminosity evaluated at the same rest-frame frequency. Substituting into the above gives
\begin{equation}
    \label{eq:ml_relation}
    \log_{10}\!\left(\frac{M_*}{M_\odot}\right)
    =
    \log_{10}C
    +
    \log_{10}\!\left[
        \frac{4\pi D_L^2(z)}{1+z}\,
        \frac{f_\nu}{L_{\nu,\odot}}
    \right],
\end{equation}
which defines the straight lines of constant $M_*/L$ overplotted in \autoref{fig: mass f444w}.

Across all redshift bins, we find a clear and approximately linear correlation between observed F444W flux and stellar mass, such that galaxies with brighter F444W emission are systematically more massive. This tight mass–flux relation persists over the full redshift range probed, remaining well defined even at the highest redshifts considered, despite the substantially reduced number statistics at early cosmic times. The overall scatter about the best-fitting relations is modest, indicating that variations in stellar population properties, such as stellar age, metallicity, SFH, and dust attenuation, introduce only limited additional dispersion in the mass–flux plane.

Furthermore, the overplotted constant-$M_*/L$ lines enable a direct physical comparison between the observed mass–flux relations and simple expectations based on fixed stellar mass-to-light ratios. At low and intermediate redshifts, we find that nearly all galaxies lie between the $M_*/L = 10$ and $M_*/L = 0.1$ relations, with the best-fitting mass–flux relations exhibiting slopes closely matching those expected for constant $M_*/L$. This behavior is consistent with F444W probing rest-frame near-infrared wavelengths, where the emission is dominated by evolved older stellar populations. In this regime, the observed light is primarily produced by long-lived G-, K-, and M-type stars and their evolved phases (e.g.\ red giant branch and asymptotic giant branch stars), which trace the time-integrated SFH and therefore closely track the total stellar mass. The relatively small scatter about the best-fitting relations reflects secondary effects such as dust attenuation, and metallicity variations, while the overall mass scale remains robustly anchored by the near-infrared emission.

At higher redshifts, the behavior of the mass–flux relation evolves systematically as F444W shifts blueward in the rest frame. As the filter begins to probe rest-frame optical wavelengths and approaches the Balmer/4000~\AA\ break, the observed flux remains a reasonably effective tracer of stellar mass, but the mass-to-light ratio becomes increasingly sensitive to stellar age and recent SFH. Once F444W moves blueward of the Balmer break, it begins to sample rest-frame near-ultraviolet emission, where the observed light is dominated by short-lived, massive O- and B-type stars, and F444W begins to trace recent star formation rather than the bulk of the total stellar population, fundamentally altering the relationship between observed flux and stellar mass. In this regime, galaxies with similar rest-frame ultraviolet fluxes can span a wide range of total stellar masses, depending on how much older, lower-luminosity stellar mass has been built up beneath the young population. As a result, this produces a slightly steeper mass–flux relation and larger scatter at high redshift.

\begin{figure*}
    \centering
    \includegraphics[width=\linewidth]{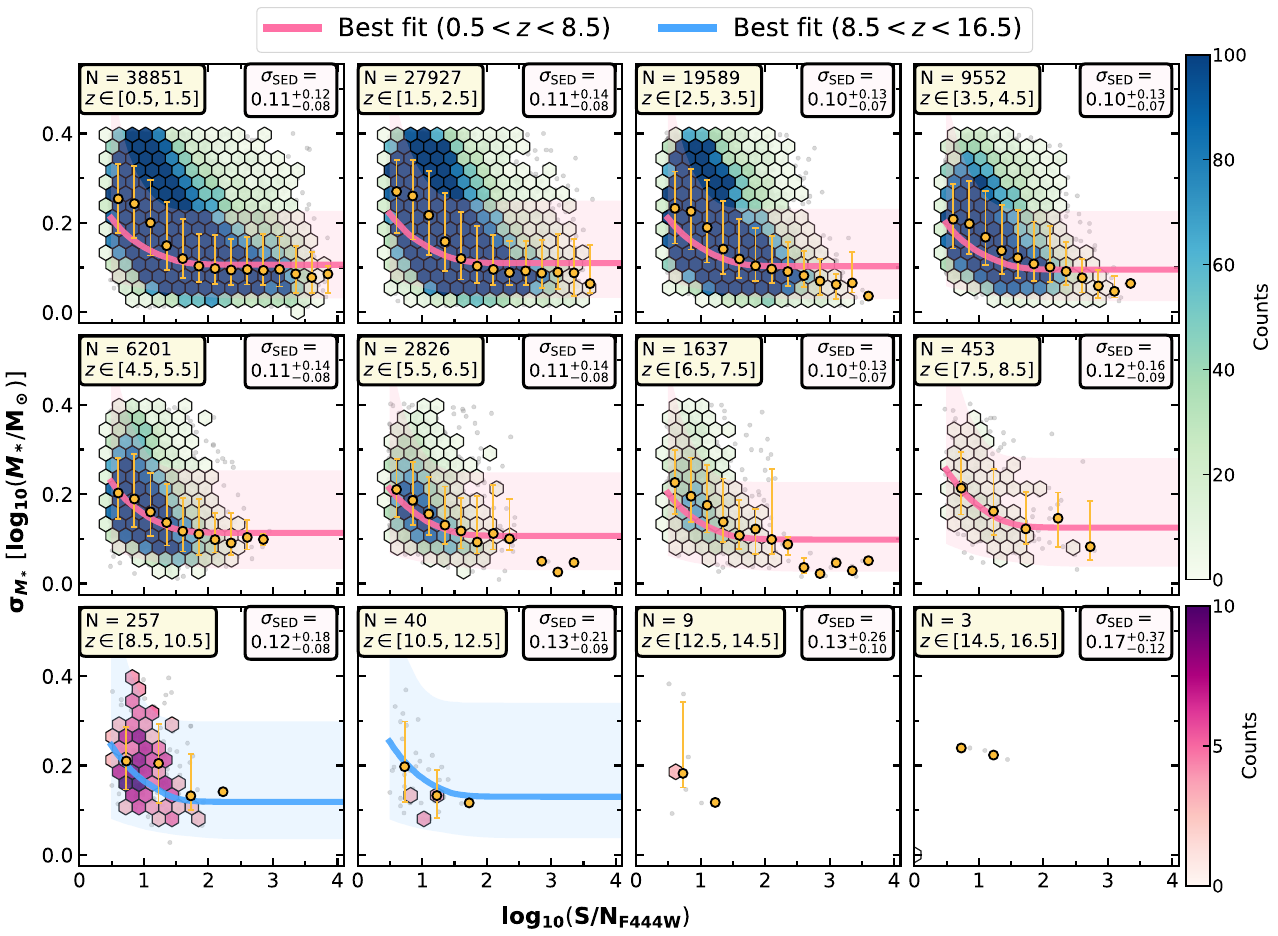}
    \caption{Hexbin plot showing the uncertainty in stellar mass, $\sigma_{M_*} = \sigma[\log_{10}(M_*/\mathrm{M}_\odot)]$, as a function of $\log_{10}(\mathrm{SNR}_{\mathrm{F444W}})$ across 12 redshift bins. Hexbins are shown wherever at least three data point is present. Galaxies with bin centers in the range $0.5 \leq z \leq 8.5$ are shown in a green–blue colormap, while those in $8.5 \leq z \leq 16.5$ are shown in purple–pink. For each redshift bin, we fit the binned data using the functional form described in \autoref{eq:sigma_m snr fit}, which consists of two components: the decrease in uncertainty with increasing SNR and the minimum uncertainty floor from the degeneracy in SED modelling. We find that $\sigma_{M_*} \, [\log_{10}(M_*/\mathrm{M}_\odot)]$ decreases with increasing SNR. At $\log_{10}(\mathrm{SNR}_{\mathrm{F444W}}) \sim 1$--$1.3$, further increases in SNR do not significantly improve $\sigma_{M_*}$, as the precision in this regime is dominated by the intrinsic uncertainty and degeneracy of the SED modelling. In this limit, the uncertainty approaches $\sigma_{\mathrm{SED}}$, whose value is indicated in the top-right corner of each panel. Analyses are performed using galaxies that satisfy the following criteria: $\mathrm{SNR}_{\mathrm{F444W}} \geq 3$, $\mathrm{N}_{\mathrm{bands,\,NIRCam}} \geq 6$, and redshift uncertainties such that $(z_{84} - z_{16}) < 0.5 \times z_{50}$.}
    \label{fig: sigma_m f444w}
\end{figure*}

\subsection{Stellar Mass Uncertainty Scaling with Signal-to-Noise}
\label{sec: stellar mass uncertainty f444}

\label{sec: snr sigma m}
It is well known and demonstrated explicitly in \autoref{sec: mass f444w} that galaxy stellar mass is tightly correlated with the observed flux in the F444W band. In this subsection, we investigate how precisely we can constrain stellar mass uncertainties as a function of increasing $\mathrm{SNR_{F444W}}$.

In \autoref{fig: sigma_m f444w}, we present the distribution of stellar-mass
uncertainty, $\sigma_{M_*}$, as a function of
$\log_{10}(\mathrm{SNR}_{\mathrm{F444W}})$. We adopt the same colour scheme as in \autoref{fig: mass f444w}: galaxies in the range $0.5<z<8.5$ are shown using a blue–green colormap, while those at $8.5<z<16.5$, where the sample size decreases rapidly, are displayed using a purple–red colormap. From visual inspection of the hexbin distributions, a
clear and physically intuitive trend emerges: stellar-mass uncertainties
systematically decrease with increasing
$\mathrm{SNR}_{\mathrm{F444W}}$. While this behaviour is expected, our goal is
to quantitatively parameterise how $\sigma_{M_*}$ evolves with increasing
$\mathrm{SNR}_{\mathrm{F444W}}$.

We begin by approximating
$\sigma_{M_*}\,[\log_{10}(M_*/\mathrm{M_\odot})]
\propto
\sigma_{\mathrm{F444W}}\,[\log_{10}(\mathrm{Flux\,[nJy]})]$,
which reduces to
$$
\label{sigma_m v.s sigma_snr}
\sigma_{M_*}\!\left[\log_{10}\!\left(\frac{M_*}{\mathrm{M_\odot}}\right)\right]
=
\frac{1}{\ln(10)}
\frac{\sigma_{\mathrm{flux}}\,[\mathrm{nJy}]}{\mathrm{Flux}\,[\mathrm{nJy}]} .
$$
Since $\sigma_{\mathrm{flux}}/\mathrm{Flux}$ is simply $1/\mathrm{SNR}_{\mathrm{F444W}}$, this implies
$$
\sigma_{M_*}\!\left[\log_{10}\!\left(\frac{M_*}{\mathrm{M_\odot}}\right)\right]
\propto
\mathrm{SNR}_{\mathrm{F444W}}^{-1}.
$$

\noindent In practice, however, stellar-mass uncertainties are not determined solely by photometric noise. Additional contributions to the uncertainties arise from intrinsic parameter degeneracies within the SED modelling. To account for these effects, we introduce an SED-driven uncertainty floor, $\sigma_{\mathrm{SED}}$, which captures the irreducible component of the stellar-mass uncertainty even at high signal-to-noise ratios.

We therefore model the total stellar-mass uncertainty as the quadratic sum of a
flux-dominated term and an SED-driven floor:
\begin{equation}
\label{eq:sigma_m snr fit}
\sigma_{M_*}\!\left[\log_{10}\!\left(\frac{M_*}{\mathrm{M_\odot}}\right)\right]
=
\sqrt{
\left(\frac{A}{\mathrm{SNR}_{\mathrm{F444W}}}\right)^2
+
\sigma_{\mathrm{SED}}^2
},
\end{equation}
\noindent where $A$ and $\sigma_{\mathrm{SED}}$ are the two free parameters. We fit this relation to the data in each redshift bin, and the best-fit relations are shown in pink for $0.5<z<8.5$ and in blue for $8.5<z<16.5$. The shaded regions indicate the 16th and 84th percentiles of the posterior distributions.

Overall, we find a clear and physically intuitive trend. At high $\mathrm{SNR}_{\mathrm{F444W}}$, the increased source brightness leads to significantly improved constraints on galaxy stellar-mass estimates. The parameter $\sigma_{\mathrm{SED}}$ acts as a floor, or asymptote, reflecting the fact that stellar masses cannot be measured with zero uncertainty due to intrinsic uncertainties and degeneracies associated with the SED modelling process itself, even in the limit of very high signal-to-noise ratios.

In addition, at higher redshifts, $\sigma_{\mathrm{SED}}$ systematically increases because the available photometric bands probe a more limited rest-frame wavelength range, increasingly sampling wavelengths blueward of the Balmer break. These wavelengths are dominated by recent star formation and provide weaker constraints on the older, long-lived stellar populations that dominate the total stellar mass, leading to more severe degeneracies in age, dust attenuation, and SFH.

Nevertheless, across our full sample and over all redshifts, even galaxies with relatively low $\mathrm{SNR}_{\mathrm{F444W}} \sim 3\text{--}5$, we find that $\sigma_{M_*}$ is typically on the order of $0.25\text{--}0.5$ dex, and decreases to $\sim 0.1$ dex at higher $\mathrm{SNR}_{\mathrm{F444W}}$. This level of precision already represents a strong constraint on stellar masses, highlighting the substantial gains enabled by the broad wavelength coverage and the inclusion of both wide and medium JWST/NIRCam bands in the SED modelling.

\begin{figure*}
    \centering
    \includegraphics[width=0.99\linewidth]{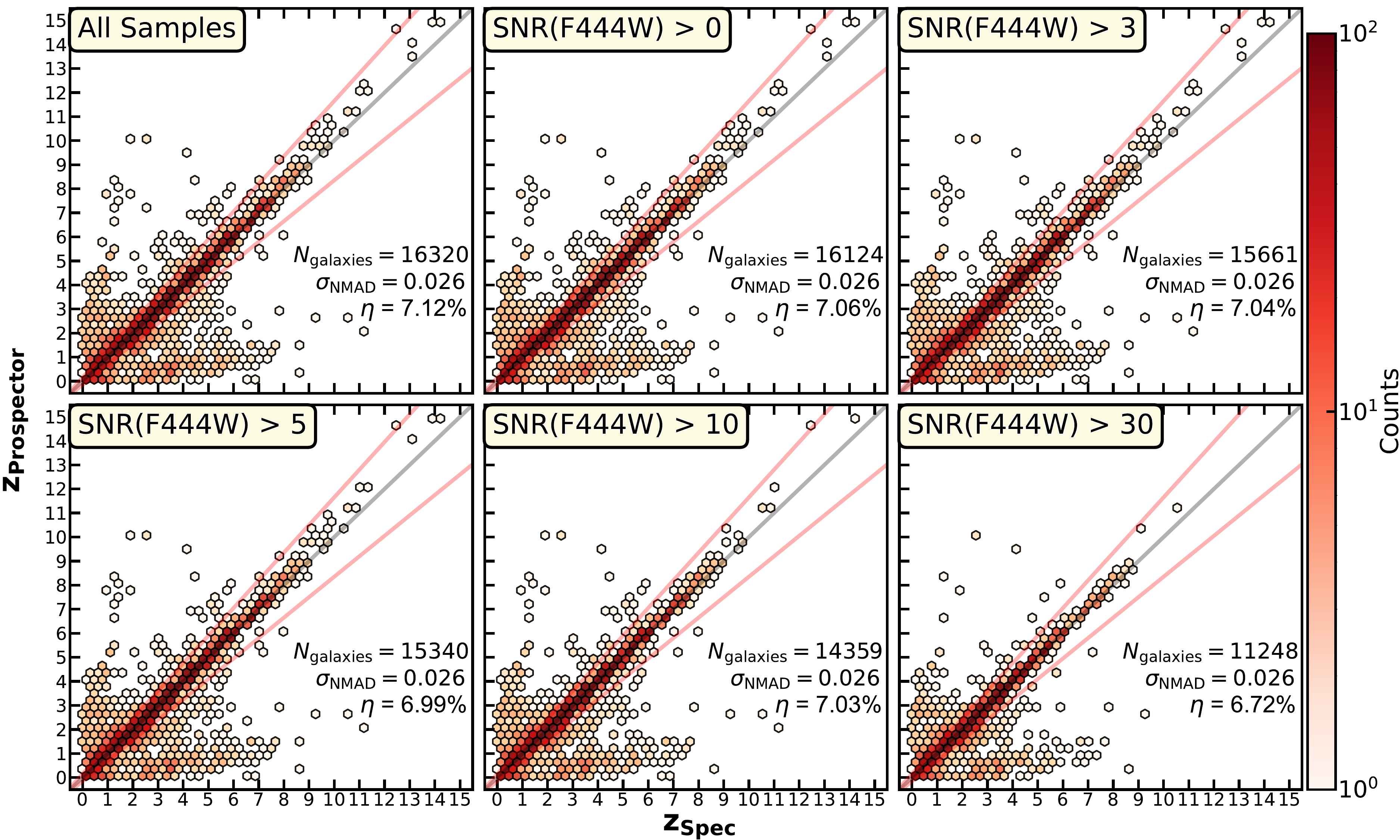}
    \vspace{0.0cm}

    \includegraphics[width=0.99\linewidth]{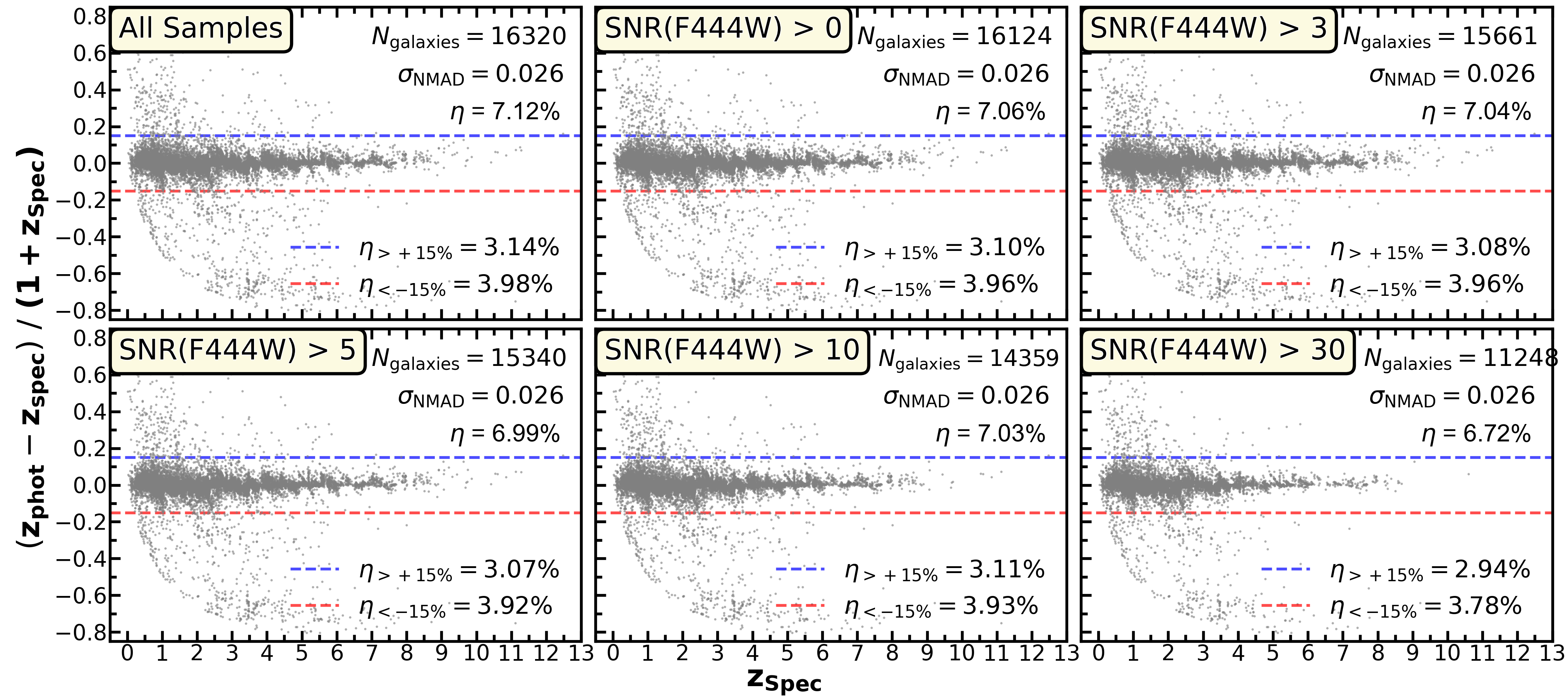}

    \caption{Two plots showing photometric redshift accuracy by comparing it with 16,320 robust spectroscopic redshifts. Hexbins are shown wherever at least one 1 point is present. \textbf{\textit{Top panel:}} Assessment of \prosp photometric redshift accuracy. Photometric redshifts are compared with spectroscopic redshifts for $16{,}320$ galaxies with highly robust and secure spectroscopic measurements. The six panels show the full sample and subsamples with progressively increasing signal-to-noise ratio in the F444W band. Two metrics are used to quantify the performance: the normalized median absolute deviation, $\sigma_\mathrm{NMAD}$, and the $15\%$ outlier fraction, $\eta$. We find excellent overall accuracy, with $\sigma_\mathrm{NMAD} = 0.026$ and $\eta = 7.12\%$ for the full sample, improving to $\eta = 6.72\%$ for the highest-SNR subsample. These indicates that reliable photometric redshifts are recovered even for sources with relatively low signal-to-noise ratios. \textbf{\textit{Bottom panel:}} Residuals between photometric and spectroscopic redshifts are shown as a function of spectroscopic redshift. We further decompose the outlier fraction into galaxies with photometric redshifts overestimated relative to the spectroscopic values ($\eta_{+15\%}$) and those with underestimated photometric redshifts ($\eta_{-15\%}$). We identify a prominent downward tail that becomes increasingly pronounced at higher spectroscopic redshifts. This trend indicates cases where the true Lyman break is misidentified as a Balmer break in the SED modelling, leading to systematically underestimated photometric redshifts. In contrast, we do not observe a significant upward tail, corresponding to scenarios in which a true Balmer break is mistaken for a Lyman break and the photometric redshift is overestimated. This asymmetry is encouraging, as it implies that our high-redshift sample is largely free from low-redshift interlopers.}
    \label{fig: zphot zspec all}
\end{figure*}

\section{Comparison with Spectroscopic Redshifts and emission-line Diagnostics}
\label{sec: compare with spec}
In this section, we compare several photometrically derived quantities with their values obtained from spectroscopic measurements. We first assess the photometric redshift quality by comparing it with spectroscopic redshifts (\autoref{sec:photometric redhisft accuracy}). We then evaluate the SFR estimates derived from SED modelling by comparing them with SFR measurements obtained from the H$\alpha$ emission-line (\autoref{sec: sfr and prosp comparison}). Lastly, we examine how the inclusion of MIRI bands improves constraints on AGN in the SED modelling by comparing our results with spectroscopically and photometrically selected Type~I and Type~II AGN (\autoref{sec:agn contribution}).

\subsection{Comparison with Spectroscopic Redshifts}
\label{sec:photometric redhisft accuracy}
Accurately inferring galaxy redshifts is one of the primary goals of SED analysis, as an incorrect redshift propagates large systematic errors into all other inferred physical parameters, ultimately rendering the derived galaxy properties unreliable. In this section, we assess the accuracy of our photometric redshift estimates by comparing them against available spectroscopic redshifts.

In \autoref{sec: spec-z}, we present a compilation of all available spectroscopic redshifts in GOODS-S and GOODS-N, as well as the reliability of the spec-z measurements and the sky matching of spec-z sources with the NIRCam DR5 photometry. For the analysis presented here, we restrict the spectroscopic sample to sources with \texttt{sky-flag} and \texttt{spec-flag} $\leq 2$ (see \autoref{sec: spec-z}). Photometric redshifts are taken from the HST+NIRCam run (\autoref{tab: number of galaxies snr}), using galaxies with at least 6 NIRCam filterbands. In total, $16{,}320$ galaxies are used in this spectroscopic versus photometric redshift comparison.

In \autoref{fig: zphot zspec all}, we assess the quality of our photometric redshift estimates by comparing them with available spectroscopic redshifts. Direct comparisons between phot and spec-z are shown in the top  panel, with residuals, $(z_\mathrm{prospector} - z_\mathrm{spec}) / (1 + z_\mathrm{spec})$, shown in the bottom panel. We quantify the photometric redshift performance using two commonly adopted metrics: the normalized median absolute deviation ($\sigma_{\mathrm{NMAD}}$), defined as
\begin{equation}
    \sigma_{\mathrm{NMAD}} = 1.48 \times \mathrm{median}\!\left(
    \frac{\left| z_{\mathrm{spec}} - z_{\mathrm{phot}} \right|}{1 + z_{\mathrm{spec}}}
    \right),
\end{equation}
and the 15\% outlier fraction, $\eta$, defined as the fraction of galaxies for which
\begin{equation}
    \eta = \frac{N\left( 
    \frac{\left| z_{\mathrm{phot}} - z_{\mathrm{spec}} \right|}
    {1 + z_{\mathrm{spec}}} > 0.15 
    \right)}{N_{\mathrm{tot}}} ,
\end{equation}

We find $\sigma_\mathrm{NMAD} = 0.026$ across all samples selected at different $\mathrm{SNR}_{\mathrm{F444W}}$ thresholds, with the outlier fraction decreasing from $7.12\%$ for all $16{,}320$ galaxies to $6.72\%$ for the $11{,}248$ galaxies with $\mathrm{SNR}_{\mathrm{F444W}} \geq 30$. Inspections of the corner plots and redshift posteriors for all outlier galaxies reveal that most exhibit bimodal behavior, with a secondary solution located close to the spectroscopic redshift, and is well representative by the photometric uncertainties. We thus would like to emphasize the importance of including redshift uncertainties and posteriors when using our data.

From the bottom panel of \autoref{fig: zphot zspec all}, we examine the residuals between the \texttt{Prospector} photometric redshifts and the spectroscopic redshifts,
$(z_\mathrm{prospector} - z_\mathrm{spec}) / (1 + z_\mathrm{spec})$.
We further decompose the outlier fraction into two components:
(i) galaxies for which the photometric redshift is overestimated relative to the spectroscopic value,
$\eta_{>+15\%}$, and
(ii) galaxies for which the photometric redshift is underestimated,
$\eta_{<-15\%}$.
As the sample selection becomes more stringent (i.e.\ with increasing $\mathrm{SNR}_{\mathrm{F444W}}$),
$\eta_{>+15\%}$ decreases modestly from $3.14\%$ for the full sample to $2.94\%$ samples with $\mathrm{SNR}_{\mathrm{F444W}} \geq 30$. Similarly, $\eta_{<-15\%}$ decreases from $3.98\%$ to $3.78\%$ over the same range of selections. We do observe some stripes and oscillations around the $y=0$ line, which originate from gaps between the photometric filter bands.

Across all samples, photometric redshifts show a systematic tendency to be underestimated relative to the spectroscopic redshifts.
This manifests as a prominent downward tail in the residual distribution that becomes increasingly pronounced toward higher spectroscopic redshifts.
Such behaviour is consistent with cases in which the true Lyman break is misidentified as a Balmer break during SED modelling, leading to systematically underestimated photometric redshifts.
By contrast, we do not observe a comparably significant upward tail, which would correspond to the misidentification of a true Balmer break as a Lyman break and hence overestimated photometric redshifts.
This asymmetry is encouraging, as it indicates that our high-redshift sample is largely free from contamination by low-redshift interlopers.

We emphasize that a uniform redshift prior is adopted in our \texttt{Prospector} modelling,
and therefore the prior does not preferentially favour lower--redshift solutions.
The observed asymmetry in the residual distribution thus provides strong evidence that our high--redshift galaxy samples are increasingly free from contamination by low--redshift interlopers as the data quality improves.

We also note that we use \texttt{KRON\_CONV} fluxes from the JADES DR5 photometry catalogue \citep{robertson2026}, which correspond to ellipsoidal aperture photometry centred on each galaxy and unavoidably include some background noise. A higher photometric redshift accuracy can be achieved by instead using \texttt{CIRC1--6\_CONV} fluxes, defined as circular aperture photometry with aperture radii of $r = [0.1'', 0.15'', 0.25'', 0.3'', 0.35'', 0.5'']$. While this approach largely avoids background noise contamination, it also excludes part of the galaxies’ intrinsic light, thereby leading to underestimated stellar masses and degraded constraints on stellar population properties. Since the primary goal of this paper is to infer galaxy properties and stellar populations, we therefore adopt the \texttt{KRON\_CONV} fluxes. For a detailed study of photometric redshift performance and high-z galaxies selections using \texttt{CIRC1--6\_CONV} fluxes, we refer the reader to \kevin. In fact, our \prosp redshift accuracy is already comparable to, and slightly better than, the photometric redshifts estimated by \texttt{EAZY} (see Appendix \ref{sec: prospector v.s eazy} for a detailed discussion).

In summary, we demonstrate the good accuracy of our redshift inference from SEDs, with low values of $\sigma_{\mathrm{NMAD}}$, outlier fraction ($\eta$), and interloper rate. Our photometric redshift performance is comparable to, and on par with, those achieved by other major \textit{JWST} surveys, including CEERS \citep{cox_ceers_catalog_2025}, EPOCHS \citep{duan2024_addingvalue, austin2024, Harvey2025, conselice2025}, and UNCOVER \citep{wang2023, wang_2024_uncover_stellar_pop}.

\subsection{Emission-Line Star-Formation Rates}
\label{sec: sfr and prosp comparison}
Massive, hot, and short-lived O- and B-type stars produce strong ionizing Lyman-continuum photons that ionize the surrounding neutral hydrogen, giving rise to classical \ion{H}{ii} regions. As this ionized gas recombines, it emits hydrogen recombination lines, in particular H$\alpha$ and H$\beta$. The luminosities of these nebular lines trace the ionizing-photon production rate and therefore provide a sensitive probe of very recent star formation on timescales of a few Myr \citep[e.g.,][]{kennicutt1998,pflamm2007,kennicutt2012,madau_dickinson_2014,tacchella2022_halpha}. 

Very detailed radiative transfer simulation studies show that SFR averaged over short timescales (5-10Myr), is in better agreement with H$\alpha$ based SFR, than SFR averaged over 50 to 100 Myr (see section 5.4 of \citealt{tacchella2022_halpha, mcclymont2025_burst_quench}). Thus, comparing our SFR inferred from SFH from \prosp with SFR from H$\alpha$ could validate the accuracy of our SFH reconstruction, and also the proposed main sequence SFH prior in this work. 

In this section, we compare SFR inferred from \prosp with H$\alpha$ based SFR. We present the methodology for calculating $\mathrm{SFR}_{\mathrm{H}\alpha}$ in \autoref{sec: sfr_halpha inferring}, and the comparison in \autoref{sec: comparing sfr_10, sfr_halpha}. 

\subsubsection{Inferring \texorpdfstring{$\mathrm{SFR}_{\mathrm{H}\alpha}$}{Halpha}}
\label{sec: sfr_halpha inferring}

To infer SFR from H$\alpha$, we use H$\alpha$ and H$\beta$ emission-line fluxes from the JADES DR4 NIRSpec grating spectroscopy \citep{emma2025_dr4,scholtz2025_dr4}. 
In low-resolution JWST/NIRSpec PRISM spectrum ($R \sim 100$), H$\alpha$ is significantly blended with neighboring emission-lines, most notably [N\,\textsc{ii}]$\lambda6583$ and, to a lesser extent, [S\,\textsc{ii}]$\lambda6716$. To avoid uncertainties associated with line blending and deblending, we therefore restrict our analysis to the higher-resolution NIRSpec grating spectra ($R\sim1000$) and do not use PRISM data for H$\alpha$-based SFR estimates.

To calculate SFR$_{\mathrm{H}\alpha}$ from the H$\alpha$ line luminosity, we use the calibration from \citet{kennicutt2012}. Specifically,
\begin{equation}
    \log_{10}\!\left(\mathrm{SFR}\;[\mathrm{M_\odot}\,\mathrm{yr}^{-1}]\right)
    = \log_{10}\!\left(L_{\mathrm{H}\alpha,\mathrm{intrinsic}}\right) - C,
\end{equation}
where $C$ is the calibration constant, with a value of $41.27$. In addition to this, metallicity plays a secondary effects on the relationship between SFR and $L_\mathrm{H\alpha}$. 

However, several physical processes can attenuate or boost the intrinsic H$\alpha$ luminosity, $L_{\mathrm{H}\alpha,\mathrm{intrinsic}}$, including dust attenuation, absorption of Lyman-continuum (LyC) photons by dust, the escape of LyC photons from the galaxy, and collisional excited gas that boosts the emission \citep{tacchella2022_halpha}.  Among these effects, dust attenuation within the interstellar medium (ISM) has the most dominant impact and is the effect we can correct directly. We therefore dust-correct the observed luminosity, $L_{\mathrm{H}\alpha,\mathrm{obs}}$, to obtain an unbiased SFR measurement.

Dust preferentially absorbs shorter-wavelength photons, producing a wavelength-dependent attenuation commonly referred to as dust reddening, which is parameterized by the color excess $E(B-V)$ \citep{calzetti1994}. For nebular emission-lines, $E(B-V)$ can be estimated using the Balmer decrement, i.e., the observed H$\alpha$/H$\beta$ flux ratio. The attenuation at the H$\alpha$ wavelength can be written as
\begin{equation}
A_{\mathrm{H}\alpha} = \kappa(\lambda_{\mathrm{H}\alpha}) \times E(B-V),
\end{equation}
where $\kappa(\lambda)$ is the dust attenuation curve. Adopting the \citet{calzetti2000} attenuation law, we use $\kappa(\lambda_{\mathrm{H}\alpha}) = 3.326$.

The color excess $E(B-V)$ inferred from the Balmer decrement is given by
\begin{equation}
E(B - V) =
\frac{2.5}{\kappa(\lambda_{\mathrm{H}\beta}) - \kappa(\lambda_{\mathrm{H}\alpha})}
\log_{10}
\left[
\frac{(\mathrm{H}\alpha / \mathrm{H}\beta)_{\mathrm{obs}}}
     {(\mathrm{H}\alpha / \mathrm{H}\beta)_{\mathrm{int}}}
\right],
\end{equation}
where $(\mathrm{H}\alpha / \mathrm{H}\beta)_{\mathrm{int}} = 2.86$ corresponds to Case~B recombination at an electron temperature $T = 10^{4}\,\mathrm{K}$ and electron density $n_e = 10^{2}\,\mathrm{cm}^{-3}$ \citep{osterbrock1989}.

With the Balmer decrement accurately measured, we can estimate the dust extinction and recover the intrinsic, dust-corrected H$\alpha$ luminosity from the observed value as
\begin{equation}
L_{\mathrm{H}\alpha,\mathrm{intrinsic}} =
L_{\mathrm{H}\alpha,\mathrm{obs}} \times 10^{0.4 A_{\mathrm{H}\alpha}} .
\end{equation}

We compute dust-corrected SFR$_{\mathrm{H}\alpha}$ only for galaxies in the JADES DR4 release with robust ($\mathrm{SNR} > 5$) measurements of both H$\beta$ and H$\alpha$. This selection yields a final sample of 2,612 galaxies after excluding spectroscopically confirmed Type-I and Type-II AGN \citep{scholtz2025_type2, ignas2026_type1}. This sample is statistically sufficient for our analysis. We present the comparison between SFR$_{\mathrm{H}\alpha}$ and SFR$_{\mathrm{10\,Myr,\,Prospector}}$ in the next section.

\subsubsection{Comparing \texorpdfstring{$\mathrm{SFR}_{10\,\mathrm{Myr}}$}{SFR10 Myr} versus \texorpdfstring{$\mathrm{SFR}_{\mathrm{H}\alpha}$}{Halpha}}
\label{sec: comparing sfr_10, sfr_halpha}

\begin{figure}
    \centering
    \includegraphics[width=\linewidth]{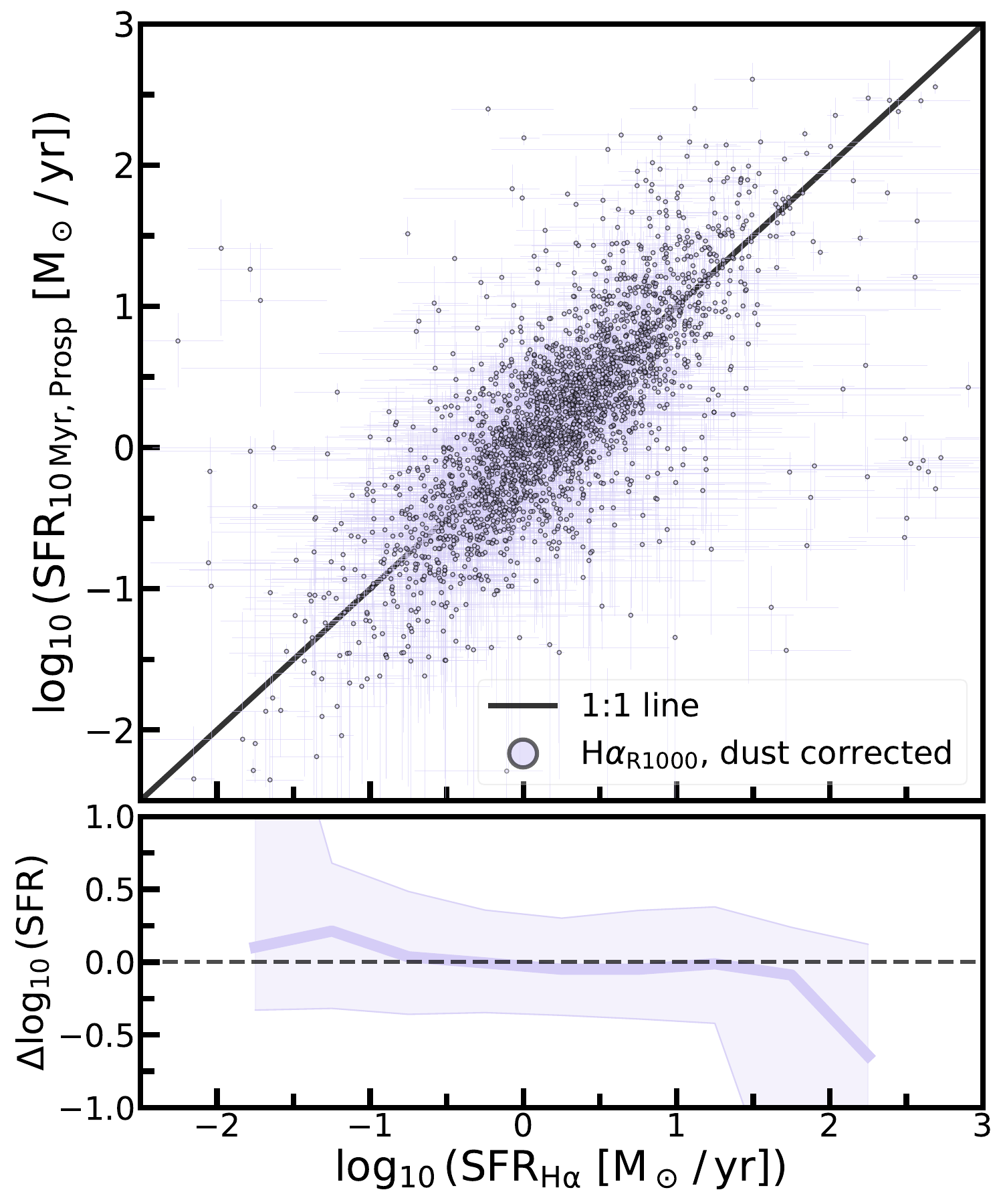}
    \caption{Comparison between SFR$_{\mathrm{H}\alpha}$, derived from H$\alpha$ emission using the R1000 grating, and SFR$_{10\,\mathrm{Myr}}$, inferred from the SFH using \texttt{Prospector}. Spectroscopically confirmed Type-I and Type-II AGN are excluded. We find a tight correlation, with a median residual (SFR$_{10\mathrm{Myr}} -$ SFR$_{\mathrm{H}\alpha}$) of $-0.017^{+0.380}_{-0.349}$ dex. This agreement demonstrates that the extensive medium-band coverage in GOODS-S and GOODS-N provided by the JADES DR5 release, combined with our \texttt{Prospector} model setup, enables accurate modelling of hydrogen recombination emission lines and recent star formation activity.}
    \label{fig: sfr halpha prosp}
\end{figure}

We compare the SFR inferred from attenuation corrected H$\alpha$, SFR$_{\mathrm{H}\alpha}$, with the SFR averaged over the past 10 Myr inferred from \texttt{Prospector}, SFR$_{10\mathrm{Myr}}$, in the top panel of \autoref{fig: sfr halpha prosp}. The residual, defined as SFR$_{10\mathrm{Myr}} -$ SFR$_{\mathrm{H}\alpha}$, is shown in the bottom panel. The one-to-one relation is indicated by the black $y=x$ line. We find that SFR$_{10\mathrm{Myr}}$ from \prosp is closely aligned with SFR$_{\mathrm{H}\alpha}$ over the range $-1.8 < \log_{10}(\mathrm{SFR},[\mathrm{M_\odot,yr^{-1}}]) < 2.0$, corresponding to $0.0158 < \mathrm{SFR} \,\,[\mathrm{M_\odot,yr^{-1}}] < 100$. This strong agreement indicates that our \prosp setup, together with the main sequence SFH prior introduced in this work, accurately models galaxy stellar populations, even on timescales as short as $\sim 10$Myr.

Looking more closely at the residual distribution, we identify two small systematic trends: an upward deviation at $\log_{10}(\mathrm{SFR}\,[\mathrm{M_\odot\,yr^{-1}}]) < -1.8$ and a downward deviation at $\log_{10}(\mathrm{SFR}\,[\mathrm{M_\odot\,yr^{-1}}]) > 2.0$. Specifically, at very low SFRs, SFR$_{10\,\mathrm{Myr}}$ inferred from \prosp tends to be higher than SFR$_{\mathrm{H}\alpha}$, whereas at very high SFRs, SFR$_{10\,\mathrm{Myr}}$ inferred from \prosp is systematically lower than SFR$_{\mathrm{H}\alpha}$.

At the high-SFR end, $\log_{10}(\mathrm{SFR}_{\mathrm{H}\alpha}\,[\mathrm{M_\odot\,yr^{-1}}]) > 2.0$, SFR$_{\mathrm{H}\alpha}$ tends to exceed SFR$_{10\,\mathrm{Myr}}$ inferred from \prosp. This behavior can be explained by very recent and intense starburst episodes \citep[e.g.,][]{tacchella2022_halpha}. A burst occurring within the past $\lesssim5$ Myr can strongly boost the H$\alpha$ luminosity, while its impact on SFR$_{10\,\mathrm{Myr}}$ is diluted by temporal averaging. In our \prosp setup, the youngest age bin spans 0-30 Myr, causing such short-timescale starbursts to be smoothed out. Moreover, at very high SFRs the ionization parameter is expected to vary more widely, and nebular emission becomes increasingly non-linear. This added complexity increases the degeneracy in SED modelling, making it more difficult to constrain very high SFR from extreme line emission.

At $\log_{10}(\mathrm{SFR}_{\mathrm{H}\alpha}\,[\mathrm{M_\odot\,yr^{-1}}]) < -1.8$, SFR$_{\mathrm{H}\alpha}$ yields significantly lower values than SFR inferred from \texttt{Prospector}. At such low SFRs, the formation of very massive O-type stars becomes increasingly stochastic, leading to large fluctuations in the production rate of ionizing photons. This effect has been observed in studies of nearby star-forming dwarf galaxies \citep[e.g.,][]{lee2009, weisz2012modeling, johnson2013measuring}, who find that the ratio SFR$_{\mathrm{FUV}}$/SFR$_{\mathrm{H}\alpha}$ deviates significantly from expectations based on continuous star formation, particularly at low SFRs and in systems with bursty SFHs. Simulation studies further support this interpretation. Using fully stochastic population synthesis models, \citet{fumagalli2011} show that even for a universal IMF, stochastic sampling can induce large variations in nebular emission relative to UV-based SFR indicators. Similarly, \cite{da_silva_2014_slug} demonstrate that, depending on the SFR tracer used, stochastic fluctuations can produce non-trivial errors at SFRs as high as $1\,\mathrm{M_\odot\,yr^{-1}}$, and biases exceeding $\sim0.5$ dex at the lowest SFRs. 

In addition, the scarcity of ionizing stars implies that secondary effects, such as dust attenuation and the absorption of Lyman-continuum photons by dust and helium, become increasingly important. These processes suppress the observed H$\alpha$ luminosity, and their relative impact is expected to be stronger in low-SFR systems than in vigorously star-forming galaxies. While dust attenuation is corrected for via the Balmer decrement, LyC photon absorption and escape are not accounted for in this work, as such corrections are highly uncertain and galaxy dependent. This further contributes to the systematic underestimation of SFR$_{\mathrm{H}\alpha}$ at low SFRs.

Nevertheless, the aforementioned biases are confined to systems with very low or very high SFRs (with SFR $< 0.016$ or $> 100\,\mathrm{M_\odot\,yr^{-1}}$). On average, our SFR$_{10\,\mathrm{Myr}}$ deviates from SFR$_{\mathrm{H}\alpha}$ by only $-0.017^{+0.380}_{-0.349}$ dex, demonstrating strong sensitivity to star formation on very short timescales and high fidelity in the reconstruction of SFHs.

\begin{figure}
    \centering
    \includegraphics[width=\linewidth]{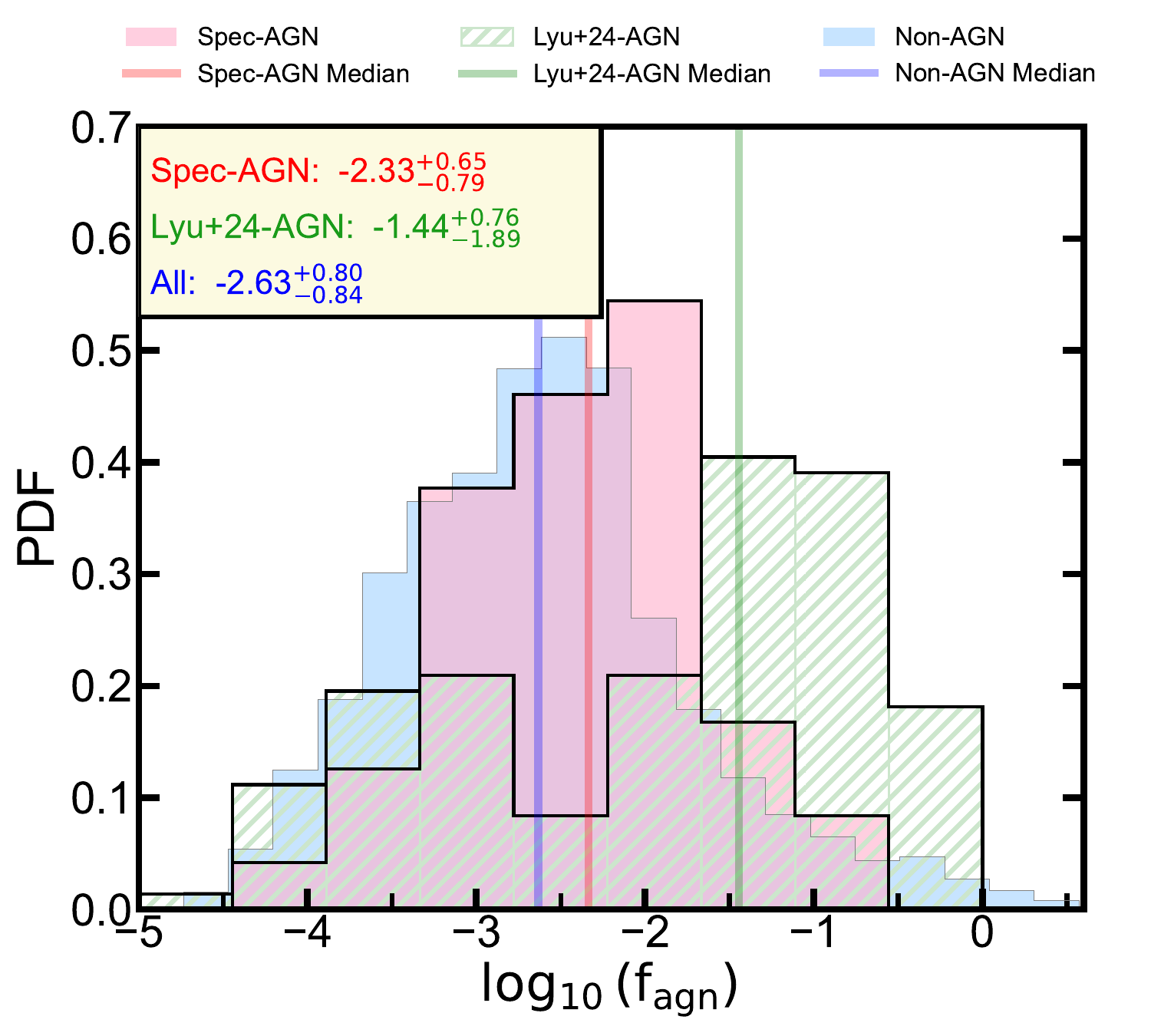}
    \caption{Histograms of $\log_{10}({\mathrm{f_{AGN}}})$ for galaxies in the HST + NIRCam + MIRI + Spec-z sample (see \autoref{tab: number of galaxies snr}), separated into three subsamples. Galaxies spectroscopically confirmed as Type I and Type II AGN are shown in red \citep{scholtz2025_type2, ignas2026_type1}. Galaxies photometrically identified as AGN by \citet{lyu2024_miri_agn} are shown in green, while galaxies not belonging to either AGN-selected sample are shown in blue. We find that literature-selected AGN exhibit median $\log_{10}(f_{\mathrm{AGN}})$ values that are higher by $0.42$ dex and $1.32$ dex compared to the non-AGN sample. This indicates that the inclusion of MIRI photometry enables our \prosp modelling framework to recover AGN contributions at a statistically significant level.}
    
    \label{fig: agn miri}
\end{figure}

\subsection{Mid-infrared AGN Contributions}
\label{sec:agn contribution}
\citet{leja2018sfh} introduced an AGN component into \texttt{Prospector}. Various tests have shown that SED models which fit MIR data without including an AGN component are susceptible to substantial biases in derived physical parameters. In our \prosp setup (see \autoref{sec: inferring galaxy properties}), AGN emission is always included as a free parameter, parameterized by $\log_{10}(f_{\mathrm{AGN}})$, the fraction of the total $4$--$20\,\mu$m luminosity contributed by the AGN relative to the bolometric luminosity, and $\log_{10}(\tau_{\mathrm{AGN}})$, the optical depth of individual dust clumps at $5500\,\text{\AA}$.

As shown in \autoref{fig: nbands all}, the JADES DR5 fields provide extensive MIRI coverage, particularly in GOODS-S, with extensive MIRI observations from SMILES (PID:1207; \citealt{alberts2024_smiles}, \citealt{george2024_smiles}, \citealt{zhu2026_smiles}), where the deepest regions include up to eight MIRI bands. Mid-infrared observations are physically critical for AGN identification, as a significant fraction of the bolometric energy output of obscured AGN is reprocessed by circumnuclear dust and emerges at mid-IR wavelengths. Optical and X-ray selections can miss heavily obscured or low-luminosity AGN, especially at high redshift, whereas the continuous $5$--$20\,\mu$m coverage from MIRI provides direct sensitivity to hot dust emission associated with accretion. Therefore, incorporating MIRI data is essential for robustly constraining AGN contributions in high-redshift galaxies.

In \autoref{fig: agn miri}, we show the probability density functions of $\log_{10}(f_{\mathrm{AGN}})$ derived from our SED modelling for GOODS-S galaxies with HST+NIRCam+MIRI+spec-$z$ coverage. Galaxies with spectroscopically confirmed Type-I and Type-II AGN from \citet{scholtz2025_type2, ignas2026_type1} are shown in red, and those matched to the MIRI-selected AGNs of \citet{lyu2024_miri_agn} are shown in green; the remaining non-AGN galaxies are shown in blue. Out of $4{,}416$ galaxies with HST+NIRCam+MIRI data and fixed spec-$z$ runs, $42$ are spectroscopically confirmed AGN and $117$ are matched to MIRI-selected AGN by \citet{lyu2024_miri_agn}. The vertical lines indicate the median values of $\log_{10}(f_{\mathrm{AGN}})$ for each distribution.

We find that spectroscopically confirmed AGN exhibit $\log_{10}(f_{\mathrm{AGN}})$ values that are, on average, $\sim
0.42$ dex higher than those of the non-AGN sample. For galaxies identified as AGN in the MIRI-selected catalogue of \citet{lyu2024_miri_agn}, the offset is even larger, reaching $\sim 1.32$ dex relative to the full sample. Although these offsets may appear moderate in absolute terms, they are physically meaningful. Numerous studies \citep[e.g.,][]{maiolino_2024, geris2026, ignas2026_type1} have shown that, particularly at high redshift and for moderate-luminosity systems, intrinsic AGN emission often contributes only a small fraction of the total bolometric luminosity. Therefore, even a $\sim 0.4$ dex enhancement in $\log_{10}(f_{\mathrm{AGN}})$ corresponds to a substantial relative increase in accretion-powered emission. The stronger offset observed for the MIRI-selected AGN further reinforces this interpretation, as these systems are independently identified based on their mid-infrared properties. Overall, this consistency supports the physical interpretability of our AGN parameterization and demonstrates that the model is sensitive to energetically sub-dominant AGN contributions.

Importantly, the robustness of these constraints relies on multi-component SED decomposition enabled by MIRI coverage. The mid-IR regime allows us to distinguish between hot dust heated by an AGN torus and warm dust emission from star formation, mitigating degeneracies that are common in optical-to-near-IR–only fits. The continuous wavelength sampling provided by MIRI improves the separation between stellar, dust, and AGN components, thereby strengthening both AGN identification and host-galaxy parameter inference. However, we note that the AGN component in our \texttt{Prospector} model is primarily constrained through mid-IR emission, and does not fully capture additional AGN contributions to the SED (e.g., accretion disk emission in the UV/optical or detailed narrow-line region physics), which may introduce residual uncertainties in some systems.

We also emphasize that AGN identification through SED modelling is inherently sensitive to survey depth, wavelength coverage, and assumed template prescriptions. Selection completeness may vary across luminosity, redshift, and obscuration regimes. In particular, low-luminosity or heavily obscured AGN may still exhibit relatively small $\log_{10}(f_{\mathrm{AGN}})$ values, while star-forming galaxies with strong dust emission could partially mimic AGN signatures in limited wavelength coverage. This degeneracy may be especially pronounced in low-mass, low-metallicity galaxies, where intense star formation can heat dust to high temperatures, producing near- and mid-infrared colours that resemble those of AGN. This effect likely reflects the limited availability of appropriate templates for such systems, and has been highlighted in recent study by \citet{Iani2026} where even with extensive photometric coverage, spectroscopic information was required to robustly confirm AGN activity. While MIRI data helps to alleviate these limitations by improving wavelength coverage, residual selection biases should be considered when interpreting the demographic differences between AGN and non-AGN populations.

Overall, the inclusion of MIRI data enables not only improved stellar population constraints, but also physically motivated and quantitatively meaningful measurements of AGN contributions within the JADES sample.

\section{Systematic Effects: Wavelength Coverage and SFH Priors}
\label{sec: systematic effects: wavelength coverage and sfh priors}
In this section, we discuss how the inclusion of \textit{JWST} NIRCam medium bands can improve the inference of galaxy properties (\autoref{sec: mass medium band}). We then assess the consistency of stellar mass estimates by comparing results obtained with and without the inclusion of \textit{JWST} MIRI photometry (\autoref{sec: mass miri}). Lastly, we evaluate how the SFMS prior introduced in this work affects stellar mass estimates relative to those obtained using other priors (\autoref{sec: mass prior comparison}).

\begin{figure*}
    \centering

    \includegraphics[width=\linewidth]{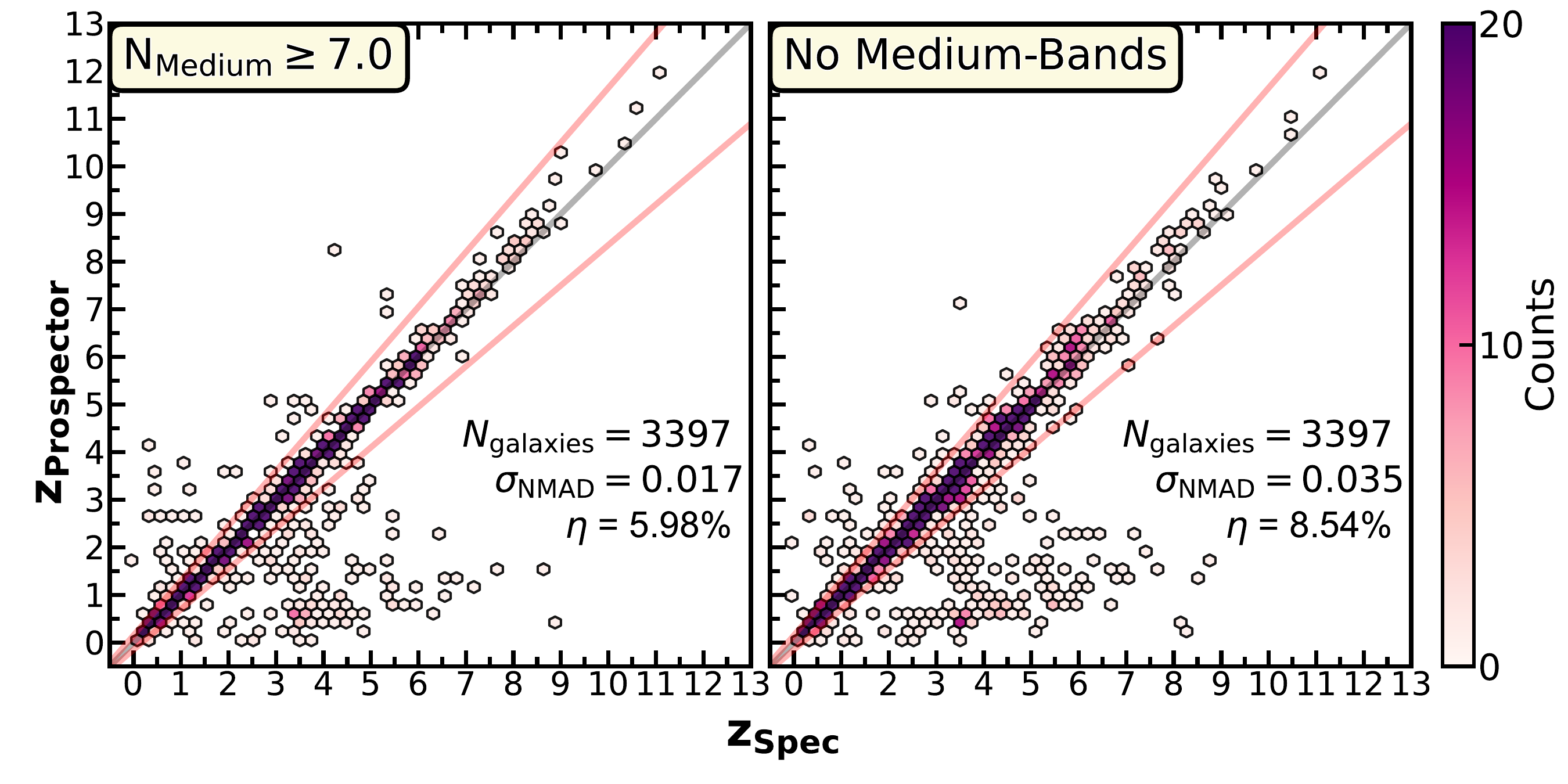}

    \vspace{0.8em}

    \includegraphics[width=0.49\linewidth]{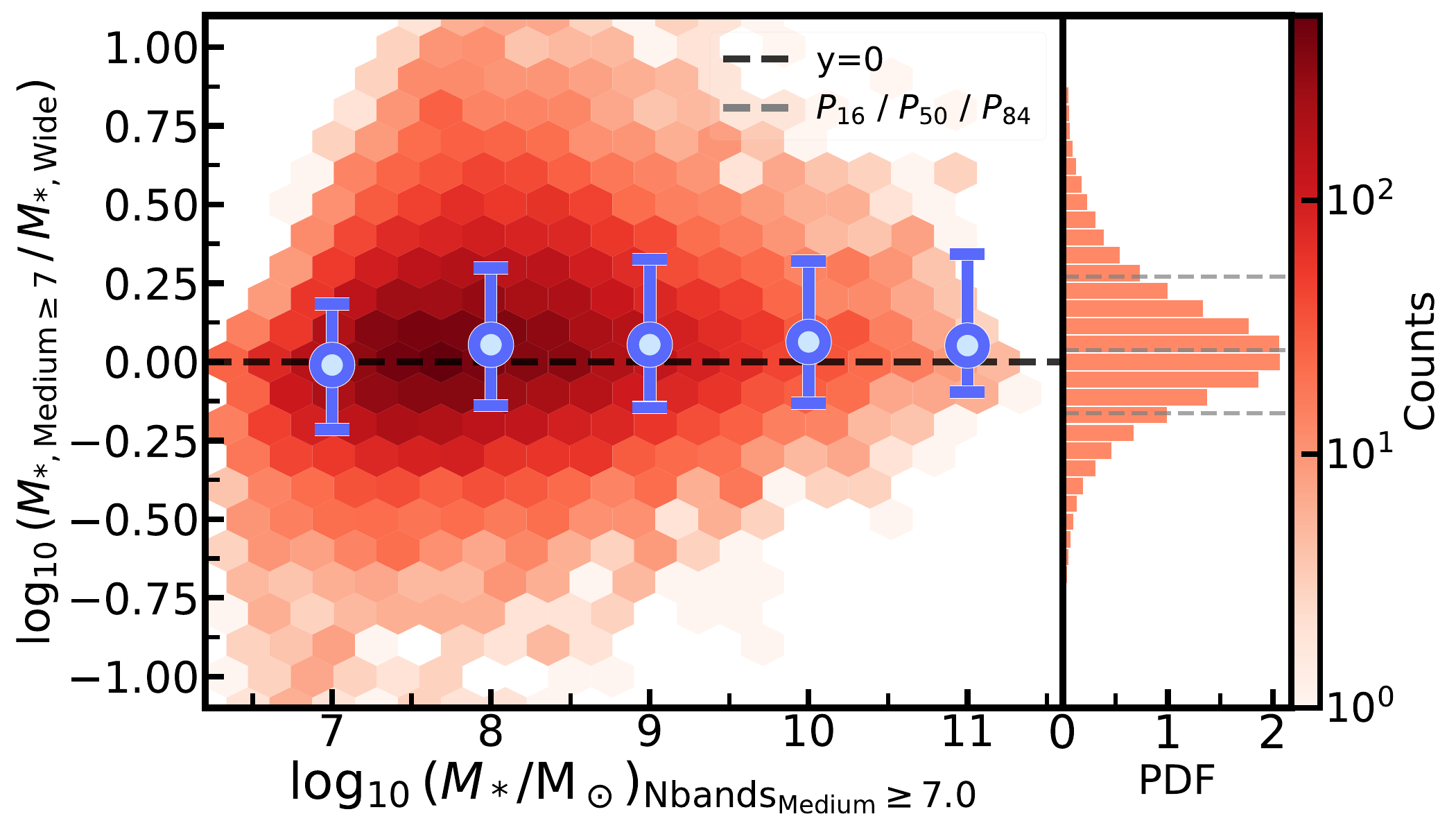}
    \hfill
    \includegraphics[width=0.49\linewidth]{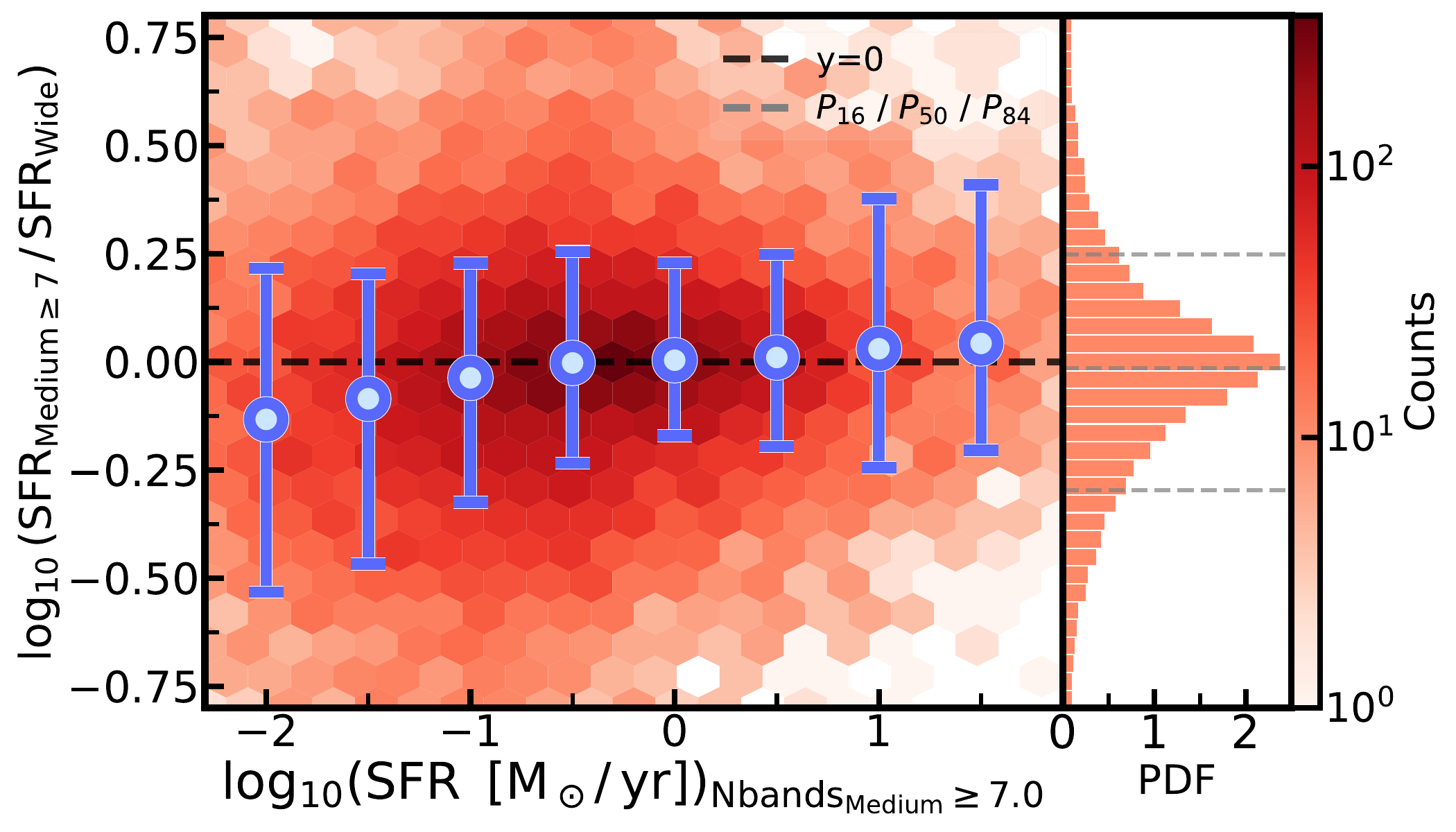}

    \caption{Comparison of redshift, stellar mass, and SFR with and without the inclusion of of medium band, using $15{,}990$ detected galaxies ($\mathrm{SNR_{F444W}} \geq 3$) with deep exposure and $ \ge 7$ medium-bands in GOODS-S. \textbf{\textit{Top panel:}} Redshift comparison with spectroscopic redshifts ($3{,}397 /15{,}990$) of these two runs. Normalised median absolute deviation, $\sigma_{\rm NMAD}$, and outlier fractions, $\eta$, are shown. Both runs have good redshift quality. With medium band photometry, we find a $\sim 1.4$--$2.1\times$ improvement in quality, and a significantly reduced scatter around the $y = x$ relation, as the medium bands help bridge the gaps between the wide bands. \textbf{\textit{Bottom panels:}} In the two bottom panels, we compare the stellar mass and $\mathrm{log}_{10}\mathrm{(SFR_\mathrm{10 Myr})}$ between runs with and without medium bands (defined as the run with medium bands minus the run without; see \autoref{eq: mass sfr med}). The results from the different runs are broadly consistent. However, when only wide-band photometry is used, emission lines tend to be very slightly overestimated. This leads to an overestimation of the SFR and a corresponding underestimation of the stellar mass.}
    \label{fig: mass med}
\end{figure*}

\begin{figure*}
    \centering
    \includegraphics[width=\linewidth]{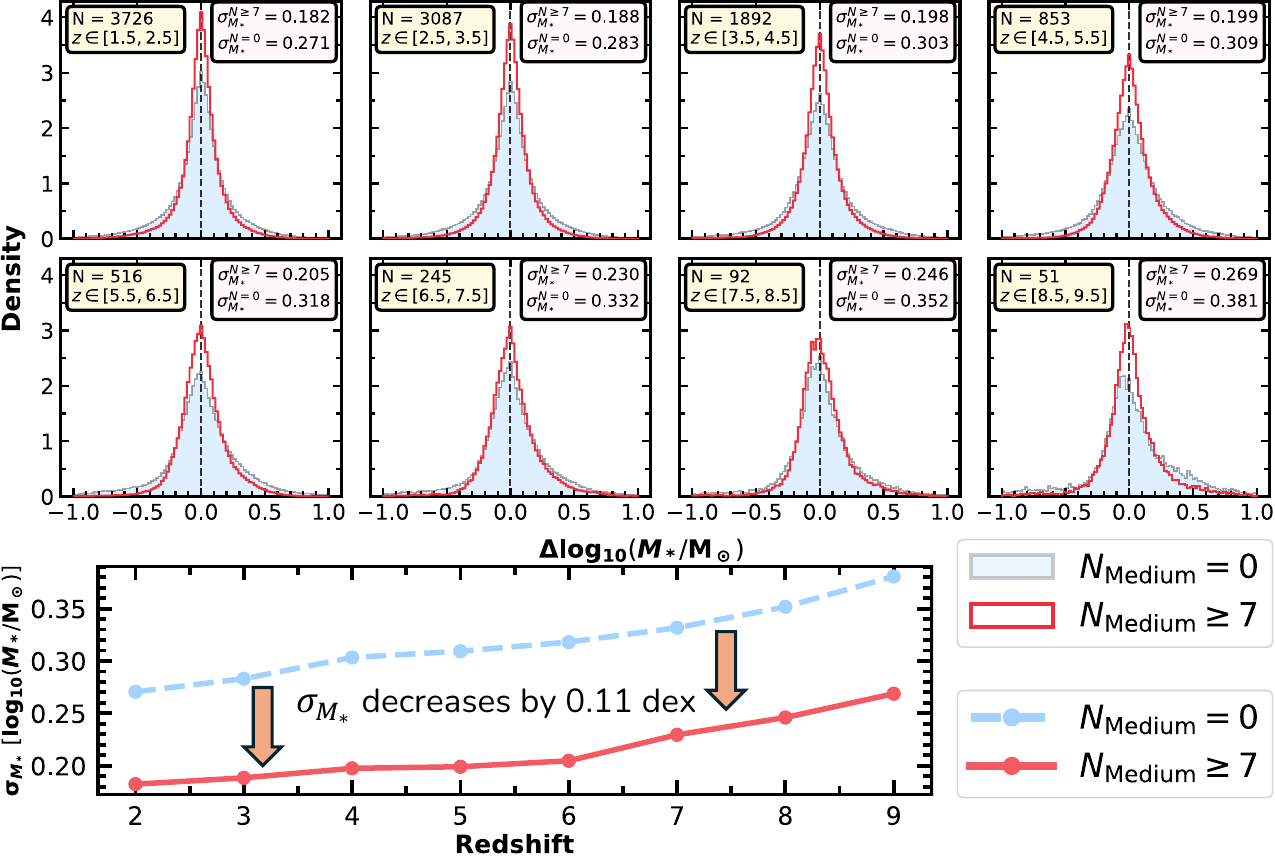}
    \caption{In this figure, we demonstrate how the inclusion of NIRCam medium-band photometry improves stellar mass uncertainty estimates, using $15{,}990$ detected galaxies ($\mathrm{SNR_{F444W}} \geq 3$) with deep exposure and $ \ge 7$ medium-bands in GOODS-S. \textbf{\textit{Top panel:}} We show the stacked stellar mass posteriors for galaxies in each redshift bin. For each galaxy, the stellar mass posterior is shifted such that its 50th percentile is centered at zero before stacking the posteriors from all samples. The red histograms correspond to galaxies with $\mathrm{N_{Medium}} \geq 7$, while the blue histograms show results using only wide-band photometry ($\mathrm{N_{Medium}} = 0$). The corresponding average stellar mass uncertainties ($\sigma_{M_*}$) are indicated in the top-right corner of each panel. \textbf{\textit{Bottom panel:}} We show the evolution of stellar mass uncertainties as a function of redshift for $\mathrm{N_{Medium}} \geq 7$ and $\mathrm{N_{Medium}} = 0$. We find that including $\geq 7$ medium bands consistently reduces stellar mass uncertainties by $0.11^{+0.01}_{-0.01}$ dex across all redshifts.}
    \label{fig: mass uncer medium}
\end{figure*}

\subsection{Impact of medium-band Photometry}
\label{sec: mass medium band}

In this section, we investigate how the inclusion of JWST/NIRCam medium-band photometry improves constraints on galaxy redshift, stellar mass, and SFR. In the JADES GOODS-S and GOODS-N fields, the number of available NIRCam medium bands can reach up to 10: F162M, F182M, F210M, F250M, F300M, F335M, F410M, F430M, F460M, and F480M. Compared to wide band photometry, medium band filters provide narrower wavelength coverage while maintaining a similar level of filter throughput, and are therefore more sensitive to the presence of strong emission-lines that would otherwise be averaged out in broad filters. This makes medium-band photometry particularly useful for constraining emission-line flux strengths and improving redshift estimates. Several studies have demonstrated the power of medium-band imaging for studying emission-line populations in large galaxy samples, benefiting from the significantly larger survey volumes accessible through imaging compared to spectroscopy \citep[e.g.,][]{williams_2023_jems,suess_2024,trussler2025_cloudy, trussler2025_candels, helton2025_filter}.

We select galaxies with at least 7 NIRCam medium bands in GOODS-S, yielding a total of $15{,}990$ galaxies. For this subsample, we perform an additional \texttt{Prospector} run in which all NIRCam medium bands are excluded, using only HST and NIRCam wide-band photometry. We then compare the inferred redshifts, stellar masses, and SFRs between runs with and without the inclusion of medium bands. These comparisons are shown in \autoref{fig: mass med}.

In the redshift comparison (top panel of \autoref{fig: mass med}), we assess how accurately photometric redshifts are recovered when medium bands are included (left panel) and excluded (right panel). Both runs yield high-quality photometric redshifts, with $\sigma_{\mathrm{NMAD}} = 0.017$ when medium bands are included and $\sigma_{\mathrm{NMAD}} = 0.035$ when they are excluded ($5.98 \%$ and $8.54 \%$ for outlier fractions). The inclusion of medium bands therefore improves the redshift accuracy by roughly a factor of $1.4$--$2.1$, both in terms of $\sigma_{\mathrm{NMAD}}$ and the outlier fraction, with the largest improvement seen in $\sigma_{\mathrm{NMAD}}$. This improvement is expected, as medium bands help bridge wavelength gaps between wide filters and more precisely capture spectral features (e.g., Lyman$-\alpha$ and Balmer break) that constrain the redshift. This is evident in the figure, where the scatter around the $y = x$ relation is significantly reduced and the points are more tightly concentrated.

In the bottom panel of \autoref{fig: mass med}, we compare the residuals in stellar mass and $\mathrm{log}_{10}\mathrm{(SFR_\mathrm{10 Myr})}$ between the runs with and without medium-band photometry, defined as
\begin{equation}
\begin{aligned}
\Delta \log_{10}(M_*/\mathrm{M_\odot}) &= \log_{10}(M_*/\mathrm{M_\odot})_{\mathrm{N_{Medium}} \geq 7} \\
&\quad - \log_{10}(M_*/\mathrm{M_\odot})_{\mathrm{N_{Medium}} = 0}, \\
\Delta \log_{10}(\mathrm{SFR} [\mathrm{M_\odot/yr}]) &= \log_{10}(\mathrm{SFR} [\mathrm{M_\odot/yr}])_{\mathrm{N_{Medium}} \geq 7} \\
&\quad - \log_{10}(\mathrm{SFR} [\mathrm{M_\odot/yr}])_{\mathrm{N_{Medium}} = 0},
\end{aligned}
\label{eq: mass sfr med}
\end{equation}

\noindent Stellar mass residuals are shown in the left panel and SFR residuals in the right panel. The results are binned in intervals of $1.0$ dex for stellar mass and $0.50$ dex for SFR. Points indicate the median ($50$th percentile) in each bin, with the $16$th--$84$th percentile range shown as error bars. Overall, we find inferred stellar masses and SFRs are broadly similar.

For stellar mass, we observe a slight positive offset, indicating that masses inferred with medium-band photometry are marginally higher, by an average of $0.037$ dex. When only broad bands are used, emission-line contributions are overestimated, which leads to an underestimate of the underlying stellar continuum and consequently a slight underestimate of the stellar mass. This shows that medium-band filters help separate emission-line flux from the underlying stellar continuum and hence providing better stellar mass constraints.

For SFR, galaxies with $\log_{10}(\mathrm{SFR} [\mathrm{M_\odot} / \mathrm{yr}])\gtrsim -1$ show broadly consistent values between the two runs. At lower SFRs ($\log_{10}(\mathrm{SFR} [\mathrm{M_\odot} / \mathrm{yr}]) \lesssim -1$), however, we find increasingly negative residuals, indicating that SFRs inferred with medium bands are slightly lower than those inferred using only wide bands. This can also be explained by the overestimation of emission-line flux in broad-band photometry we see above, which can lead to overestimated SFR. This effect is particularly noticeable in low-SFR systems, where emission lines are weak or absent and can be averaged out within the broad wavelength coverage of wide bands. Medium bands are sensitive to this and therefore provide a more accurate measurement of the emission-line contribution.

Furthermore, in \autoref{fig: mass uncer medium}, we demonstrate how the inclusion of medium-band photometry reduces the uncertainty in stellar mass estimates by comparing galaxies with $\mathrm{N_{Medium}} \geq 7$ (red) to those fitted using only wide-band photometry ($\mathrm{N_{Medium}} = 0$, blue).

In the top panel of \autoref{fig: mass uncer medium}, we show the stacked stellar mass posterior distributions in different redshift bins, together with the corresponding $\sigma_{M_*}$ values (the standard deviation of the stacked distributions) for both the $\mathrm{N_{Medium}} \geq 7$ and wide-band-only samples. To construct the stacked posterior distributions, we first shift the stellar mass posterior of each galaxy such that its 50th percentile is centered at zero by subtracting the median from all posterior samples. The shifted posterior distributions of all galaxies within the same redshift bin are then stacked, and the resulting distributions are plotted.

We find that galaxies with $\mathrm{N_{Medium}} \geq 7$ exhibit significantly narrower and more concentrated stellar mass posterior distributions compared to those inferred using only wide-band photometry, demonstrating the ability of medium-band photometry to improve stellar mass estimates. In the bottom panel, we further examine the redshift evolution of $\sigma^{\mathrm{N_{Medium}} \geq 7}_{M_*}$ and $\sigma^{\mathrm{N_{Medium}} = 0}_{M_*}$. On average, the stellar mass uncertainties inferred with $\mathrm{N_{Medium}} \geq 7$ are approximately $0.67$ times those obtained with $\mathrm{N_{Medium}} = 0$, corresponding to an improvement of $\sim 33\%$. In absolute terms, $\sigma^{\mathrm{N_{Medium}} \geq 7}_{M_*}$ is consistently smaller by $0.11^{+0.01}_{-0.01}$ dex than $\sigma^{\mathrm{N_{Medium}} = 0}_{M_*}$ across all redshift bins.

In summary, when only NIRCam wide-band photometry is used, emission-line flux can be slightly overestimated, leading to marginally lower stellar masses and higher inferred SFRs. However, the magnitude of these effects is small, and we conclude that stellar mass and SFR estimates remain broadly consistent with and without the inclusion of medium-band photometry. The inclusion of medium-band photometry improves the precision of stellar mass estimates, reducing the uncertainty by $\sim 0.11$ dex across all redshifts.

\subsection{Impact of MIRI Coverage}
\label{sec: mass miri}
As introduced in \autoref{sec: inferring galaxy properties}, we perform four separate runs. The primary run includes galaxies with HST and NIRCam photometry, effectively covering all galaxies. For galaxies with additional MIRI band coverage, we perform extra runs that incorporate the additional MIRI data. In GOODS-S, $96{,}730$ out of $303{,}907$ galaxies ($32\%$) have MIRI coverage and have the additional run. In GOODS-N, the number is significantly smaller: $4{,}861$ out of $181{,}061$ ($3\%$).

\begin{figure}
    \centering
    \includegraphics[width=\linewidth]{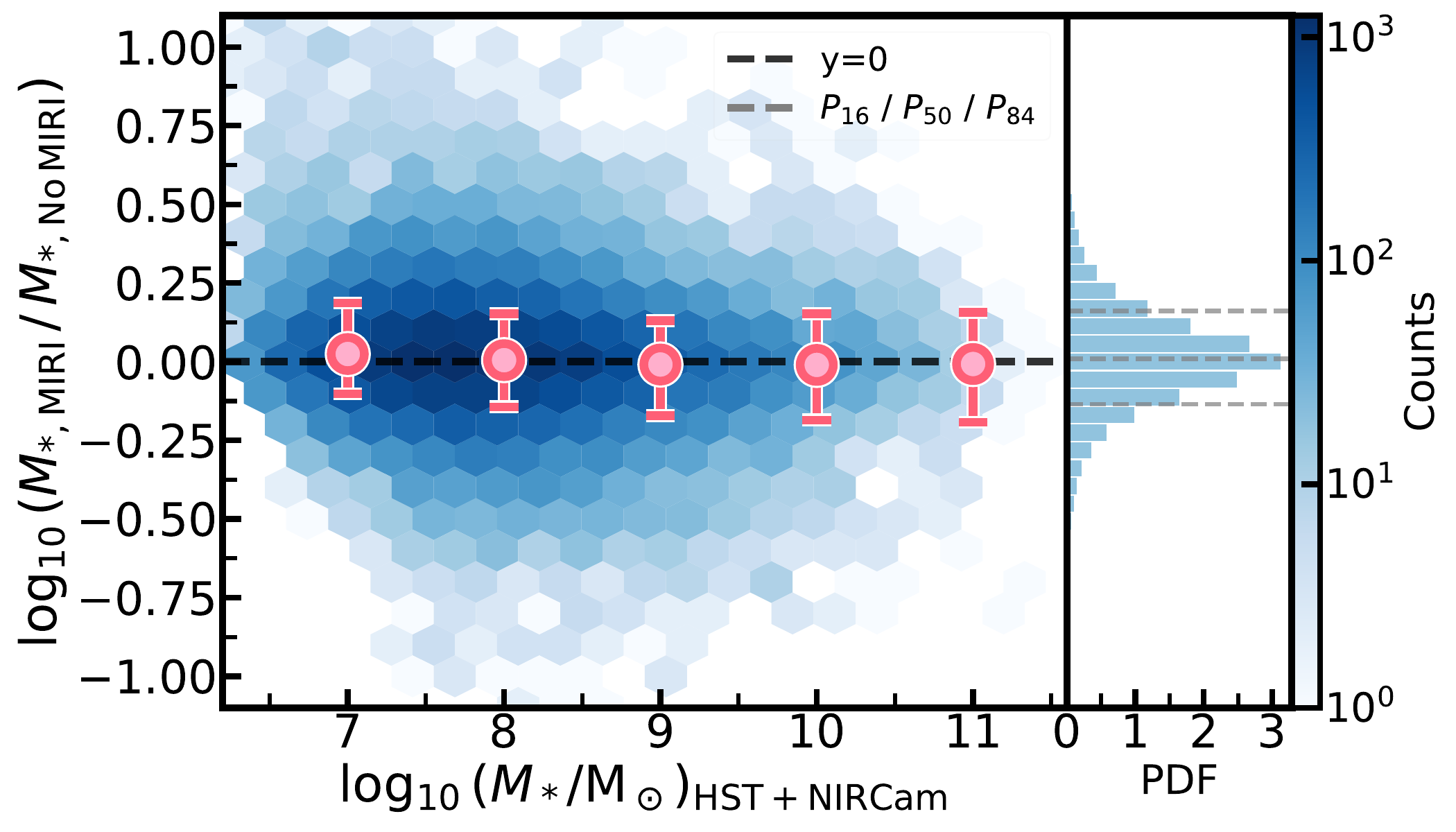}
    \caption{Stellar mass residuals shown as a function of stellar mass derived from HST + NIRCam alone. The residuals are defined as (HST + NIRCam + MIRI minus HST + NIRCam, \autoref{eq: mass miri}). We find excellent agreement across the full stellar mass range, with an average offset of $+0.008^{+0.139}_{-0.147}$ dex, well within the typical stellar mass uncertainties. We further examine potential redshift dependence and find that this strong consistency persists across all redshifts and stellar masses.}
    \label{fig: mass miri residual}
\end{figure}

The inclusion of the MIRI bands improves constraints on dust emission and the AGN contribution to the SED, but it is not expected to strongly affect stellar mass estimates. In this section, we therefore validate this expectation. We compare the stellar masses derived from fits using HST + NIRCam bands, $\log_{10}(M_*/\mathrm{M_\odot})_{\mathrm{HST+NIRCam}}$, to those derived using HST + NIRCam + MIRI, $\log_{10}(M_*/\mathrm{M_\odot})_{\mathrm{HST+NIRCam+MIRI}}$. This comparison is shown in \autoref{fig: mass miri residual}. The y-axis of the plot shows the residual 
\begin{equation}
\label{eq: mass miri}
\begin{aligned}
\Delta \log_{10}(M_*/\mathrm{M_\odot}) &= \log_{10}(M_*/\mathrm{M_\odot})_{\mathrm{HST+NIRCam+MIRI}} \\
&\quad - \log_{10}(M_*/\mathrm{M_\odot})_{\mathrm{HST+NIRCam}}
\end{aligned}
\end{equation}
plotted as a function of $\log_{10}(M_*/\mathrm{M_\odot})_{\mathrm{HST+NIRCam}}$.

In \autoref{fig: mass miri residual}, we show the hexbin density plot with bin counts indicated by the colorbar. To the right, we show the histogram of residuals. Additionally, we compute binned statistics in stellar mass, using bin widths of $\Delta \log_{10}(M_*/\mathrm{M_\odot}) = 1.0$, centered on $\log_{10}(M_*/\mathrm{M_\odot}) = [7, 8, 9, 10, 11]$. The 16th, 50th, and 84th percentiles are plotted with error bars on pink circular markers.

We find excellent agreement in stellar mass estimates. The median residual between mass estimates with and without MIRI across the sample is $+0.008^{+0.139}_{-0.147}$ dex. Even when we skip the redshift consistency cut and include all galaxies regardless of their $\mathrm{SNR}_{\mathrm{F444W}}$, the agreement remains tight, with a median residual of $-0.010^{+0.159}_{-0.165}$ dex. This result aligns well with our expectations. In the redshift range of interest, the key rest-frame wavelengths that dominate stellar mass estimates, typically $\lambda_\mathrm{rest} \sim4000 - 10{,}000$ \AA, are already well covered by NIRCam bands (especially F356W and F444W). In addition, due to the extensive NIRCam filter coverage in JADES, which already provides very robust and accurate stellar mass estimates. Therefore, the addition of MIRI is not expected to significantly improve them. 

We further investigate whether there is any redshift dependence in $\Delta \log_{10}(M_*/\mathrm{M_\odot})$ by examining its evolution across different stellar mass bins. We find that only at very high redshift ($z \sim 9$), and in the most massive bin ($\log_{10}(M_*/\mathrm{M_\odot}) \sim 9.5$), there appears to be a negative offset, with $\Delta \log_{10}(M_*/\mathrm{M_\odot}) = -0.12^{+0.34}_{-0.33}$. However, this offset is associated with substantially larger scatter, and only a very few galaxies at this high redshift have such high masses. Given the large uncertainties and enhanced scatter in this regime, we conclude that the apparent negative offset is not statistically significant at all, and no evidence for a redshift dependence mass residual is observed.

\begin{figure*}
    \centering
    \includegraphics[width=\linewidth]{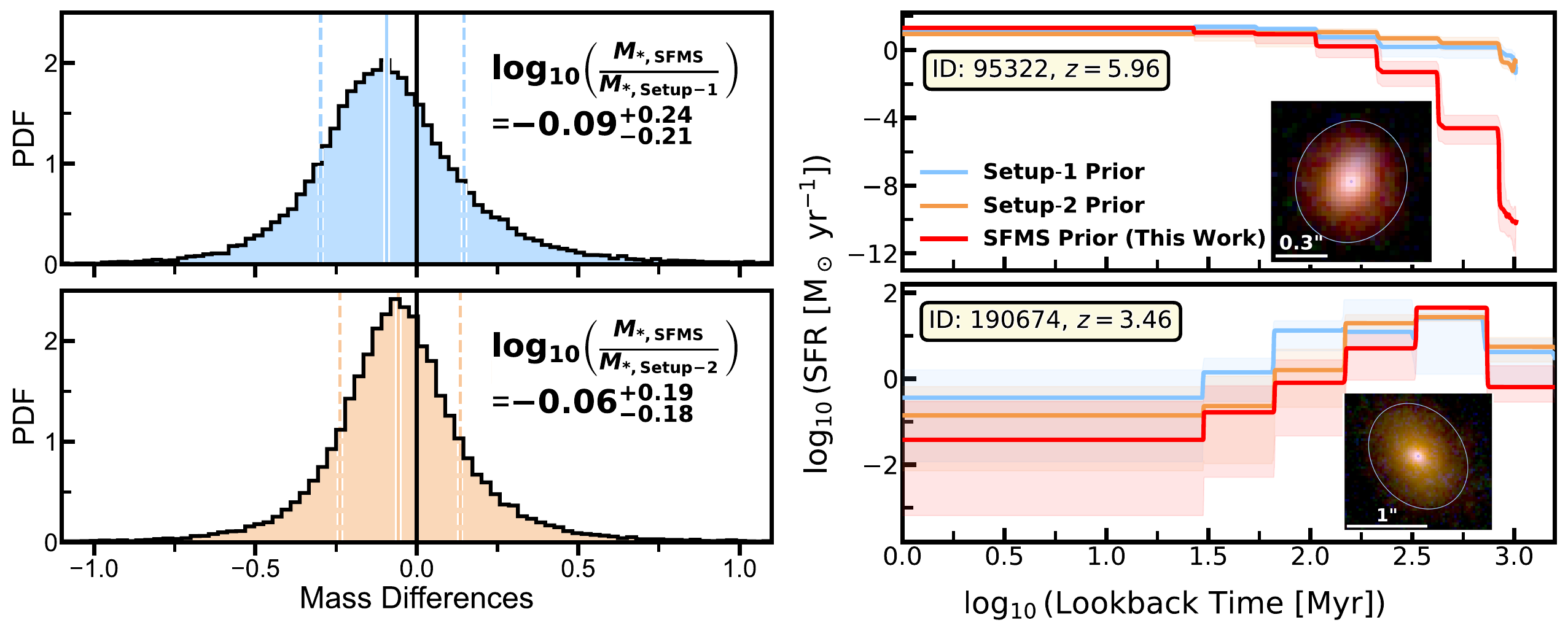}
    \caption{In this figure, we compare the inferred stellar masses and SFHs obtained using the SFMS prior introduced in this work with those derived from two SFH priors commonly adopted in the literature. The comparison is performed using a sample of $40{,}641$ galaxies with $\mathrm{SNR}_{\mathrm{F444W}} > 5.0$ and at least 20 HST+NIRCam bands. \textbf{Setup-1 Prior:} $\sigma = 0.3$, $\mu = 0$ \citep{leja2017sfh, tacchella2022, simmonds_2024_ionising}; \textbf{Setup-2 Prior:} $\sigma = 0.3$, with $\mu$ matched to the cosmic SFR density evolution from \citet{Behroozi2019} \citep{wang2023, wang_2024_uncover_stellar_pop}. \textbf{\textit{Left panels:}} The top panel shows the stellar mass ratios between the SFMS prior and the Setup-1 prior, while the bottom panel shows the ratios between the SFMS and Setup-2 priors. Vertical lines indicate the 16th, 50th, and 84th percentiles of the distributions. We find that stellar masses inferred using the SFMS prior are typically $\sim 0.06$--$0.09$ dex lower than those derived using the other two priors. \textbf{\textit{Right panels:}} The top and bottom panels present the SFHs of a representative star-forming galaxy and a quiescent galaxy, inferred using the three different priors. The SFMS prior introduced in this work produces an earlier rising SFH, leading to a slightly lower integrated stellar mass. Overall, however, the SFHs inferred from the three priors are broadly consistent.}
    \label{fig: sfh compare}
\end{figure*}

\subsection{Dependence on the SFH Prior}
\label{sec: mass prior comparison}
In \autoref{sec: rising sfh prior}, we introduced the star-forming main sequence (SFMS) prior for the non-parametric SFH. In this section, we investigate how the inferred stellar masses change under the SFMS prior by comparing it with two commonly adopted priors within the Continuity SFH framework. As discussed in \autoref{sec: rising sfh prior}, different studies have adopted different choices for the scaling factor $\sigma$ and the expected value $\mu$ of the Student’s $t$ distribution (\autoref{eq: students't}) that governs the prior on $\log_{10}\mathrm{(SFR_\mathrm{ratio})}$, resulting in different assumptions about the allowed level of SFH variability.

For clarity, we briefly summarise the SFMS prior adopted in this work, together with the comparison configurations described below:
\begin{itemize}
    \item \textbf{SFMS prior (This work):} $\sigma = 0.5$, with $\mu$ matched to the expected values from the SFMS relation of \citet{simmonds_2025}.
    \item \textbf{Setup-1:} $\sigma = 0.3$, $\mu = 0$ \citep[see][]{leja2017sfh, tacchella2022, simmonds_2024_ionising}.
    \item \textbf{Setup-2:} $\sigma = 0.3$, with $\mu$ matched to the cosmic SFR density evolution from \citet{Behroozi2019} \citep[see][]{wang2023, wang_2024_uncover_stellar_pop}.
\end{itemize}

Out of the 303,907 galaxies in GOODS-S with HST+NIRCam coverage, we select galaxies with $\mathrm{SNR}_{\mathrm{F444W}} > 5.0$ and at least 20 HST+NIRCam bands, yielding a sample of 40,641 galaxies. For this sample, we perform two additional \texttt{Prospector} runs, using the same model and priors as in our \prosp setup (\autoref{sec: inferring galaxy properties}), except for the SFH prior, which we change to Setup-1 and Setup-2 described above, respectively.

The comparison is shown in \autoref{fig: sfh compare}. The left panels present the stellar mass ratio between the SFMS prior and the Setup-1 prior (top), and between the SFMS prior and the Setup-2 prior (bottom). The right panels show example SFHs for two representative galaxies: a star-forming galaxy (top) and a quiescent galaxy (bottom).

We find that stellar masses inferred with the SFMS prior are, on average, lower by $-0.09^{+0.24}_{-0.21}$ dex relative to the Setup-1 prior and $-0.06^{+0.19}_{-0.18}$ dex relative to the Setup-2 prior. In other words, the SFMS prior yields stellar masses that are lower by approximately $\sim 0.1$ dex. This offset can be understood from the rising nature of the SFMS prior, which favours lower SFRs at the largest lookback times compared to the other two priors, thereby reducing the total accumulated stellar mass. A typical example of this is the SFH of the galaxy $\mathrm{ID_{DR5}} = 95322$, shown in the middle top panel.

Importantly, despite its rising expected SFH, the SFMS prior is still capable of modelling rapid quenching events, as has been discussed extensively in \autoref{sec: quiescent galaxies sfh}, \autoref{fig: combined quiescent sfh}, and illustrated in the middle-bottom panel of \autoref{fig: sfh compare}. Although from the SFMS prior, the expected value of the Student’s-$t$ distribution lies on the SFMS (see \autoref{sec: rising sfh prior} and \autoref  {fig:sfh prior combine}), the posterior can strongly deviate toward low SFR solutions when supported by the data. In fact, quiescent classifications under the SFMS prior are arguably more physically motivated: since the prior expectation is that galaxies lie on the SFMS, convergence toward low SFR values reflects quenching driven by the observed data, rather than being implicitly favored by the prior or stochastic.

\section{The JADES DR5 Stellar Population Catalogue}
\label{sec: prospector catalog}
In this section we present the JADES DR5 stellar population catalogue derived using the \prosp SED modelling code. Descriptions of all catalogue columns are provided in \autoref{tab: prospector vac tab}. In \autoref{sec: catalogue description} we give an overview of the catalogue and its structure, and provide recommendations for using the catalogue in \autoref{sec: usage recommendations}.

\subsection{Catalogue Description}
\label{sec: catalogue description}

Our \texttt{Prosepctor} modelling is performed on all $\sim 500{,}000$ detected galaxies in the JADES DR5 photometric catalogue \citep{robertson2026}. The fiducial run is based on HST+NIRCam photometry. For galaxies with available spectroscopic redshifts (Spec-$z$) and/or MIRI coverage, we additionally perform complementary fits. A detailed description of each modelling configuration is provided in \autoref{sec: Input Photometry and SED-Modelling Setup}. The number of galaxies included in each run is summarised in \autoref{tab: number of galaxies snr}.

Descriptions of catalogue columns (\autoref{tab: prospector vac tab}) are organised into five main categories. The first two categories describe (i) the observed properties of each galaxy and (ii) the physical parameters inferred by \texttt{Prospector}. In particular, we report the number of available bands in HST, $\mathrm{NIRCam_{medium}}$, $\mathrm{NIRCam_{wide}}$, and MIRI, as well as SNR in key \textit{JWST} medium and wide band filters. We also provide the full set of galaxy physical properties inferred from our 18-parameter \prosp model (see \autoref{sec: inferring galaxy properties} and \autoref{tab:prospector_priors} for details of the model and priors).

The third category contains galaxies' rest-frame AB absolute magnitudes across a range of filter systems, including Bessell $UBVRI$, 2MASS $JHK_s$, and SDSS $g_0$, $r_0$, and $i_0$, as well as colours constructed from these bands. We additionally provide the UV absolute magnitude $\mathrm{M_{UV}}$, the UV continuum slope $\beta$, and the dust attenuation in these filter bands.

Finally, the last two categories consist of quality and flags. These indicate whether a galaxy has spectroscopic redshift and/or MIRI coverage, and the quality of constraints on redshift and stellar mass from \texttt{Prospector}. We also include flags identifying spectroscopically confirmed Type~I and Type~II AGNs \citep{ignas2026_type1, scholtz2025_type2}, as well as photometrically selected LRD candidates from \citet{rinaldi2026_lrd}.

\subsection{Data Access and Usage Recommendations}
\label{sec: usage recommendations}

The data products will be made publicly available at the time of publication. In this subsection, we provide guidance on how to use the catalogue.

We encourage users to define their own sample-selection criteria from the JADES DR5 stellar population catalogue, tailored to their specific scientific goals, using the various quality flags provided. As a convenient starting point, we introduce a fiducial quality flag, \texttt{use\_phot}, which selects galaxies that are well detected and have reliably constrained redshifts and stellar masses from \texttt{Prospector}. Specifically, \texttt{use\_phot} = 1 corresponds to galaxies that simultaneously satisfy the following four criteria: $\mathrm{SNR}_{\mathrm{F444W}} \geq 3$, $\mathrm{N}_{\mathrm{bands,\,NIRCam}} \geq 6$, \texttt{flag\_good\_redshift\_50}, and \texttt{flag\_good\_mass\_dex08}. These constitute a relatively loose, fiducial selection. In \autoref{sec: n galaxies selection}, we present the number of sources and their fraction of the total sample in each run, under several different selection criteria.

Given we have \texttt{Prospector} modelling results for all detected galaxies in the JADES GOODS-S and GOODS-N fields, we strongly recommend that users tailor their selections to the specific needs of their analyses. For example, studies focusing on emission-line properties inferred from photometry may benefit from requiring a higher number of NIRCam medium bands (e.g.\ larger \texttt{Nbands\_nircam\_medium}). Similarly, analyses that demand the most robust constraints on galaxy properties may consider selecting galaxies with \texttt{flag\_good\_redshift\_10} = 1 and \texttt{flag\_good\_mass\_dex05} = 1.

Overall, the catalogue is designed to be flexible, enabling users to construct samples optimised for a wide range of scientific applications.

\section{Conclusions}
\label{sec: conclusions}
In this work, we present a galaxy stellar population catalogue for $\sim 500{,}000$ galaxies in the latest JADES DR5 release across GOODS-S and GOODS-N, covering approximately $500$ arcmin$^2$, with up to 35 and 26 HST ACS/WFC3 + NIRCam + MIRI filter bands in GOODS-S and GOODS-N, respectively. We employ the Bayesian forward-modelling framework \prosp to infer galaxy physical properties, together with the \texttt{Parrot} neural network emulator to accelerate \texttt{FSPS} model evaluations. We also introduce a new physically motivated prior (SFMS prior) on non-parametric SFHs. Using this framework, we robustly infer a wide range of galaxy properties, including redshift, stellar mass, SFR, sSFR, dust attenuation, dust emission, and AGN contribution. Our main results are summarised below:

\textbf{I.} We introduce a physically motivated prior for non-parametric SFHs, referred to as the SFMS prior. This prior is based on the assumption that galaxies predominantly grow and evolve along the SFMS, while still being able to model and capture sudden starburst and quenching events. Extensive tests demonstrate that stellar masses inferred using this prior are, on average, lower by $\sim 0.06$--$0.09$ dex compared to those obtained with currently popular priors (\texttt{Prospector}-$\alpha$ \& $\beta$). The inferred SFHs remain broadly consistent across priors (see \autoref{fig: sfms}, \autoref{fig:sfh prior combine}, \autoref{fig: combined quiescent sfh}, \autoref{fig: sfh compare}).

\textbf{II.} We compare the observed and model-predicted photometry and find excellent agreement across all filter bands, with no significant systematic offsets, including in bands sensitive to strong emission-lines (see \autoref{fig: chi all}). In addition, we use the Kullback–Leibler (KL) divergence metric to quantify the extent to which the \prosp free parameters are constrained by the data relative to their priors. We find that redshift, stellar mass, metallicity, and dust attenuation are strongly data-driven. Metallicity and SFH parameters are also well constrained, while still incorporating physically informative priors. The inclusion of MIRI photometry significantly improves constraints on AGN contributions and mid-infrared dust emission, which would otherwise remain largely prior-dominated (see \autoref{fig: kl divergence}).

\textbf{III.} We evaluate the robustness of our stellar mass measurements by examining the relationship between stellar mass and $\mathrm{F444W}$ flux, as well as the dependence of stellar mass uncertainties on $\mathrm{SNR}_{\mathrm{F444W}}$. We find tight and consistent scaling relations between stellar mass and flux across all redshifts. As expected, increasing SNR reduces the uncertainty in stellar mass. At very high $\mathrm{SNR}_{\mathrm{F444W}}$, further increases in SNR provide only negligible improvements in precision, as the uncertainties in this regime are dominated by intrinsic degeneracies within the SED modelling framework rather than by photometric noise (see \autoref{fig: mass f444w}, \autoref{fig: sigma_m f444w}).

\textbf{IV.} We assess the quality of our photometric redshift inference by comparing against spectroscopic redshifts for 16,320 galaxies with secure spec-$z$ measurements. We obtain an outlier fraction of $7.04\%$ and a normalized median absolute deviation ($\sigma_{\rm NMAD}$) of 0.026, indicating robust and accurate redshift performance. We note that this redshift precision is competitive with, and in some regimes superior to, that obtained using \texttt{EAZY}. In our analysis, we adopt \texttt{KRON\_CONV} elliptical apertures to capture the total galaxy light and avoid underestimating stellar masses. Even higher redshift precision could be achieved by using smaller circular apertures, which reduce background noise contamination and improve photometric signal quality (see \autoref{fig: zphot zspec all}, \autoref{fig: prosp eazy redshift}).

\textbf{V.} We compare SFR$_{\mathrm{10Myr}}$ inferred from our SED modelling with SFRs derived from R1000 H$\alpha$ grating measurements, obtaining a median residual of $-0.017^{+0.380}_{-0.349}$ dex. This difference is well within the intrinsic uncertainties of the SED modelling, indicating strong consistency between the two independent SFR estimators (see \autoref{fig: sfr halpha prosp}). We further assess the reliability of the \prosp inferred AGN fraction, defined as the ratio of bolometric AGN luminosity to the galaxy’s total $4$–$20\,\mu$m luminosity. We find systematically higher AGN fractions in samples that are spectroscopically confirmed Type-I and Type-II AGNs, as well as in MIRI photometrically selected AGNs, demonstrating that our framework provides meaningful and physically consistent measurements of AGN contributions to (see \autoref{fig: agn miri}).

\textbf{VI.} We examine the consistency of stellar mass estimates when including NIRCam medium-band or MIRI photometry. The inclusion of medium-band filters significantly improves redshift constraints, reducing $\sigma_{\rm NMAD}$ by a factor of $\sim 2.1$ compared to fits using only wide-band photometry. The inferred stellar masses and SFRs obtained with and without medium-band photometry remain broadly consistent (\autoref{fig: mass med}). The inclusion of medium-band photometry reduces stellar mass uncertainties by 0.11 dex across all redshifts, compared to wide-band-only photometry (\autoref{fig: mass uncer medium}). We also test the impact of including MIRI photometry in the SED modelling. With the addition of MIRI bands, inferred stellar mass are on averaged,  $+0.008^{+0.139}_{-0.147}$ dex higher relative to fits using HST + NIRCam photometry alone (\autoref{fig: mass miri residual}). This difference is well within, and smaller than, the typical stellar mass uncertainties. Furthermore, we find no significant redshift-dependent trend in the stellar mass offsets, indicating the robustness and consistency of the stellar mass estimates presented in this work.

In conclusion, leveraging the deepest and most extensive multi-filter \textit{JWST} dataset from JADES, together with the sophisticated Bayesian SED modelling framework \texttt{Prospector}, we robustly and precisely infer the physical properties of $\sim 500{,}000$ galaxies in GOODS-S and GOODS-N. This work delivers one of the most comprehensive galaxy stellar population catalogue to date, spanning redshift, stellar mass, SFHs, dust attenuation and emission, and AGN contributions. The depth, wavelength coverage, and statistical scale make this catalogue a powerful resource for a wide range of studies in galaxy formation and evolution.

\section*{Acknowledgements}
Qiao Duan acknowledges support from a PhD studentship awarded by Trinity College, University of Cambridge. ST acknowledges support by the Royal Society Research Grant G125142. BDJ acknowledges support from the NIRCam Science Team contract to the University of Arizona, NAS5-02105, and \textit{JWST} Program 3215. BER acknowledges support from the NIRCam Science Team contract to the University of Arizona, NAS5-02105, and \textit{JWST} Program 3215. CS acknowledges support from the Science and Technology Facilities Council (STFC), by the ERC through Advanced Grant 695671 “QUENCH”, by the UKRI Frontier Research grant RISEandFALL. WMB gratefully acknowledges support from DARK via the DARK fellowship. This work was supported by a research grant (VIL54489) from VILLUM FONDEN. AJB acknowledges funding from the “FirstGalaxies” Advanced Grant from the European Research Council (ERC) under the European Union’s Horizon 2020 research and innovation program (Grant agreement No. 789056). S.C acknowledges support by European Union’s HE ERC Starting Grant No. 101040227 - WINGS. CC acknowledges support from the JWST/NIRCam Science Team contract to the University of Arizona, NAS5-02105, and JWST Programs 3215 and 5015. ECL acknowledges support of an STFC Webb Fellowship (ST/W001438/1). A.L.D. thanks the University of Cambridge Harding Distinguished Postgraduate Scholars Programme and Technology Facilities Council (STFC) Center for Doctoral Training (CDT) in Data intensive science at the University of Cambridge (STFC grant number 2742605) for a PhD studentship. FDE acknowledges support by the Science and Technology Facilities Council (STFC), by the ERC through Advanced Grant 695671 ``QUENCH'', and by the UKRI Frontier Research grant RISEandFALL. DJE is supported as a Simons Investigator and by JWST/NIRCam contract to the University of Arizona, NAS5-02105. Support for program 3215 was provided by NASA through a grant from the Space Telescope Science Institute, which is operated by the Association of Universities for Research in Astronomy, Inc., under NASA contract NAS 5-03127. SG is supported by a Woolf Fisher Scholarship from the Woolf Fisher Trust of New Zealand and Cambridge Commonwealth and European Trust. RC acknowledges funds by Johns Hopkins University, Institute for Data Intensive Engineering and Science (IDIES). J.M.H. is supported by JWST Program 8544. PIN thanks the LSST-DA Data Science Fellowship Program, which is funded by LSST-DA, the Brinson Foundation, the WoodNext Foundation, and the Research Corporation for Science Advancement Foundation; her participation in the program has benefited this work. PIN acknowledge financial support from the State Research Agency of the Spanish Ministry of Science and Innovation (AEI-MCINN) under the grants “Galaxy Evolution with Artificial Intelligence” with reference PGC2018-100852-A-I00 and “BASALT” with reference PID2021-126838NB-I00. YI is supported by JSPS KAKENHI Grant No. 24KJ0202. ZJ acknowledges JWST/NIRCam contract to the University of Arizona, NAS5-02105. MK acknowledges support from a PhD studentship awarded by the University of Cambridge Harding Distinguished Postgraduate Scholars Programme, UK Science and Technology Facilities Council (STFC) Center for Doctoral Training (CDT) in Data Intensive Science, and Girton College Cambridge. TJL gratefully acknowledges support from the Swiss National Science Foundation through a SNSF Mobility Fellowship and from the NASA/JWST Program OASIS (PID 5997). RM acknowledges support by the Science and Technology Facilities Council (STFC), by the ERC through Advanced Grant 695671 “QUENCH”, and by the UKRI Frontier Research grant RISEandFALL. RM also acknowledges funding from a research professorship from the Royal Society. RGP acknowledges funding support from a STFC PhD studentship. PGP-G acknowledges support from grant PID2022-139567NB-I00 funded by Spanish Ministerio de Ciencia e Innovaci\'on MCIN/AEI/10.13039/501100011033, FEDER, UE. DP acknowledges support by the Huo Family Foundation through
a P.C. Ho PhD Studentship. MR acknowledges JWST/NIRCam contract to the University of Arizona NAS5-02105. B.R.P acknowledges support from grant PID2024-158856NA-I00 funded by Spanish Ministerio de Ciencia e Innovación MCIN/AEI/10.13039/501100011033 and by “ERDF 4A way of making Europe”. J.S. acknowledges support by the Science and Technology Facilities Council (STFC), ERC Advanced Grant 695671 "QUENCH". JAAT acknowledges support from the Simons Foundation and JWST program 3215. Support for program 3215 was provided by NASA through a grant from the Space Telescope Science Institute, which is operated by the Association of Universities for Research in Astronomy, Inc., under NASA contract NAS 5-03127. H\"U acknowledges support by the Max Planck Society through the Lise Meitner Excellence Program. H\"U acknowledges funding by the European Union (ERC APEX, 101164796). Views and opinions expressed are however those of the authors only and do not necessarily reflect those of the European Union or the European Research Council Executive Agency. Neither the European Union nor the granting authority can be held responsible for them. The research of CCW is supported by NOIRLab, which is managed by the Association of Universities for Research in Astronomy (AURA) under a cooperative agreement with the National Science Foundation. LW acknowledges support from the Gavin Boyle Fellowship at the Kavli Institute for Cosmology, Cambridge and from the Kavli Foundation. YZ acknowledges JWST/NIRCam contract to the University of Arizona NAS5-02105.

This work made use of the DiRAC High Performance Computing facility through grant APP53421 (PI: Tacchella), From Darkness to Light: \textit{The First \textit{JWST} Consensus of Early Galaxies}.

This work is based [in part] on observations made with the NASA/ESA/CSA James Webb Space Telescope. The data were obtained from the Mikulski Archive for Space Telescopes at the Space Telescope Science Institute, which is operated by the Association of Universities for Research in Astronomy, Inc., under NASA contract NAS 5-03127 for JWST.

\section*{Data Availability}
This work is part of the JADES Data Release 5 (DR5). It uses JADES DR5 data products, including imaging products from \cite{johnson2026, alberts2026}, photometric detections and catalogue from \cite{robertson2026}, the morphology catalogue from \cite{carreira2026}, and high-$z$ galaxy selections from \cite{Hainline2026}.  

The stellar population catalogue presented in this work will be made available for download at the time of publication.



\bibliographystyle{mnras}
\bibliography{main} 



\appendix
\section{Prospector Value-Added Stellar Population Catalogue}
\label{sec: prospector catalog appendix}
We present the Prospector stellar population catalogue in \autoref{tab: prospector vac tab}. For the full description of the catalogue, see \autoref{sec: prospector catalog}.

Note that we perform \texttt{Prospector} SED modelling for all detected sources in the JADES DR5 photometric catalogue. The fiducial run uses \textbf{HST + NIRCam} data and is applied to all galaxies. Additional complementary runs are carried out when \textbf{MIRI} and \textbf{Spec-\boldmath$z$} information are available.

In the following subsection, we present how one can apply different quality and selection criteria to select galaxy samples based on their scientific goals.

\subsection{Sample Sizes Across Different Selection Criteria}
\label{sec: n galaxies selection}
\begin{figure*}
    \centering
    \includegraphics[width=0.97\linewidth]{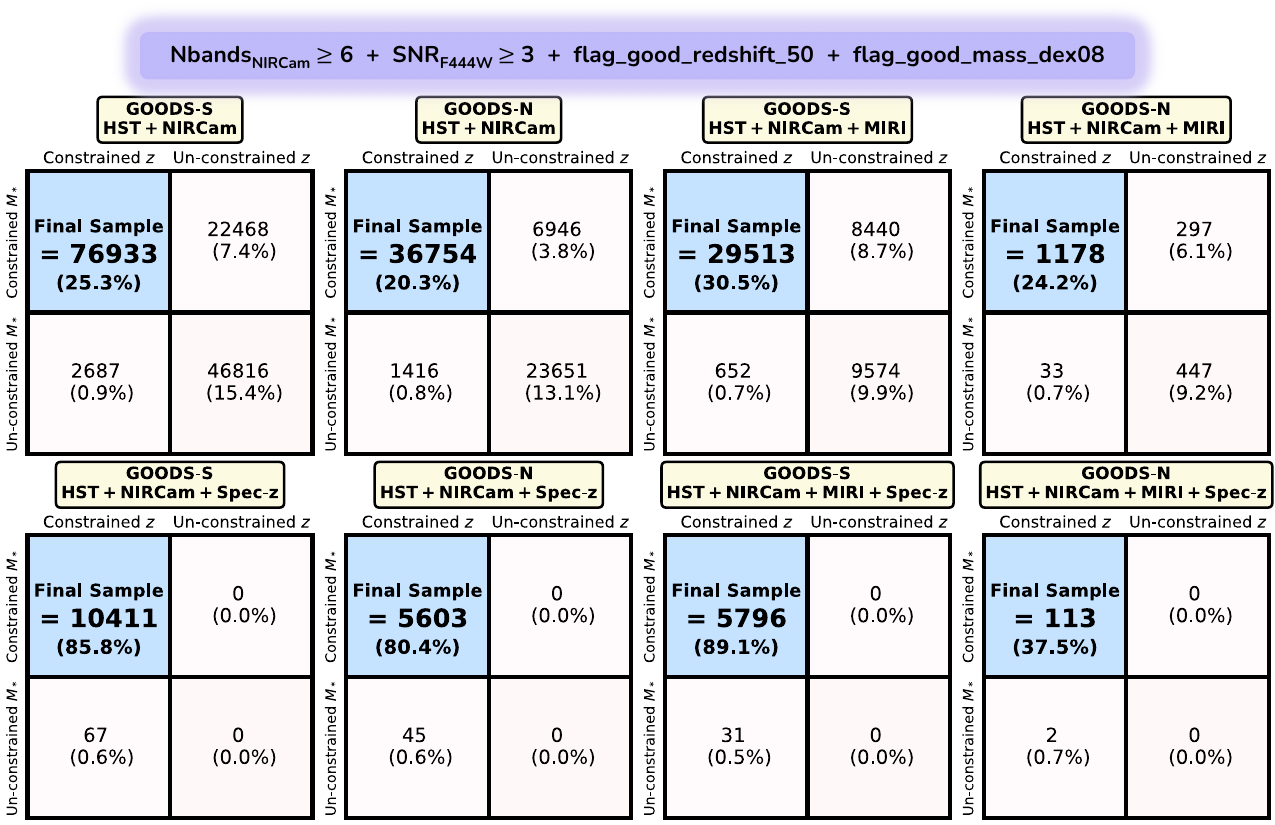}
    \includegraphics[width=0.97\linewidth]{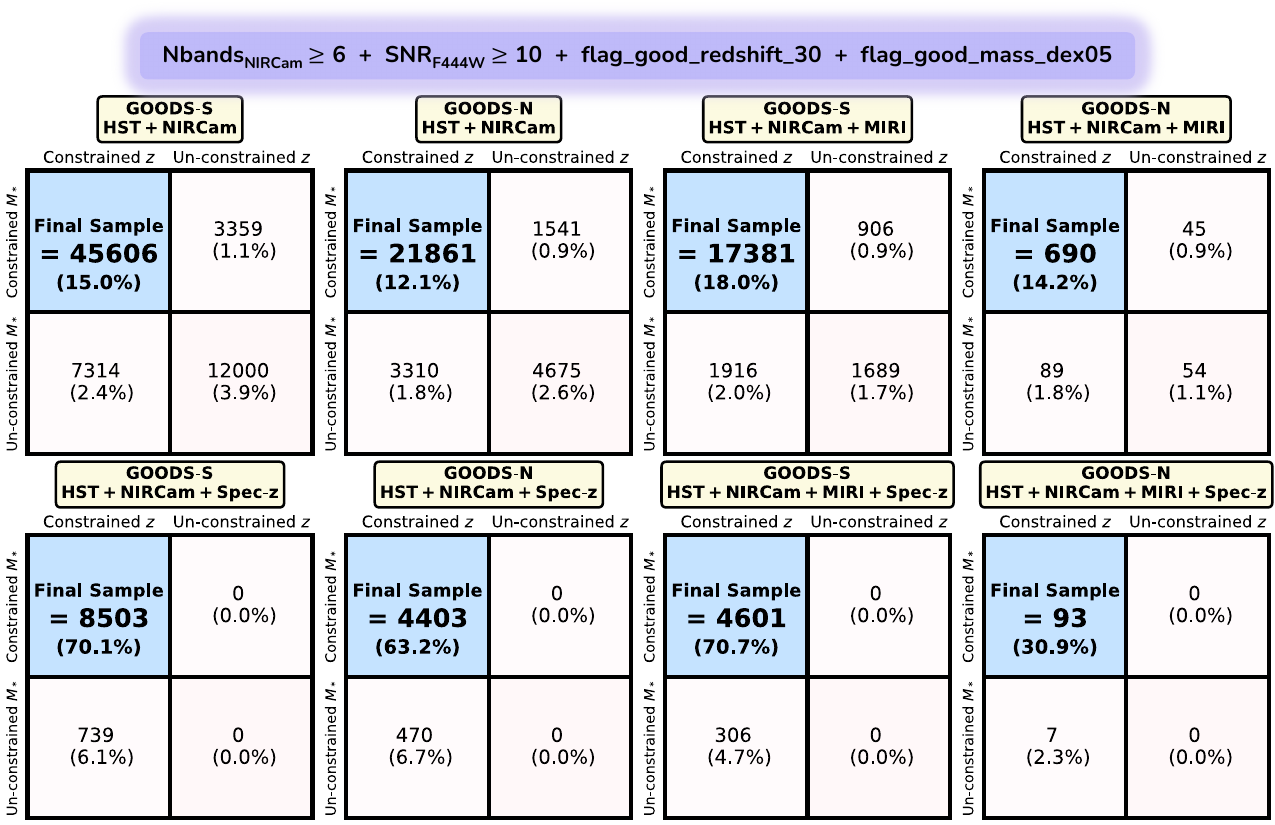}
    \caption{Sample sizes for each of the four \texttt{Prospector} runs in GOODS-S and GOODS-N (see \autoref{tab: number of galaxies snr}). We first apply photometric quality cuts, based on $\mathrm{SNR}_{\mathrm{F444W}}$ and $\mathrm{N}_{\mathrm{bands,\,NIRCam}}$, to define a robustly detected galaxy sample. These diagrams then illustrate how additional quality criteria on redshift and stellar mass further refine this selection. The quoted fractions represent the number of selected galaxies relative to the total number of sources in each filterband configuration. For example, in the top-left panel, 76{,}933 (25.3\%) corresponds to $76{,}933 / 303{,}907$, where 303{,}907 is the total number of galaxies in GOODS-S with HST + NIRCam coverage.}
    \label{fig: box diagram}
\end{figure*}

In this subsection, we present two illustrative example selection criterias, along with the corresponding number of galaxies that remain and their fraction relative to the full sample. A complete visualisation of how sample sizes change under different cuts is shown in \autoref{fig: box diagram}.

The first panel of \autoref{fig: box diagram} shows our fiducial selection, which constitutes \texttt{use\_phot} flag in the catalog. This selection requires $\mathrm{SNR}_{\mathrm{F444W}} \geq 3$, $\mathrm{N}_{\mathrm{bands,\,NIRCam}} \geq 6$, and the inference quality flags \texttt{flag\_good\_redshift\_50} and \texttt{flag\_good\_mass\_dex08}. The second panel presents a more stringent example, where we impose $\mathrm{SNR}_{\mathrm{F444W}} \geq 10$, retain $\mathrm{N}_{\mathrm{bands,\,NIRCam}} \geq 6$, and adopt tighter inference criteria, \texttt{flag\_good\_redshift\_30} and \texttt{flag\_good\_mass\_dex05}.

From the top panel, corresponding to the fiducial selection, we find that for a relatively broad initial sample selections of $\mathrm{SNR}_{\mathrm{F444W}} \geq 3$, approximately $\sim 20\%$ of galaxies do not have well-constrained redshifts and stellar masses. In contrast, for the more stringent initial selection with $\mathrm{SNR}_{\mathrm{F444W}} > 10$, even when applying tighter constraints on both redshift and stellar mass, this fraction decreases to $\sim 7\%$. This shows a higher SNR could significantly improve inference quality, which has already been discussed extensively in the paper. 

Lastly, these two examples are intended for illustration only. We recommend that users define their own sample selection criteria based on their scientific goals when using the JADES DR5 stellar population catalogue.

\begin{table*}
\centering
\caption{Description of all columns in the JADES DR5 stellar population catalogue.}
\label{tab: prospector vac tab}
\begin{tabular}{l c c p{10.5cm}}
\toprule \toprule
\textbf{Field} & \textbf{Data type} & \textbf{Unit} & \textbf{Description} \\
\midrule
\texttt{field} & str & -- & GOODS-S / GOODS-N \\[1.0pt]
\texttt{ID} & int & -- & Galaxy ID in JADES DR5 Release \\[1.0pt]
\texttt{Nbands} & int & -- & Total number of photometric bands used in the SED modelling \\[1.0pt]
\texttt{Nbands\_hst} & int & -- & Number of \textit{HST} bands used in the SED modelling \\[1.0pt]
\texttt{Nbands\_nircam} & int & -- & Number of \textit{JWST/NIRCam} bands used in the SED modelling \\[1.0pt]
\texttt{Nbands\_nircam\_wide} & int & -- & Number of \textit{JWST/NIRCam} wide-bands used in the SED modelling \\[1.0pt]
\texttt{Nbands\_nircam\_medium} & int & -- & Number of \textit{JWST/NIRCam} medium-bands used in the SED modelling \\[1.0pt]
\texttt{Nbands\_miri} & int & -- & Number of \textit{JWST/MIRI} bands used in the SED modelling \\[1.0pt]
\texttt{chi\_tot} & float & -- & Total absolute chi: $\chi_{\mathrm{tot}} = \sum |\mathrm{obs} - \mathrm{model}| / \sigma_{\mathrm{obs}}$ \\[1.0pt]
\texttt{chi2\_tot} & float & -- & Total chi-squared: $\chi^2_{\mathrm{tot}} = \sum (\mathrm{obs} - \mathrm{model})^2 / \sigma_{\mathrm{obs}}^2$ \\[1.0pt]
\texttt{snr\_F\{xxxx\}} & float & -- & Signal-to-noise ratio in major wide and medium \textit{JWST} NIRCam bands: F090W, F150W, F200W, F277W, F335M, F356W, F410M, F444W \\[1.0pt]
{\color{blue}\texttt{stellar\_mass\_\{xx\}th}} & float & $\log_{10}(M_*/\mathrm{M_\odot})$ &
\parbox[t]{\linewidth}{
$16^{\mathrm{th}}/50^{\mathrm{th}}/84^{\mathrm{th}}$ percentile of \texttt{stellar\_mass}.\\
\textbf{{\color{blue} Note}}: \texttt{stellar\_mass} refers to the stellar mass at the epoch of observation, accounting for mass loss from dying stars.} \\[1.0pt]
{\color{blue}\texttt{mass\_weighted\_age\_\{xx\}th}} & float &  $\log_{10}(\mathrm{yr})$ & $16^{\mathrm{th}}/50^{\mathrm{th}}/84^{\mathrm{th}}$ percentile of mass-weighted age \\[1.0pt]
{\color{blue}\texttt{sfr\_\{xx\}myr\_\{yy\}}} & float & $\mathrm{M_\odot\,yr^{-1}}$ & SFR averaged over the past \texttt{xx}~Myr of the SFH (\texttt{yy} = 16th/50th/84th percentiles)\\[1.0pt]
{\color{blue}\texttt{ssfr\_\{xx\}myr\_\{yy\}}} & float & $\log_{10}(\mathrm{yr^{-1}})$ & Specific SFR averaged over the past \texttt{xx}~Myr of the SFH (\texttt{yy} = 16th/50th/84th percentiles) \\[1.0pt]

\texttt{agebin\_1} & list & $\log_{10}(\mathrm{yr})$ & Time boundaries of the 1st SFH bin, $[\mathrm{start},\,\mathrm{end}]$ \\[1.0pt]
\texttt{agebin\_2} & list & $\log_{10}(\mathrm{yr})$ & Time boundaries of the 2nd SFH bin, $[\mathrm{start},\,\mathrm{end}]$ \\[1.0pt]
\texttt{agebin\_3} & list & $\log_{10}(\mathrm{yr})$ & Time boundaries of the 3rd SFH bin, $[\mathrm{start},\,\mathrm{end}]$ \\[1.0pt]
\texttt{agebin\_4} & list & $\log_{10}(\mathrm{yr})$ & Time boundaries of the 4th SFH bin, $[\mathrm{start},\,\mathrm{end}]$ \\[1.0pt]
\texttt{agebin\_5} & list & $\log_{10}(\mathrm{yr})$ & Time boundaries of the 5th SFH bin, $[\mathrm{start},\,\mathrm{end}]$ \\[1.0pt]
\texttt{agebin\_6} & list & $\log_{10}(\mathrm{yr})$ & Time boundaries of the 6th SFH bin, $[\mathrm{start},\,\mathrm{end}]$ \\[1.0pt]
\texttt{agebin\_7} & list & $\log_{10}(\mathrm{yr})$ & Time boundaries of the 7th SFH bin, $[\mathrm{start},\,\mathrm{end}]$ \\[1.0pt]
\texttt{sfr\_agebin\_1\_\{xx\}th} & float & $\mathrm{M_\odot\,yr^{-1}}$ & $16^{\mathrm{th}}/50^{\mathrm{th}}/84^{\mathrm{th}}$ percentile of the SFR in the 1st SFH age bin \\[1.0pt]
\texttt{sfr\_agebin\_2\_\{xx\}th} & float & $\mathrm{M_\odot\,yr^{-1}}$ & $16^{\mathrm{th}}/50^{\mathrm{th}}/84^{\mathrm{th}}$ percentile of the SFR in the 2nd SFH age bin \\[1.0pt]
\texttt{sfr\_agebin\_3\_\{xx\}th} & float & $\mathrm{M_\odot\,yr^{-1}}$ & $16^{\mathrm{th}}/50^{\mathrm{th}}/84^{\mathrm{th}}$ percentile of the SFR in the 3rd SFH age bin \\[1.0pt]
\texttt{sfr\_agebin\_4\_\{xx\}th} & float & $\mathrm{M_\odot\,yr^{-1}}$ & $16^{\mathrm{th}}/50^{\mathrm{th}}/84^{\mathrm{th}}$ percentile of the SFR in the 4th SFH age bin \\[1.0pt]
\texttt{sfr\_agebin\_5\_\{xx\}th} & float & $\mathrm{M_\odot\,yr^{-1}}$ & $16^{\mathrm{th}}/50^{\mathrm{th}}/84^{\mathrm{th}}$ percentile of the SFR in the 5th SFH age bin \\[1.0pt]
\texttt{sfr\_agebin\_6\_\{xx\}th} & float & $\mathrm{M_\odot\,yr^{-1}}$ & $16^{\mathrm{th}}/50^{\mathrm{th}}/84^{\mathrm{th}}$ percentile of the SFR in the 6th SFH age bin \\[1.0pt]
\texttt{sfr\_agebin\_7\_\{xx\}th} & float & $\mathrm{M_\odot\,yr^{-1}}$ & $16^{\mathrm{th}}/50^{\mathrm{th}}/84^{\mathrm{th}}$ percentile of the SFR in the 7th SFH age bin \\[3.0pt]

\hline \\[-3.0pt]
\multicolumn{4}{c}{\normalsize{\textbf{Inferred Prospector free parameters as defined in \autoref{tab:prospector_priors}.}}} \\[6.0pt]
{{\color{blue}\texttt{zred\_\{xx\}th}}} & float & -- & $16^{\mathrm{th}}/50^{\mathrm{th}}/84^{\mathrm{th}}$ percentile of \texttt{zred} \\[1.0pt]

{\color{blue}\texttt{total\_formed\_mass\_\{xx\}th}} & float & $\log_{10}(M_*/\mathrm{M_\odot})$ &
\parbox[t]{\linewidth}{
$16^{\mathrm{th}}/50^{\mathrm{th}}/84^{\mathrm{th}}$ percentile of galaxy total formed stellar mass\\
\textbf{{\color{blue} Note: }} \texttt{total\_formed\_mass\_\{xx\}th} refers to the total formed stellar mass, computed from the integral of the SFH. Differs from \texttt{stellar\_mass} defined below
} \\[1.0pt]

\texttt{metallicity\_\{xx\}th} & float & $\log_{10}(Z_*/\mathrm{Z_\odot})$ & $16^{\mathrm{th}}/50^{\mathrm{th}}/84^{\mathrm{th}}$ percentile of galaxy stellar metallicity  \\[1.0pt]

\texttt{logsfr\_ratios\_1\_\{xx\}th} & float & $\log_{10}(Z_*/\mathrm{Z_\odot})$ & 
\parbox[t]{\linewidth}{
$16^{\mathrm{th}}/50^{\mathrm{th}}/84^{\mathrm{th}}$ percentile of $\log_{10}(\mathrm{SFR}_{1}/\mathrm{SFR}_{2})$,\\
\textbf{{\color{blue} Note: }} bin~1 corresponds to the earliest bin in lookback time, 
} \\[1.0pt]

\texttt{logsfr\_ratios\_2\_\{xx\}th} & float & -- & $16^{\mathrm{th}}/50^{\mathrm{th}}/84^{\mathrm{th}}$ percentile of $\log_{10}(\mathrm{SFR}_{2}/\mathrm{SFR}_{3})$ \\[1.0pt]
\texttt{logsfr\_ratios\_3\_\{xx\}th} & float & -- & $16^{\mathrm{th}}/50^{\mathrm{th}}/84^{\mathrm{th}}$ percentile of $\log_{10}(\mathrm{SFR}_{3}/\mathrm{SFR}_{4})$ \\[1.0pt]
\texttt{logsfr\_ratios\_4\_\{xx\}th} & float & -- & $16^{\mathrm{th}}/50^{\mathrm{th}}/84^{\mathrm{th}}$ percentile of $\log_{10}(\mathrm{SFR}_{4}/\mathrm{SFR}_{5})$ \\[1.0pt]
\texttt{logsfr\_ratios\_5\_\{xx\}th} & float & -- & $16^{\mathrm{th}}/50^{\mathrm{th}}/84^{\mathrm{th}}$ percentile of $\log_{10}(\mathrm{SFR}_{5}/\mathrm{SFR}_{6})$ \\[1.0pt]
\texttt{logsfr\_ratios\_6\_\{xx\}th} & float & -- & $16^{\mathrm{th}}/50^{\mathrm{th}}/84^{\mathrm{th}}$ percentile of $\log_{10}(\mathrm{SFR}_{6}/\mathrm{SFR}_{7})$ \\[1.0pt]
\texttt{dust2\_\{xx\}th} & float & -- & $16^{\mathrm{th}}/50^{\mathrm{th}}/84^{\mathrm{th}}$ percentile of diffuse dust optical depth \citep{charlot2000} \\[1.0pt]
\texttt{dust\_index\_\{xx\}th} & float & -- & $16^{\mathrm{th}}/50^{\mathrm{th}}/84^{\mathrm{th}}$ percentile of the power-law index in the \citet{calzetti2000} dust attenuation\\[1.0pt]
\texttt{dust1\_fraction\_\{xx\}th} & float & -- & $16^{\mathrm{th}}/50^{\mathrm{th}}/84^{\mathrm{th}}$ percentile of birth cloud to diffuse dust optical depth ratio \citep{charlot2000}\\[1.0pt]
\texttt{log\_fagn\_\{xx\}th} & float & -- & $16^{\mathrm{th}}/50^{\mathrm{th}}/84^{\mathrm{th}}$ percentile of $\log_{10}$ of the AGN-to-bolometric luminosity ratio \\[1.0pt]
\texttt{log\_agn\_tau\_\{xx\}th} & float & -- & $16^{\mathrm{th}}/50^{\mathrm{th}}/84^{\mathrm{th}}$ percentile of AGN torus dust optical depth \\[1.0pt]
\texttt{gas\_logz\_\{xx\}th} & float & $\log_{10}(Z_*/\mathrm{Z_\odot})$ & $16^{\mathrm{th}}/50^{\mathrm{th}}/84^{\mathrm{th}}$ percentile of Gas-phase metallicity \\[1.0pt]
\texttt{duste\_qpah\_\{xx\}th} & float & -- & $16^{\mathrm{th}}/50^{\mathrm{th}}/84^{\mathrm{th}}$ percentile of fraciton of PAH mass \citep{draine2007} \\[1.0pt]
\texttt{duste\_umin\_\{xx\}th} & float & -- & $16^{\mathrm{th}}/50^{\mathrm{th}}/84^{\mathrm{th}}$ percentile of the minimum radiation field for dust emission ($U_\mathrm{min}$)\\[1.0pt]
\texttt{log\_duste\_gamma\_\{xx\}th} & float & -- & $16^{\mathrm{th}}/50^{\mathrm{th}}/84^{\mathrm{th}}$ percentile of fraction of dust mass exposed to $U_\mathrm{min}$ \\[3.0pt]
\hline 

\end{tabular}
\end{table*}

\begin{table*}
\centering
\begin{tabular}{l c c p{10.5cm}}
\toprule 
\textbf{Field} & \textbf{Data type} & \textbf{Unit} & \textbf{Description} \\
\midrule \\[-6.0pt]
\multicolumn{4}{c}{\normalsize{\textbf{Absolute Magnitudes and Colors}}} \\[3.0pt] 

\texttt{M\_UV\_\{xx\}th} & float & -- &
$16^{\mathrm{th}}/50^{\mathrm{th}}/84^{\mathrm{th}}$ percentiles of the rest-frame absolute magnitude at $1500$\AA \\[1.0pt]

\texttt{M\_bessell\_U\_\{xx\}th} & float & -- &
$16^{\mathrm{th}}/50^{\mathrm{th}}/84^{\mathrm{th}}$ percentiles of the rest-frame absolute magnitude in the Bessell \texttt{U} band \\[1.0pt]

\texttt{M\_bessell\_B\_\{xx\}th} & float & -- &
$16^{\mathrm{th}}/50^{\mathrm{th}}/84^{\mathrm{th}}$ percentiles of the rest-frame absolute magnitude in the Bessell \texttt{B} band \\[1.0pt]

\texttt{M\_bessell\_V\_\{xx\}th} & float & -- &
$16^{\mathrm{th}}/50^{\mathrm{th}}/84^{\mathrm{th}}$ percentiles of the rest-frame absolute magnitude in the Bessell \texttt{V} band \\[1.0pt]

\texttt{M\_bessell\_R\_\{xx\}th} & float & -- &
$16^{\mathrm{th}}/50^{\mathrm{th}}/84^{\mathrm{th}}$ percentiles of the rest-frame absolute magnitude in the Bessell \texttt{R} band \\[1.0pt]

\texttt{M\_bessell\_I\_\{xx\}th} & float & -- &
$16^{\mathrm{th}}/50^{\mathrm{th}}/84^{\mathrm{th}}$ percentiles of the rest-frame absolute magnitude in the Bessell \texttt{I} band \\[1.0pt]

\texttt{M\_twomass\_J\_\{xx\}th} & float & -- &
$16^{\mathrm{th}}/50^{\mathrm{th}}/84^{\mathrm{th}}$ percentiles of the rest-frame absolute magnitude in the 2MASS \texttt{J} band \\[1.0pt]

\texttt{M\_twomass\_H\_\{xx\}th} & float & -- &
$16^{\mathrm{th}}/50^{\mathrm{th}}/84^{\mathrm{th}}$ percentiles of the rest-frame absolute magnitude in the 2MASS \texttt{H} band \\[1.0pt]

\texttt{M\_twomass\_Ks\_\{xx\}th} & float & -- &
$16^{\mathrm{th}}/50^{\mathrm{th}}/84^{\mathrm{th}}$ percentiles of the rest-frame absolute magnitude in the 2MASS \texttt{Ks} band \\[1.0pt]

\texttt{M\_sdss\_g0\_\{xx\}th} & float & -- &
$16^{\mathrm{th}}/50^{\mathrm{th}}/84^{\mathrm{th}}$ percentiles of the rest-frame absolute magnitude in the SDSS \texttt{g} band \\[1.0pt]

\texttt{M\_sdss\_r0\_\{xx\}th} & float & -- &
$16^{\mathrm{th}}/50^{\mathrm{th}}/84^{\mathrm{th}}$ percentiles of the rest-frame absolute magnitude in the SDSS \texttt{r} band \\[1.0pt]

\texttt{M\_sdss\_i0\_\{xx\}th} & float & -- &
$16^{\mathrm{th}}/50^{\mathrm{th}}/84^{\mathrm{th}}$ percentiles of the rest-frame absolute magnitude in the SDSS \texttt{i} band \\[1.0pt]

\texttt{UV\_color\_\{xx\}th} & float & -- & 
$16^{\mathrm{th}}/50^{\mathrm{th}}/84^{\mathrm{th}}$ percentile of rest-frame $\texttt{U}-\texttt{V}$ color \\[1.0pt]

\texttt{VJ\_color\_\{xx\}th} & float & -- & 
$16^{\mathrm{th}}/50^{\mathrm{th}}/84^{\mathrm{th}}$ percentile of rest-frame $\texttt{V}-\texttt{J}$ color \\[3.0pt]

\hline \\[-3.0pt]
\multicolumn{4}{c}{\normalsize{\textbf{Quality Flags}}} \\[5.0pt] 
{{\color{blue}\texttt{use\_phot}}}  & int & -- & Set to 1 if the galaxy is robustly detected ($\mathrm{SNR_{F444W}} \geq 3.0$ and $\mathrm{N_{bands, nircam}} \geq 6.0$) and its properties are robustly constrained (i.e., both \texttt{flag\_good\_redshift\_50} and \texttt{flag\_good\_mass\_dex08} are satisfied); 0 otherwise. \\ [1.0pt]
\texttt{flag\_spec-z}  & int & -- &
\parbox[t]{\linewidth}{
Set to 1 if the galaxy has a robust spectroscopic redshift; 0 otherwise.\\
\textbf{{\color{blue} Note: }} For these galaxies, an additional Prospector run was performed with redshift fixed to the spec-$z$ value, see the corresponding VAC table for those results :)} \\[1.0pt]

\texttt{flag\_MIRI}  & int & -- &
\parbox[t]{\linewidth}{
Set to 1 if the galaxy has at least one MIRI band included; 0 otherwise.\\
\textbf{{\color{blue} Note: }} For these galaxies, an additional Prospector run was performed with MIRI photometry included, see the corresponding VAC table for those results :)} \\[1.0pt]

\texttt{flag\_snr\_3} & int & -- & Set to 1 if $\mathrm{SNR}_\mathrm{F444W} \geq 3$, else 0 \\[1.0pt]
\texttt{flag\_snr\_5} & int & -- & Set to 1 if $\mathrm{SNR}_\mathrm{F444W} \geq 5$, else 0 \\[1.0pt]
\texttt{flag\_snr\_10} & int & -- & Set to 1 if $\mathrm{SNR}_\mathrm{F444W} \geq 10$, else 0 \\[1.0pt]
\texttt{flag\_Nbands\_3} & int & -- & Set to 1 if \texttt{Nbands} $\geq 3$, else 0 \\[1.0pt]
\texttt{flag\_Nbands\_5} & int & -- & Set to 1 if \texttt{Nbands} $\geq 5$, else 0 \\[1.0pt]
\texttt{flag\_Nbands\_10} & int & -- & Set to 1 if \texttt{Nbands} $\geq 10$, else 0 \\[1.0pt]
\texttt{flag\_Nbands\_15} & int & -- & Set to 1 if \texttt{Nbands} $\geq 15$, else 0 \\[1.0pt]
\texttt{flag\_Nbands\_20} & int & -- & Set to 1 if \texttt{Nbands} $\geq 20$, else 0 \\[1.0pt]

\texttt{flag\_good\_redshift\_10} & int & -- & Set  1 if the redshift is well constrained, defined as $(z_{84} - z_{16}) \leq 0.10 \times z_{50}$; otherwise 0 \\[1.0pt]
\texttt{flag\_good\_redshift\_20} & int & -- & Set to 1 if the redshift is well constrained, defined as $(z_{84} - z_{16}) \leq 0.20 \times z_{50}$; otherwise 0 \\[1.0pt]
\texttt{flag\_good\_redshift\_30} & int & -- & Set to 1 if the redshift is well constrained, defined as $(z_{84} - z_{16}) \leq 0.30 \times z_{50}$; otherwise 0 \\[1.0pt]
\texttt{flag\_good\_redshift\_40} & int & -- & Set to 1 if the redshift is well constrained, defined as $(z_{84} - z_{16}) \leq 0.40 \times z_{50}$; otherwise 0 \\[1.0pt]
\texttt{flag\_good\_redshift\_50} & int & -- & Set to 1 if the redshift is well constrained, defined as $(z_{84} - z_{16}) \leq 0.50 \times z_{50}$; otherwise 0 \\[1.0pt]

\texttt{flag\_good\_mass\_dex05} & int & -- & Set to 1 if the stellar mass posterior width $(\mathrm{84^{th}} - \mathrm{16^{th}})$ is less than 0.50 dex; otherwise 0 \\[1.0pt]
\texttt{flag\_good\_mass\_dex08} & int & -- & Set to 1 if the stellar mass posterior width $(\mathrm{84^{th}} - \mathrm{16^{th}})$ is less than 0.80 dex; otherwise 0 \\[1.0pt]
\texttt{flag\_good\_mass\_dex1} & int & -- & Set to 1 if the stellar mass posterior width $(\mathrm{84^{th}} - \mathrm{16^{th}})$ is less than 1 dex; otherwise 0 \\[1.0pt]
\texttt{flag\_good\_mass\_dex2} & int & -- & Set to 1 if the stellar mass posterior width $(\mathrm{84^{th}} - \mathrm{16^{th}})$ is less than 2 dex; otherwise 0 \\[1.0pt]
\texttt{flag\_good\_mass\_dex3} & int & -- & Set to 1 if the stellar mass posterior width $(\mathrm{84^{th}} - \mathrm{16^{th}})$ is less than 3 dex; otherwise 0 \\[3.0pt]
\hline \\[-3.0pt]
\multicolumn{4}{c}{\normalsize{\textbf{AGN \& LRD Flags}}} \\[6.0pt]
\texttt{flag\_AGN} & int & -- & Set to 1 if spectroscopically confirmed Type-I or Type-II AGN, else 0 \\[1.0pt]
\texttt{flag\_Type-I} & int & -- & Set to 1 if spectroscopically confirmed Type-I AGN \citep{ignas2026_type1}, else 0 \\[1.0pt]
\texttt{flag\_Type-II} & int & -- & Set to 1 if spectroscopically confirmed Type-II AGN \citep{scholtz2025_type2}, else 0 \\[1.0pt]
\texttt{flag\_LRD} & int & -- & Set to 1 if galaxies are in photometrically selected LRD samples from \citet{rinaldi2026_lrd}, else 0 \\[1.0pt]

\bottomrule \bottomrule
\end{tabular}
\end{table*}

\FloatBarrier

\clearpage
\clearpage

\section{Photometric Redshift Quality Assessment: \texttt{Prospector} and \texttt{EAZY}}
\label{sec: prospector v.s eazy}
In this subsection, we compare and discuss the redshift accuracy of \texttt{Prospector} and \texttt{EAZY} by comparing both to spectroscopic redshifts. We follow the same sample selection method as in the \texttt{Prospector} redshift quality assessment section (\autoref{sec: compare with spec} and \autoref{fig: zphot zspec all}). Specifically, we select galaxies with $\mathrm{N_{bands,\,NIRCam}} \geq 6$ and robust spectroscopic redshifts, yielding a sample of $16{,}320$ galaxies for the comparison. The \texttt{EAZY} redshifts are computed using the same \texttt{KRON\_CONV} aperture photometry.

\begin{figure*}
    \centering

    \includegraphics[width=0.49\linewidth]{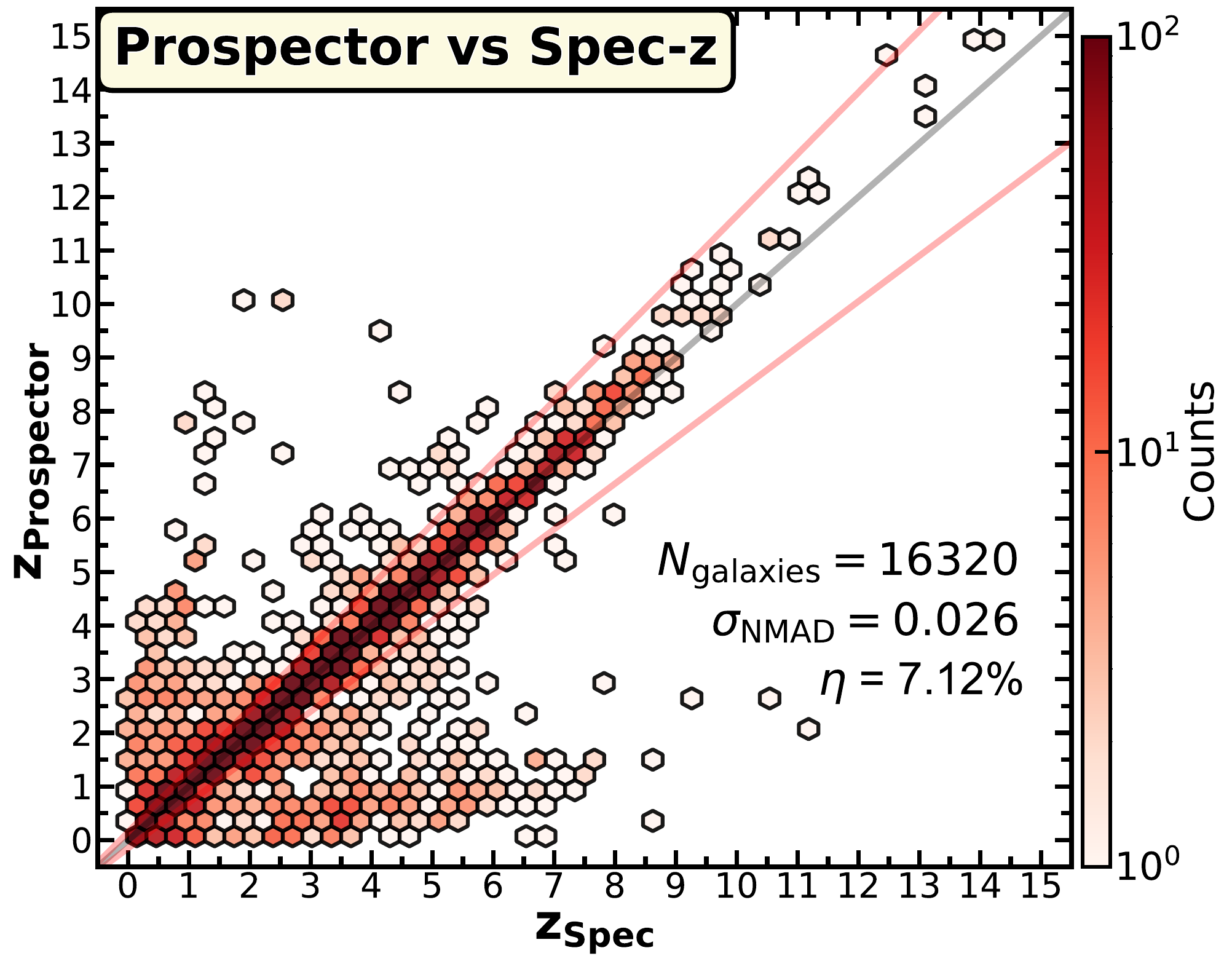}
    \hfill
    \includegraphics[width=0.49\linewidth]{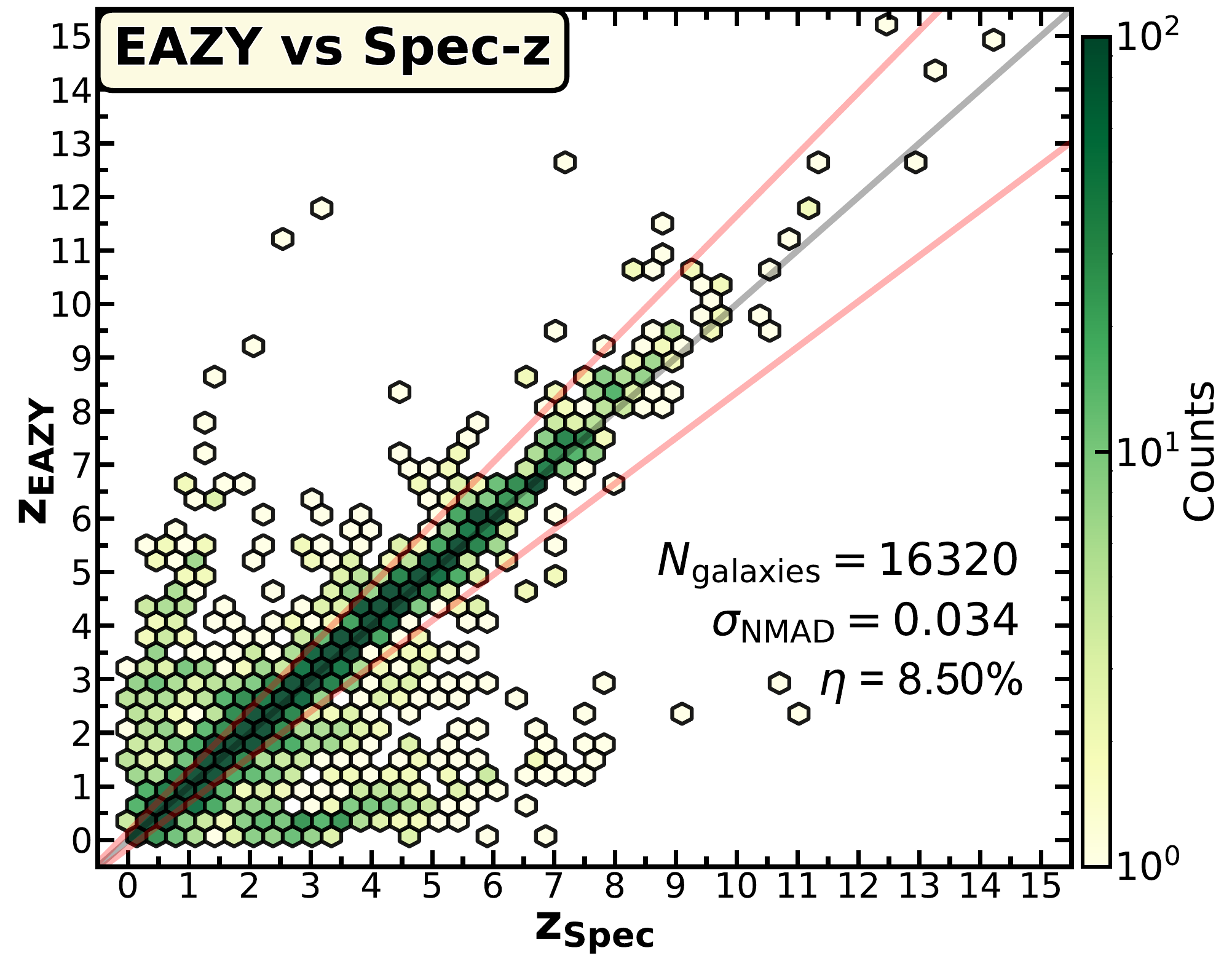}
    
    \vspace{0.8em}
    \includegraphics[width=0.49\linewidth]{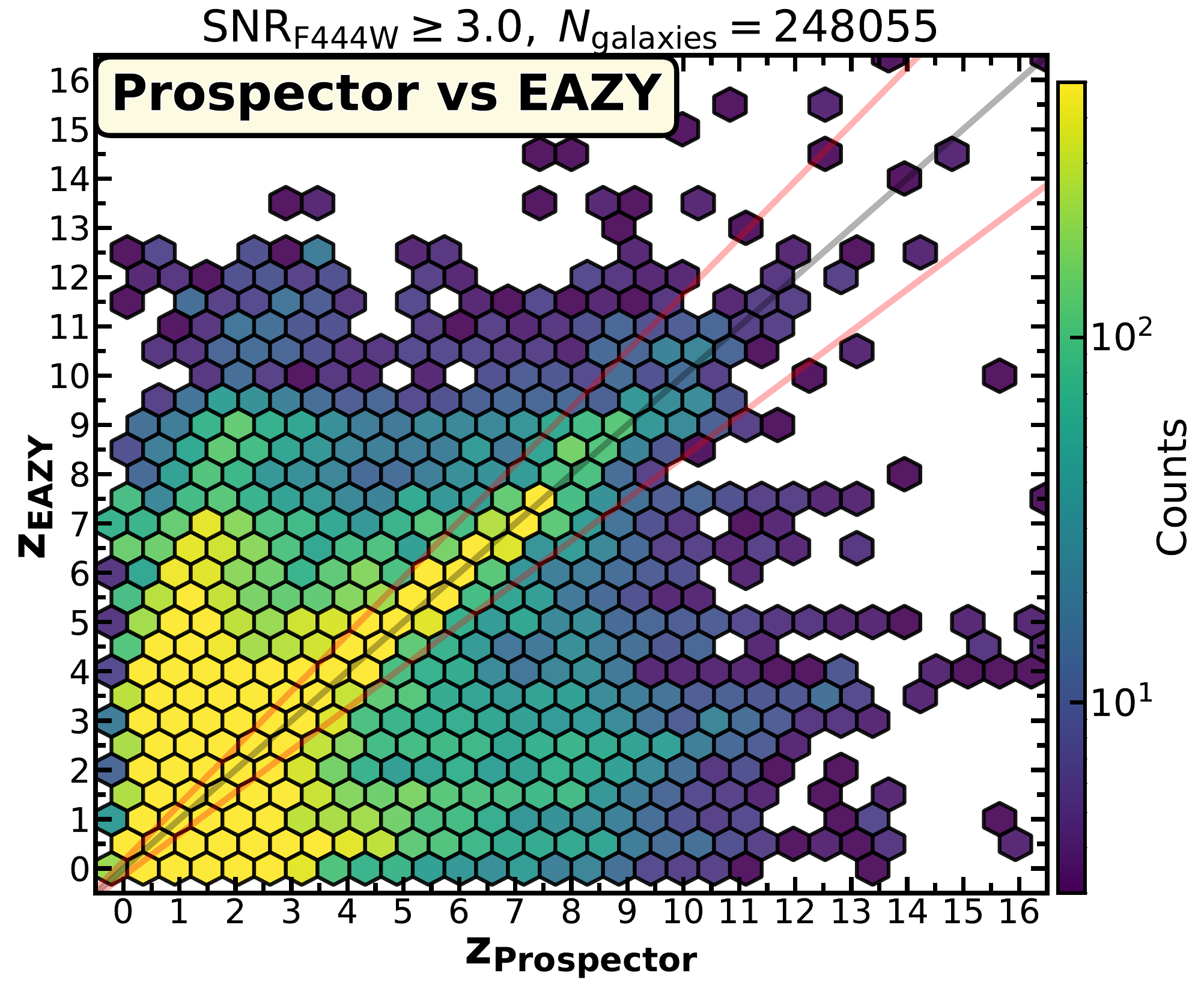}
    \hfill
    \includegraphics[width=0.49\linewidth]{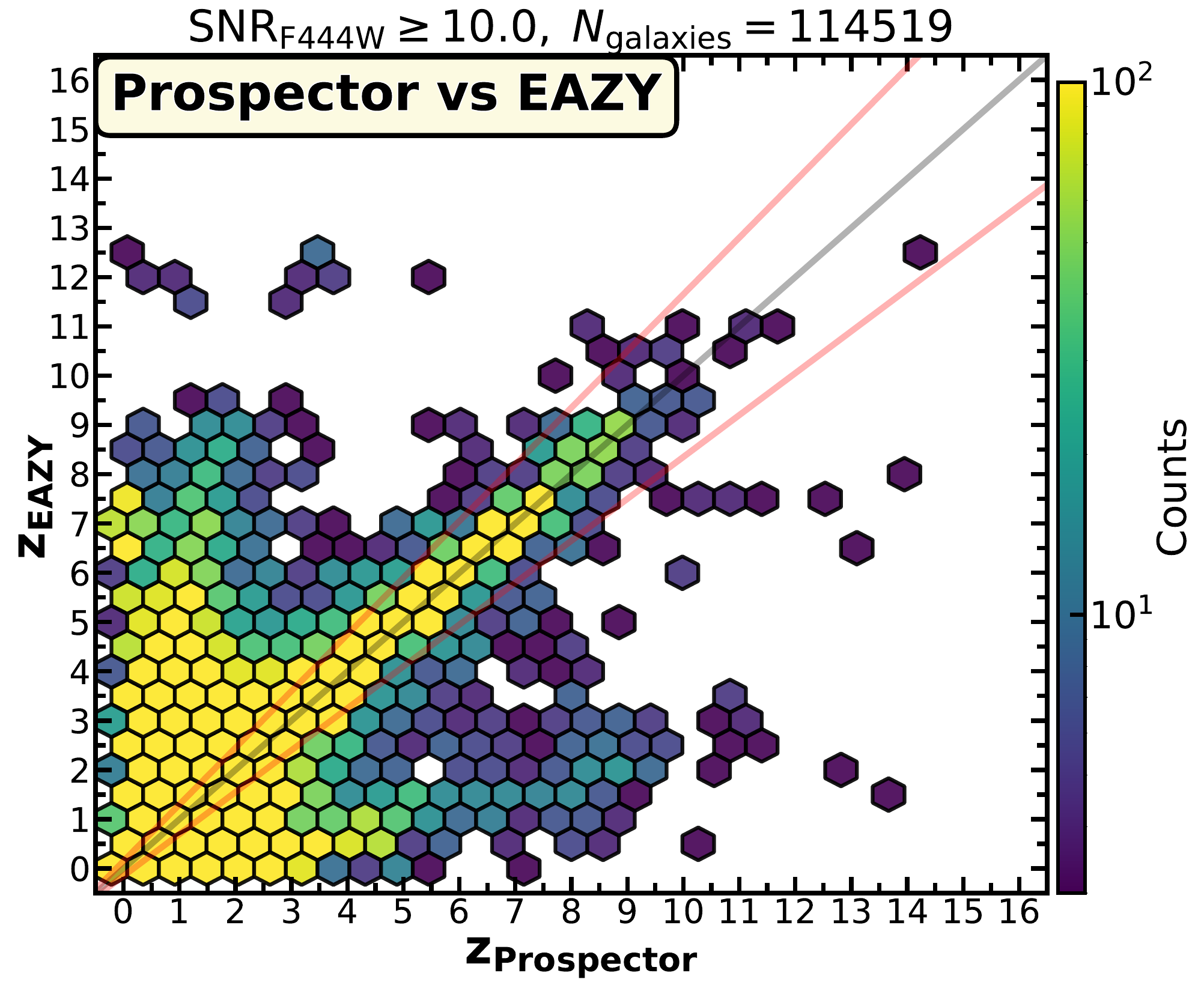}
    \caption{Redshift comparisons between \texttt{Prospector} and \texttt{EAZY}. \textbf{\textit{Top panels:}} We assess the redshift accuracy of \texttt{Prospector} and \texttt{EAZY} against spectroscopic redshifts for a sample of $16{,}320$ galaxies with $\geq 6$ NIRCam bands. \texttt{Prospector} achieves higher redshift accuracy. Hexbins are shown wherever at least 1 data point is present. \textbf{\textit{Bottom panel:}} Direct comparison between redshifts inferred by \texttt{Prospector} and \texttt{EAZY}. Hexbins are shown wherever at least 3 data points are present. \texttt{EAZY} systematically infers higher redshifts than \texttt{Prospector}. This discrepancy is not solely due to the misidentification of the Balmer break as the Lyman break, but also arises from a combination of factors, including differences in model assumptions, limitations of template fitting, sensitivity to photometric noise, and the adopted priors on free parameters.}
    \label{fig: prosp eazy redshift}
\end{figure*}

We present the redshift accuracy of \texttt{Prospector} and \texttt{EAZY} in the top two panels of \autoref{fig: prosp eazy redshift}. Hexbins are shown wherever at least 1 data point is present. From these panels, we find that \texttt{Prospector} achieves better accuracy, with $\sigma_{\rm NMAD}$ lower by a factor of 1.3 and an outlier fraction ($\eta$) lower by $1.38\%$ compared to \texttt{EAZY}. We also find that \texttt{Prospector} is less likely to infer galaxies at higher redshift (i.e., low-$z$ interlopers) compared to \texttt{EAZY}.

In the bottom panels of \autoref{fig: prosp eazy redshift}, we compare the inferred redshifts from \texttt{Prospector} and \texttt{EAZY} for all $248{,}055$ galaxies with $\mathrm{SNR_{F444W}} \geq 3.0$ and $114{,}519$ galaxies with $\mathrm{SNR_{F444W}} \geq 10.0$. Hexbins are shown wherever at least 3 data points are present. We find a prominent trend in which \texttt{EAZY} systematically infers higher redshifts than \texttt{Prospector}. This offset is most pronounced for galaxies with $z_{\texttt{Prospector}} < 3.0$, where a substantial fraction of galaxies could reach $z_{\texttt{EAZY}}$ as high as $\sim 8$.

We find that this discrepancy is not primarily driven by the degeneracy in which the Balmer break is misidentified as the Lyman break (and hence a higher inferred redshift). Instead, this scenario accounts for only a small fraction of the outliers. The dominant source of disagreement arises from differences in how spectral features are interpreted in the two frameworks. In many cases, \texttt{EAZY} identifies apparent breaks (either Lyman or Balmer), whereas in \texttt{Prospector} these features are instead attributed to strong emission-line contributions. Additionally, for galaxies with smoothly rising spectral energy distributions (e.g., those with older stellar populations and/or dust or AGN reddening), \texttt{Prospector} typically favors lower-redshift solutions. In contrast, \texttt{EAZY} can interpret segments of these continuum as spectral breaks, leading to systematically higher inferred redshifts.

These mismatches reflect a combination of factors, including differences in model flexibility, template limitations, sensitivity to photometric noise, and the adopted priors. Furthermore, the absence of MIR constraints in \texttt{EAZY} may contribute to this behaviour, as such data help to distinguish between continuum shapes and genuine spectral breaks.

To sum up, remarkably, \texttt{Prospector} achieves higher redshift accuracy than \texttt{EAZY}. While \texttt{EAZY} is primarily designed to infer redshifts, \texttt{Prospector} simultaneously constrains 18 free parameters, including redshift, stellar population properties, nebular emission, dust attenuation and emission, and MIR AGN contributions. Despite this substantially higher dimensionality, it still delivers improved redshift performance.

\section{Fitting a Linear Relation with Intrinsic Scatter}
\label{sec: linear with scatter fit}

In \autoref{sec: mass f444w}, we model the relationship between the \texttt{Prospector}-inferred stellar mass and the observed $\log_{10}(\mathrm{F444W}_{\mathrm{KRON\text{-}Conv}})$ as a function of $\log_{10}(\mathrm{F444W}_{\mathrm{KRON\text{-}Conv}})$ flux using a linear relation with intrinsic scatter. In this section, we describe the fitting methodology in detail.

Treating $\log_{10}(\mathrm{F444W}_{\mathrm{KRON\text{-}Conv}})$ as $x$, galaxy stellar mass as $y$, we parameterise the relation between $x$ and $y$ as
\begin{equation}
\label{eq: mass flux linear scatter}
y
= m \times \log_{10}\!\left(x\right)
+ b + \mathcal{N}\!\left(0, \sigma_{\mathrm{int}}^2\right),
\end{equation}
where $m$ and $b$ are the slope and intercept of the linear relation, respectively, and $\sigma_{\mathrm{int}}$ represents the intrinsic scatter about the best-fit line.

To simultaneously account for measurement uncertainties in both $x$ and $y$, as well as intrinsic scatter about the relation, we adopt the elegant likelihood formalism proposed by \citet{Hogg2010}. The log-likelihood is given by
\begin{equation}
\label{eq: likelihood}
\ln \mathcal{L}
= -\frac{1}{2}\sum_{i=1}^{N} \ln\!\left(\Sigma_i^2 + V\right)
- \frac{1}{2}\sum_{i=1}^{N} \frac{\Delta_i^2}{\Sigma_i^2 + V},
\end{equation}
where $\Delta_i$ is the orthogonal distance of the $i$-th data point $(x_i, y_i)$ from the model line,
\begin{equation}
\Delta_i = \hat{\mathbf{v}}^{\,T}
\begin{bmatrix}
x_i \\ y_i
\end{bmatrix}
- \frac{b}{\sqrt{1 + m^2}},
\end{equation}
and the unit vector normal to the line is
\begin{equation}
\hat{\mathbf{v}} = \frac{1}{\sqrt{1 + m^2}}
\begin{bmatrix}
-m \\ 1
\end{bmatrix}.
\end{equation}

\noindent The term $\Sigma_i^2$ represents the measurement uncertainty projected orthogonally to the model line,
\begin{equation}
\Sigma_i^2 =
\hat{\mathbf{v}}^{\,T}
\begin{bmatrix}
\sigma_{x,i}^2 & \sigma_{xy,i} \\
\sigma_{xy,i} & \sigma_{y,i}^2
\end{bmatrix}
\hat{\mathbf{v}},
\end{equation}
where we assume $\sigma_{xy,i} = 0$. The contribution from intrinsic scatter is
\begin{equation}
V =
\hat{\mathbf{v}}^{\,T}
\begin{bmatrix}
0 & 0 \\
0 & \sigma_{\mathrm{int}}^2
\end{bmatrix}
\hat{\mathbf{v}}.
\end{equation}

We sample the posterior distribution of the free parameters ($m$, $b$, and $\sigma_{\mathrm{int}}$) using the nested sampling code \texttt{Nautilus} \citep{lange2023}. We adopt broad, uniform priors:
$$
0 < m < 5,\quad -5 < b < 10,\quad 0 < \sigma_{\mathrm{int}} < 1.5.
$$


\bsp	
\label{lastpage}
\end{document}